\documentclass[12pt]{article}
\usepackage{amsmath}
\usepackage{mathtools}
\usepackage{amssymb} 
\usepackage{amsthm}
\usepackage{bbm}
\usepackage{graphicx,epsf}
\usepackage{epstopdf}
\usepackage{enumerate}
\usepackage{subcaption}
\usepackage{caption}
\usepackage{array}
\usepackage{natbib}
\usepackage{hyperref}
\usepackage{bm}
\usepackage{color}
\usepackage{fourier}

\newtheorem{remark}{Remark}
\newtheorem{proposition}{Proposition}
\DeclareMathOperator*{\argmax}{arg\,max}
\DeclareMathOperator*{\argmin}{arg\,min}
\DeclareMathOperator{\kmax}{k_{\mbox{max}}}
\DeclareMathOperator{\Var}{\mbox{Var}}
\DeclareMathOperator{\sd}{\mbox{sd}}
\DeclareMathOperator{\psd}{\mbox{psd}}
\DeclareMathOperator{\pmad}{\mbox{pmad}}
\DeclareMathOperator{\mad}{\mbox{mad}}
\DeclareMathOperator{\median}{\mbox{median}}
\DeclareMathOperator{\Gap}{\mbox{Gap}}

\author{Jakob Raymaekers$^{1*}$ \and Ruben H. Zamar$^2$}
\date{%
    $^1$ KU Leuven, Department of Mathematics,\\
		Celestijnenlaan 200B, 3001 Leuven, Belgium.\\%
    $^2$ University of British Columbia, Department of Statistics,\\
		Earth Sciences Building, 2207 Main Mall, Vancouver, Canada.\\[2ex]%
    \today
}

\begin{document}
\title{\Large Pooled variable scaling for cluster analysis}
\maketitle
		
\begin{abstract}
We propose a new approach for scaling prior to cluster analysis based on the concept of pooled variance. Unlike available scaling procedures such as the 
standard deviation and the range, our proposed scale avoids dampening the beneficial effect of informative clustering variables. We confirm through an extensive simulation study and applications to well known real data examples that the proposed scaling method is safe and generally useful. Finally, we use our approach to cluster a high dimensional genomic dataset consisting of gene expression data for several specimens of breast cancer cells tissue obtained from human patients.\\
\end{abstract}

\clearpage

\section{Introduction}

\noindent Every time cluster analysis is used to find homogeneous groups in the data, we face the issue of how to scale the variables. The choice of scaling (including the option of not scaling at all) has practical implications because in general clustering methods are not scale-invariant. Even those clustering methods that are scale-invariant, become scale-dependent when they are combined with variable selection or regularization. As a consequence, we may get a different data partition if the variables in the data are rescaled. Unfortunately, there is not a generally accepted way to scale the variables for clustering \citep{jain1988algorithms,jain2010}.\\

\noindent There are conflicting recommendations regarding the scaling of
the data in the clustering literature. For example, \cite{Milligan1988}
- often cited as the main benchmark study of this topic in the context of
hierarchical clustering - recommend scaling the variables by their range.
\cite{Steinley2004a} came to the same conclusion using k-means. On the other
hand, \cite{Vesanto2001} recommends the use of the standard deviation for k-means clustering. \cite{Schaffer1996} studies clustering on real data
examples and argues against scaling with the range or the standard deviation. \cite{Stoddard1979} also argued against the use of the standard deviation
in the analysis of laboratory procedures.\\

\noindent While scaling may not always be necessary, it seems that in general the most reliable approach is to use some sort of scaling for two main reasons. The first reason is that scaling gets rid of the measurement scales of the variables. These measurement scales may have a strong influence on the clustering results, to the extent that a single very large variable can solely determine the whole clustering outcome. Furthermore, it can be of practical importance to get rid of measurement scales, e.g. a variable measuring ``height'' should have the same effect on the clustering procedure when measured in centimeters or in meters. The second reason is that scaling becomes practically mandatory in the context of high-dimensional clustering with variable selection. Variable selection is usually achieved through the addition of a penalization term to the objective function. This penalization is typically not scale invariant and thus yields different outcomes depending on the variables' relative sizes. Therefore, without scaling, the penalty acts differently on each variable which will often lead to ineffective variable selection. For these reasons, we consider feature scaling as a necessary pre-processing step in cluster analysis.
\\

\noindent Though the arguments in favor of scaling before cluster analysis are clear, the issue of how to scale is delicate. The reason is that there are two types of variables: informative variables and noise variables. Informative variables help to separate clusters in the data whereas noise variables do not. If an informative variable is scaled with a very large scale, it will be compressed to a small size and thus there is a risk that much of its clustering power will disappear. In contrast, if a noise variable is scaled with a very small scale, it can potentially be blown up to where it solely determines the clustering outcome. Note that this issue is even more pronounced in high-dimensional datasets as they typically contain many noise variables.\\

\noindent Interestingly, most of the commonly used scale estimators such as the standard deviation and the range have a tendency to yield large scale estimates on potentially very informative variables. More precisely, these estimators may produce unduly large scales for a variable with a multimodal distribution because of the dispersion between the groups. Hence a drawback of feature scaling by these estimates is the possible undesirable reduction of the relative clustering importance of features that clearly separate different groups. While these features are not guaranteed to reveal the whole clustering structure of the data, they show good potential and should be treated carefully.\\

\noindent The discussion so far suggest that there is a need for further study of both the effect of scaling and the best choice of scale estimator in cluster analysis. Despite the considerable potential impact of scaling on cluster outcomes, papers on this topic are scant, far apart, and lack consensus.\\

\noindent In Section 2 we introduce two new scale estimators specially designed to scale variables before clustering: the {\it pooled standard deviation} and the {\it pooled absolute deviation}. 
By using pooled scale estimators, we aim to scale variables without destroying their clustering power if they have any. If the marginal distribution of a feature shows evidence of clustering, our proposed scaling will enhance its clustering importance. 
If the marginal distribution of a feature $X$ does not show evidence of clustering, but the joint distribution of some subset of features including $X$ carries important clustering information, our proposed scaling approach will have a neutral effect on this variable. The proposed scaling approach can be used as pre-processing before applying any clustering method, including those with variable selection and subspace clustering, as illustrated with several examples.
To calculate our proposed scale, we first run k-means clustering on each of the variables. In order to choose the number of clusters used to estimate the pooled scales, we use the Gap statistic \citep{Tibshirani2001} with a sped-up bootstrapping procedure that bypasses the otherwise unaffordable computational burden of this approach.\\

\noindent 
 The rest of the paper is organized as follows. The complete algorithm for scaling all features in a dataset is given in Section \ref{sec:algo}.
In Section \ref{sec:sim} we compare the new scale estimators with existing scaling procedures in an extensive simulation study. In Section \ref{sec:realdata} we show an application of the pooled scale estimators prior to the hierarchical clustering of gene expressions of breast cancer sample tissues (additional real-data examples are presented in the Supplementary Material). Finally, section \ref{sec:conclusion} concludes.

\section{Methodology}
\subsection{Pooled scale estimators}

Our starting point is the univariate k-means clustering. Assume we have a univariate data set $x_1,\ldots, x_n$. The well-known k-means clustering looks for the $k$ cluster centers which minimize the squared deviations for every point in the data set to the closest cluster center. More precisely, the vector of cluster centers is defined as:
\begin{equation*}
\bm \mu = (\mu_1, \ldots, \mu_k) = \argmin_{\bm t =(t_1, \ldots, t_k)}{S_k(\bm t)},
\end{equation*}
where 
\begin{equation}\label{eq:kmeansscale}
S_k(\bm t) = \sqrt{\frac{1}{n} \sum_{i=1}^{n}{d_{2, i}(\bm t)}}
\end{equation}
and $\displaystyle d_{2, i}(\bm t) = \min_{1 \leq j \leq k}{||x_i - t_j||_2^2}$. Note that $S_k( \bm \mu)$ can be interpreted as an estimator of scale. In particular, if $k= 1$, then $\bm \mu$ is the classical sample mean and $S_1(\bm \mu)$ reduces to the classical standard deviation. If $k>1$, the squared scale can be interpreted as a pooled variance of the points around their cluster centers. As an example, suppose that $k = 2$ and that the sets $C_1$ and $C_2$ contain the indices of the points in the two clusters. Moreover, let $|A|$ represent the number of elements in the set $A$, so that $|C_1|+|C_2|=n$. We then have $S_2^2(\bm \mu) = \frac{1}{n} \sum_{i=1}^{n}{d_{2,i}(\bm \mu)} = \frac{1}{n} \sum_{i \in C_1}{||x_i - \mu_1||_2^2} + \frac{1}{n} \sum_{i \in C_2}{||x_i - \mu_2||_2^2}= \frac{|C_1|}{n}  \Var(C_1) + \frac{|C_2|}{n} \Var(C_2)$, where $\Var(C_j)$ denotes the sample variance of the elements belonging to cluster $j$. We thus obtain a weighted mean of the within-cluster variances, with weights proportional to the number of observations in each cluster.\\

\noindent Our idea is to use $S_k$, with an appropriate (variable depending) value of $k$, for scaling the variables prior to the application of a clustering procedure. We will refer to this scale estimator as the \textit{pooled standard deviation}. The intuition for this scale estimator is the following. If a certain variable appears to not separate any clusters in the data, we consider the variability of this variable to be uninformative and we use the largest scale, $S_1$ (i.e. the classical standard deviation), to scale it before clustering. However, when a variables does seem to separate for example $k$ clusters, then $S_k$ will tend to be relatively small compared with $S_1$, and using $S_k$ in that case will avoid dampening the variability (i.e. information) in this variable. This way, we hope to preserve as much of the clustering information in each variable as possible, while still making the variables unitless. \\

\noindent As an alternative to k-means, k-medians clustering can also be used for the scaling of the variables.
Using the notation introduced above, the vector of cluster centers in k-medians is defined as:
\begin{equation*}
\bm \mu = (\mu_1, \ldots, \mu_k) = \argmin_{\bm t =(t_1, \ldots, t_k)}{M_k(\bm t)},
\end{equation*}
where 
\begin{equation*}
M_k(\bm t) = \frac{1}{n} \sum_{i=1}^{n}{d_{1, i}(\bm t)}
\end{equation*}
and $\displaystyle d_{1, i}(\bm t) = \min_{1 \leq j \leq k}{|x_i - t_j|}$. The cluster centers now correspond to the median of the observations in each cluster. Note that once again, $M_k$ can be interpreted as an estimator of scale. If $k= 1$, then $\bm \mu$ is the classical sample median and $M_1(\bm \mu)$ is the mean absolute deviation (from the median). If $k>1$, $M_k$ can be interpreted as the pooled mean absolute deviation of every point around its cluster center, where the pooling is done by a weighted average with weights determined by the number of observations in each cluster. We will call $M_k$ the \textit{pooled mean absolute deviation}. The pooled mean absolute deviation is expected to be more robust against outliers compared with the pooled standard deviation. This is consistent with the results of the simulation study of Section \ref{sec:sim}.\\

\begin{remark}\label{remark:otheroptions}
Since we are using univariate k-means and k-medians to obtain the pooled standard deviation and pooled mean absolute deviation, it is worth noting that there is an algorithm which guarantees the convergence to a global optimum instead of a local one. This algorithm uses dynamic programming, runs in $\mathcal{O}(kn^2)$ time and is implemented in the \textsf{R}-package \texttt{Ckmeans.1d.dp}, see \cite{Wang2011Ckmeans}.
\end{remark}

\subsection{Determining k}

\noindent An important question regarding our proposed pooled scale estimators is how to choose the appropriate number of clusters $k^*$ for each variable in the dataset. There is a vast literature on
how to choose the appropriate number of clusters in cluster analysis. Relevant comparative studies giving a good overview of the most common techniques are found in \cite{milligan1985examination}, \cite{halkidi2001clustering}, \cite{maulik2002performance}, \cite{brun2007model}, and \cite{Arbelaitz2013}. In the setting of pooled scale estimators, we need a criterion that is relatively fast to compute and not too sensitive to spurious clusters. Most importantly, we need a criterion that can distinguish between the ``null-case'' of one cluster versus the alternative case of two or more clusters. This is an important point, since especially in high dimensional datasets, we expect that many variables may not have interesting information for clustering the data and thus should be scaled by the largest scale, i.e. the standard deviation or mean absolute deviation. Note that this requirement makes many popular cluster validation indices unsuitable for us, since many of them do not yield a reasonable comparative assessment of the one-cluster-case.\\ 

\noindent A well known criterion that satisfies our needs is the Gap statistic \citep{Tibshirani2001,hastie2009elements}.
Given a certain clustering algorithm and the resulting partition of the data, the gap statistic works as follows. For each value of the number of clusters under consideration, the ``tightness'' of the clusters in the found partition is compared with the tightness of clusters obtained by clustering random datasets using the same algorithm. If this difference is large for a given number of clusters, it indicates ``stronger than random'' clustering structure and the gap statistic will pick that number of clusters as the true number. A mathematical description of the gap statistic is as follows. 
 Suppose we have clustered the data into $k$ clusters, $C_1, \ldots, C_k$, where $C_j$ denotes the indices of the observations in cluster $j$. Let $W_k$ be the sum of the within-cluster sums of squares around their corresponding cluster means, i.e. $W_k = \sum_{j=1}^{k}{ \sum_{i \in C_j}{(x_i - \bar{x}_j)^2}}$ where $\bar{x}_j = \frac{1}{|C_j|}\sum_{i \in C_j}{x_i}$ is the mean of the observations in cluster $j$. In order to identify the number of clusters, the value of $\log\left(W_k\right)$ is compared to its expected value $E_n^*\left[\log\left(W_k\right)\right]$ under a uniform reference distribution on the range of the dataset. If this value deviates too much from its expected value under a uniform distribution, it indicates the existence of clusters in the data. The intuition for the comparison with the uniform distribution is that it is the distribution which is most likely to generate spurious clusters (within the family of log-concave densities) and will thus on average provide the strongest evidence against the alternative hypothesis.\\

\noindent In principle, the reference distribution of $\log\left(W_k\right)$ is determined by generating bootstrap samples from the uniform distribution. As we would like scale every variable in the dataset, this appears to lead to a prohibitive computational cost which would scale poorly with the number of variables. Fortunately, there is an efficient way to bypass this hindrance based on the results of proposition 1 below.\\

\noindent In our case, the clusters are coming from the k-means (or k-medians) clustering algorithm. This means that for a given clustering $C_1, \ldots, C_k$, we have
\begin{equation*}
W_k = \sum_{r=1}^{k}{d_{2, i}(\mu_1, \ldots, \mu_k)} = n \; S_k^2.
\end{equation*}
in the computation of the pooled standard deviation. For the pooled mean absolute deviation, we redefine $W_k$ such that it corresponds to the pooled within-cluster sum of absolute deviations around the cluster medians:
\begin{equation*}
W_k = \sum_{r=1}^{k}{d_{1, i}(\mu_1, \ldots, \mu_k)} = n \; M_k.
\end{equation*}
Therefore, in our setting the gap statistic will be large when there is a ``significant difference'' in the estimated pooled scale compared with the pooled scaled estimate on data coming from a uniform distribution. \\

\noindent In order to estimate $E_n^*\left[\log\left(W_k\right)\right]$ and $\mbox{Var}_n^*\left[\log\left(W_k\right)\right]$, we apply $k$-means to $B$ bootstrap samples of size $n$. The mean of these samples serves as the estimate for $E_n^*\left[\log\left(W_k\right)\right]$ and the appropriate scale, which accounts for the simulation error in $E_n^*\left[\log\left(W_k\right)\right]$, is then the standard deviation of the bootstrap samples multiplied by $\sqrt{1+ 1/B}$.\\

\noindent We now turn to speeding up the bootstrapping procedure for scaling all the variables in a dataset. The speed-up is achieved by exploiting the fact that we are in the univariate setting and by using the properties of the proposed scaling methods \noindent established in the following proposition.

\begin{proposition}
Let $ \bm x = x_1, \ldots, x_n$ be a sample of univariate observations and let $C_1, \ldots, C_k$ be a partition of $\bm x$ resulting from solving the k-means clustering problem. Denote the value of the objective function with $S_k$ as in Equation \ref{eq:kmeansscale}. Let $s>0$ and $t \in \mathbb{R}$ and consider $\bm z = z_1 ,\ldots, z_n$ where $z_i = (x_i - t) / s$ for $i = 1,\ldots, n$. We then have: 
\begin{enumerate}

\item Shift and Scale invariance of k-means clusters:\\
$C_1, \ldots, C_k$ is a solution to the k-means clustering problem on $\bm z$.
\item Shift invariance and Scale equivariance of k-means objective function:\\
The value of the objective function of this clustering is $S_k / s$.
\item Shift and Scale invariance of gap statistic:\\
The number of clusters selected by the gap statistic is the same for $\bm x$ and $\bm z$.
\end{enumerate}
\end{proposition}

\noindent The exact same result holds for clustering with k-medians. These results are intuitive and follow from the fact that Manhattan and euclidean distances are scale equivariant. We give the proof in the Supplementary Material. Using this proposition, it becomes clear that we need to bootstrap the reference distribution of $\log\left(W_k\right)$ only once, instead of for every variable separately. The reason is that we can first rescale each variable in the dataset by the range, so that all of them have the same range of length one. Now the reference distribution of $W_k$ is the same for each of these variables, allowing us to bootstrap once from the uniform distribution on $[0,1]$ to obtain the reference distribution for all variables. Note that it does not matter on which interval of length 1 the variable takes its values, since the whole procedure is shift invariant as well.

\begin{remark}
The uniform distribution is not the only possible reference distribution one can use. Another option briefly suggest by \cite{Tibshirani2001} is to use log-concave density estimation, which is possible for univariate distributions. This would yield a fit of a log-concave density to every variable in the data from which bootstrap samples can then be drawn. While this takes into account the individual distributions of every variable, the drawback is that in this case we have to take the slow approach of generating separate reference distributions for every variable.
\end{remark}

\begin{remark}
Instead of the gap statistic, the jump statistic \citep{Sugar2003} may be used as an alternative. In the univariate setting, the jump statistic considers the ``distortions'' $\widehat{d}_k = S_k^2$ for several numbers of clusters $k$. They then define the jumps as $J_k = \widehat{d}_k^{-1/2} - \widehat{d}_{k-1}^{-1/2}$, where $\widehat{d}_0 \equiv 0$. Finally, the number of clusters is estimated by taking $K^* = \argmax_k{J_k}$. One advantage of the jump statistic over the gap statistic is that it faster to compute, since we do not need to bootstrap a reference distribution. In our simulations however, the gap statistic performed better.
\end{remark}

\section{Algorithm}\label{sec:algo}
We are now ready to describe the procedure we propose for scaling a dataset prior to clustering. For a p-variate dataset $X_1, \ldots, X_p$ with $n$ observations per variable, we apply the following steps to scale all the variables:\\

\noindent\fbox{%
    \parbox{0.93\columnwidth}{%
\begin{enumerate}
\item Generate $B$ bootstrap samples of size $n$ from the uniform distribution on $[0, 1]$ and cluster each of them using k-means. Retain all the values $W_{k, b}^*$ for $k = 1, \ldots, \kmax$ and $b = 1, \ldots, B$. Denote with $m_k = (1/B)\sum_b \log\left(W_{k, b}^*\right)$, $\mbox{sd}_k = \left[(1/B)\sum_b{\left(\log\left(W_{k, b}^*\right) - m_k\right)^2}\right]^{1/2}$, the estimates for the location and scale of $\log\left(W_k\right)^*$ respectively. Finally, put $s_k =  \sqrt{1+ 1/B} \; \mbox{sd}_k$.

\item For all variables $X_j$, $j = 1,\ldots, p$, do:
\begin{enumerate}
\item Rescale the variable with its range to obtain $Z_j = X_j / r_j$, where $r_j =  \mbox{range}\left(X_j\right)$.
\item Cluster $Z_j$ using k-means for $k =1,...,\kmax$ and retain the values $W_{k, j} = n \;S_{k, j}^2$.
\item Calculate the values of the gap statistic: $\Gap_j\left(k\right) = m_k - \log\left(W_{k, j}\right)$.
\item Choose the number of clusters $k^*$, by setting\\ $k^* = \mbox{ smallest k such that } \Gap_j\left(k\right)\geq \Gap_j\left(k+1\right) - c \;s_{k+1}$.
\item Rescale the value of the objective function of the appropriate $k$, $r_j \; S_{k^*, j} \; $, and use this pooled standard deviation to scale $X_j$.
\end{enumerate}
\end{enumerate}
}
}\\

\noindent The constant $c$ in the above procedure controls the rejection of the null model. As $c$ goes up, it becomes less likely to reject the null model of zero clusters. A default value is $c = 1$, which works well according to \cite{breiman1984} and \cite{Tibshirani2001}. For the number of bootstrap samples, a default of $B = 1000$ yields almost no variance in the resulting scale estimates in our experience. Replacing k-means with k-medians yields the scaling procedure with $M_k$, the pooled mean absolute deviation.\\

\noindent\textbf{Example 1: Fisher's Iris data\\}
\noindent As an illustrative example we consider Fisher's well-known Iris dataset \citep{Fisher1936}, collected by \cite{Anderson1935}. This dataset contains fifty samples from each of three types of iris: Iris setosa, versicolor, and virginica. Each flower is described by four variables which describe the dimensions of its sepal and petal. Table \ref{table:iris} illustrates the effect of scaling with the proposed pooled standard deviation on this data and compares with the standard deviation and the range. For the first two variables, the pooled standard deviation is equal to the standard deviation, because these variables do not seem to clearly distinguish any groups in the data. However variables three and four seem to both distinguish two clear groups, resulting in significantly lower pooled standard deviations. The last row of the table reports the adjusted rand index (ARI) \citep{Hubert1985} with respect to the true classification when performing k-means with $k = 3$ on the dataset after scaling. The ARI takes on values between -1 and 1, where an ARI of 1 indicates perfect agreement between partitions and the lower the ARI, the higher the disagreement between the partitions. It is clear that scaling with the pooled standard deviation gives better results than scaling with the standard deviation or the range, since neither of these take into account the individual separative power of the variables. As a reference, we mention that k-means clustering without scaling gives an ARI of 0.73.

\begin{table}[h!]
\begin{center}
\begin{tabular}{>{\centering\arraybackslash} m{4cm} >{\centering\arraybackslash} m{2cm} >{\centering\arraybackslash} m{2cm}  >{\centering\arraybackslash} m{2cm}}
\hline
variable & standard deviation & range & pooled standard deviation \\ 
\hline
\includegraphics[width=0.25\textwidth]{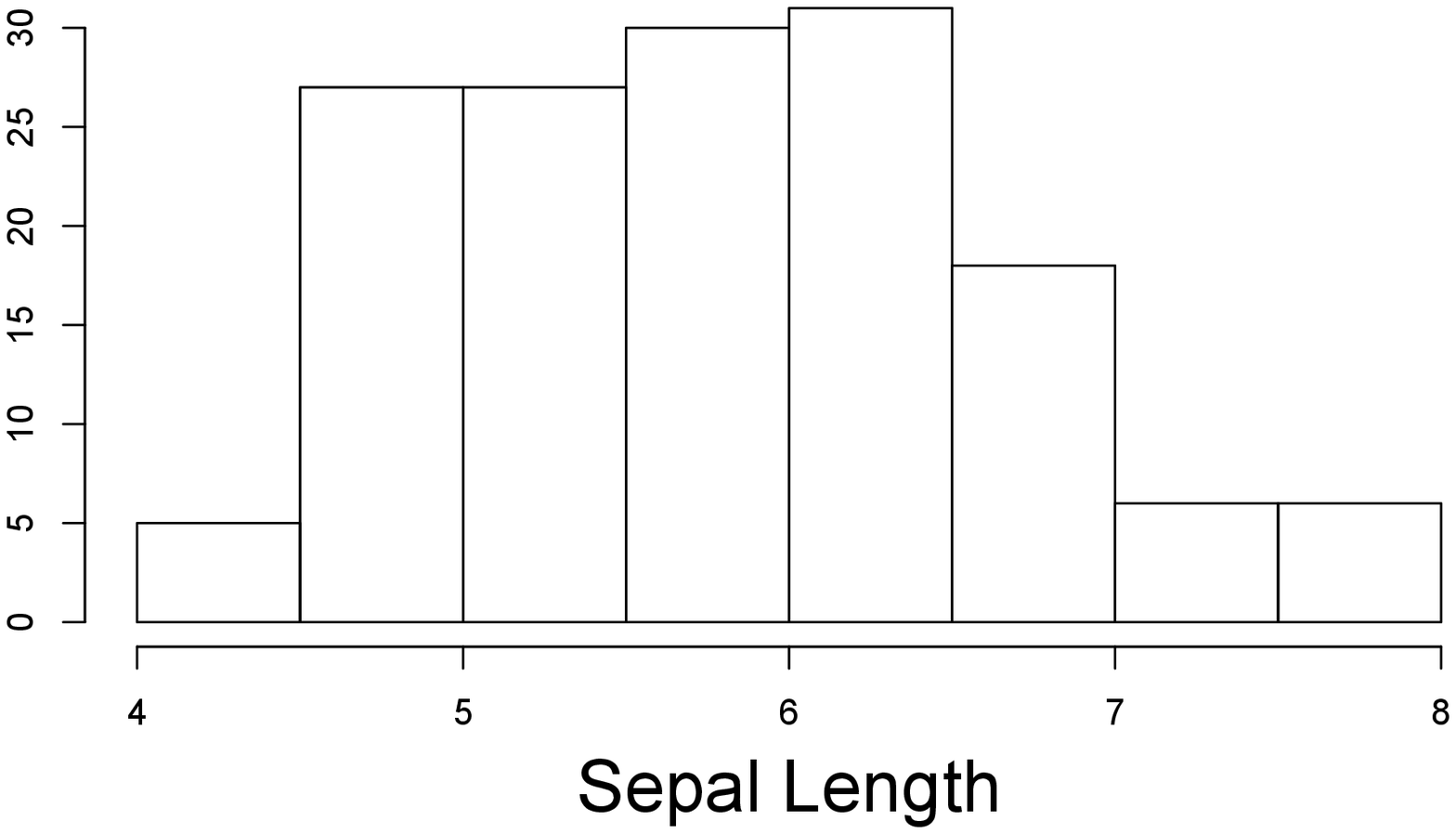} & 0.83 & 3.6 & 0.83\\
\includegraphics[width=0.25\textwidth]{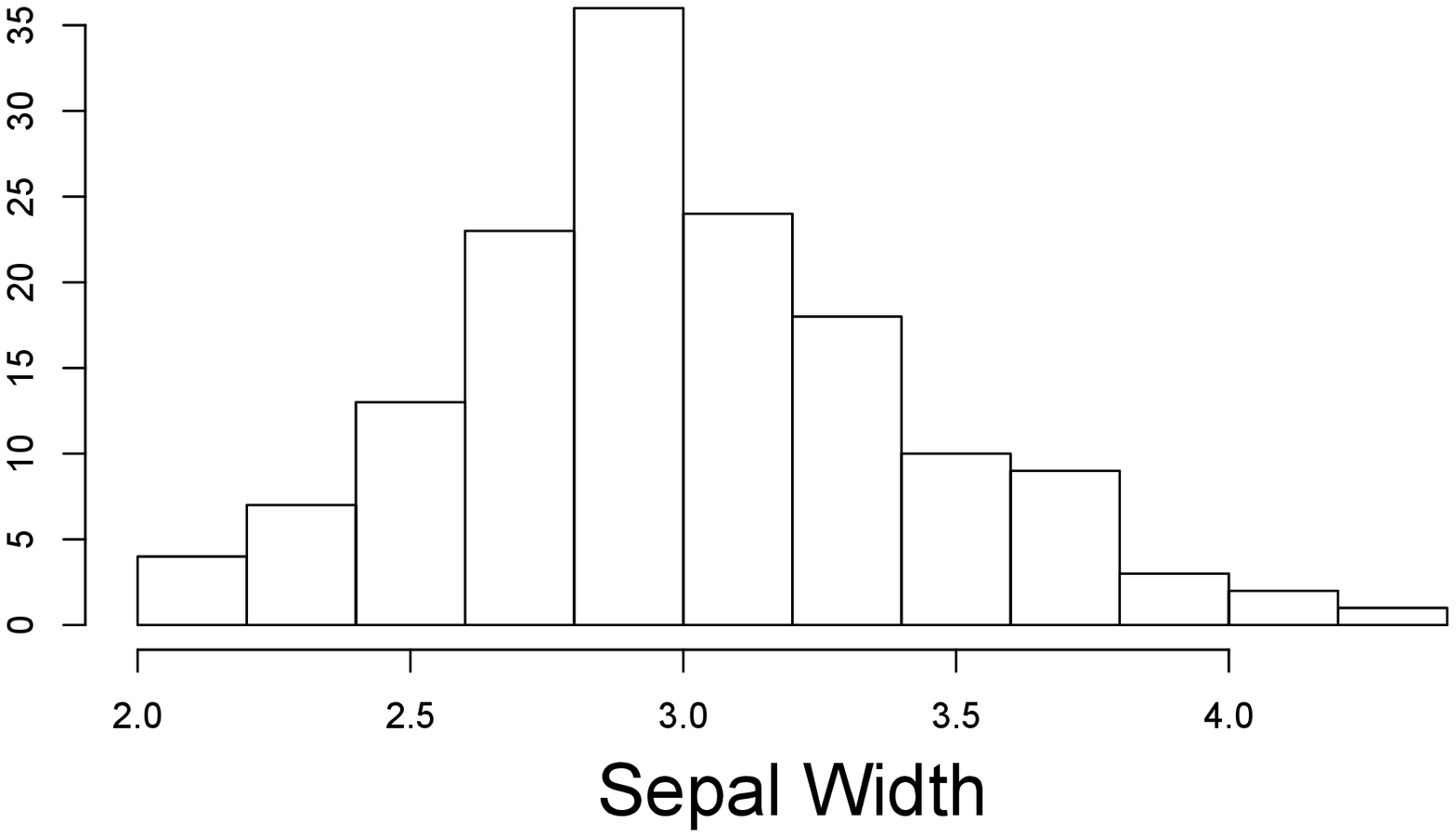} & 0.44 & 2.4 & 0.44\\
\includegraphics[width=0.25\textwidth]{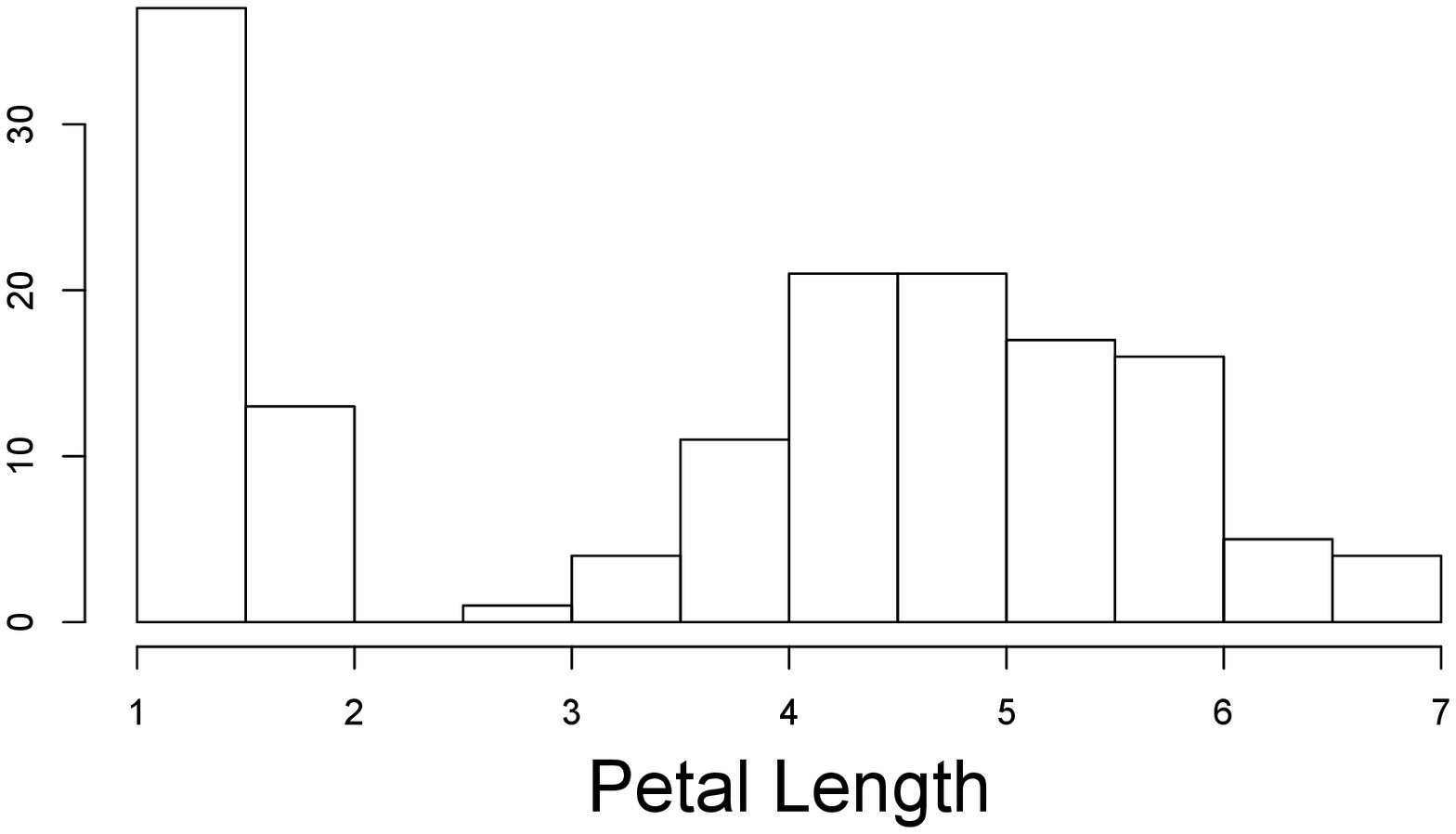} & 1.77 & 5.9 & \textbf{0.40}\\
\includegraphics[width=0.25\textwidth]{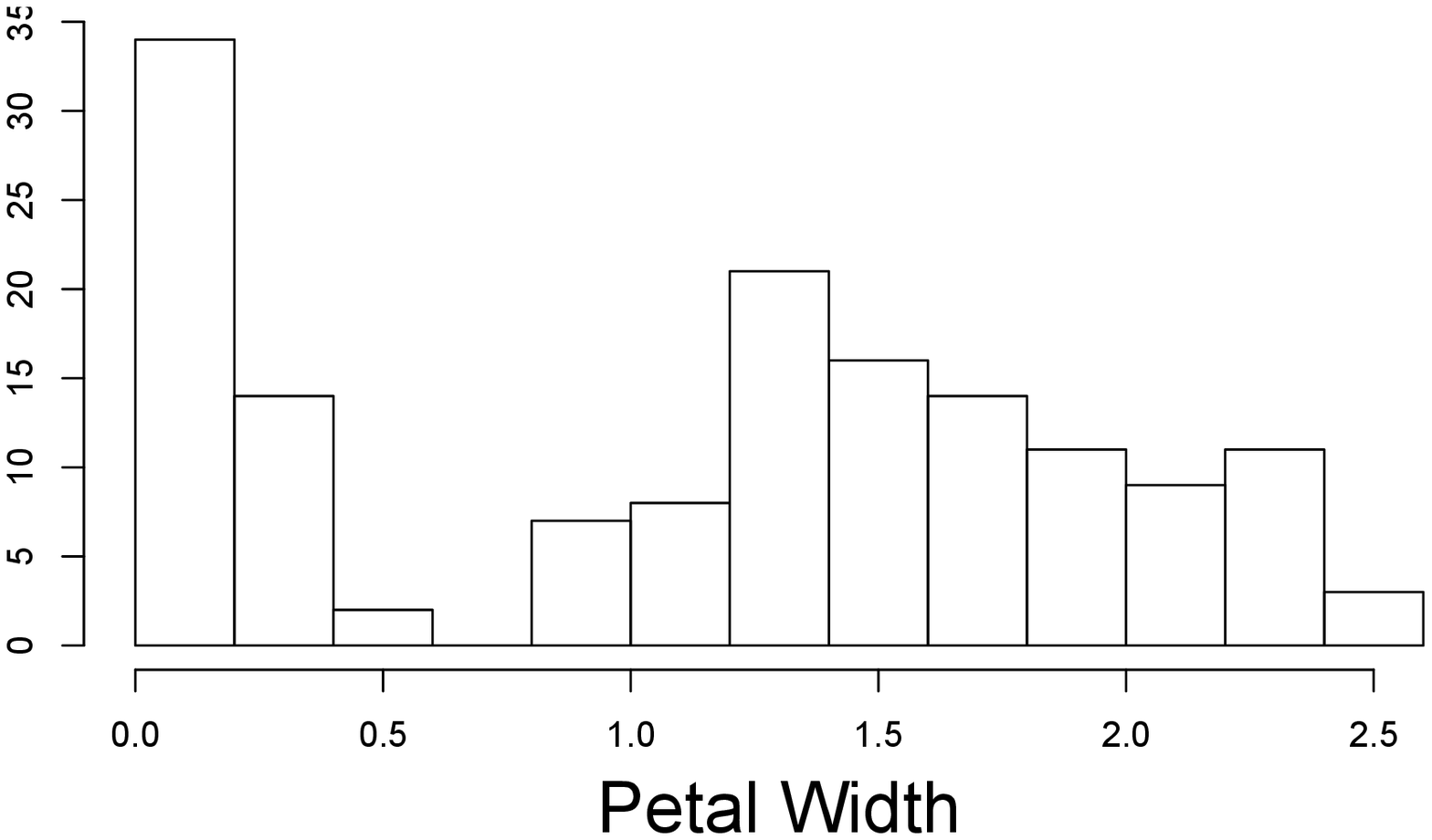} & 0.76 & 2.4 & \textbf{0.18}\\
\hline
\\[-1em]
k-means ARI & 0.62 & 0.72 &	\textbf{0.89} \\
\end{tabular}
\end{center}
\caption{The effect of variable scaling on Fisher's Iris data. The pooled standard deviation produces smaller scales for variable 3 and 4, resulting in a higher ARI when clustering the data with the k-means algorithm after scaling.}
\label{table:iris}
\end{table}

\section{Simulation study}\label{sec:sim}
The most well-known comparative study on the scaling of variables in clustering is arguably the one by \cite{Milligan1988}, building on \cite{Milligan1985}. Recently, \cite{Qiu2006} used the design of this simulation study as a basis for a new algorithm to generate clusters with a specified degree of separation. The R package \texttt{clusterGeneration} \citep{Qiu2015} contains an implementation of their algorithm and it will be the basis of our simulation study.\\

\noindent We compare the following types of scaling in our simulation study:
\begin{enumerate}
\item No scaling
\item The standard deviation: $\sd = \frac{1}{n - 1}\sqrt{\sum_{i=1}^{n}{(x_i - \bar{x})^2}}$
\item The range: $\mbox{range} = x_{(n)} - x_{(1)}$
\item The mean absolute deviation: $\mad = \frac{1}{n - 1}\sum_{i=1}^{n}{|x_i - \median_i{x_i}|}$
\item The pooled standard deviation: $\psd = S_{k^*}$
\item The pooled mean absolute deviation: $\pmad = M_{k^*}$
\end{enumerate}

\noindent In order to get a complete picture of the different scaling methods, we perform an extensive simulation study. Each generated dataset has an even number of clean variables which we vary between 2 and 10. The clusters are generated from the multivariate standard normal distribution. We consider equally sized clusters of size 100 each. The degree of separation between the clusters is either separated (0.21) or well-separated (0.34), see \cite{Qiu2006}. We then add a percentage of noise variables to the dataset varied between 0 and 2000 \% of the number of clean variables. The noise variables can either be multivariate standard normal or uniform over the range of the clean variables. The uniform noise variables are generated by adding a small gaussian perturbation to an equally spaced grid over the range of the signal variables to ensure that they don't have any separative power. We did the simulation both on clean data as well as contaminated data. For the contaminated data, 5 \% of the observations of each of the signal variables are replaced with points sampled randomly from the uniform distribution on $\left[\overline{X}_j - 4 \; s\left(X_j\right), \overline{X}_j + 4 \; s\left(X_j\right)\right]$, where $\overline{X}_j$ and $s\left(X_j\right)$ denote the mean and standard deviation of the signal variable. Table \ref{tab:simsetup} summarizes the factors of our simulation study and their levels.

\begin{table}[!h]
\begin{tabular}{|c|c|c|}
\hline
factor & levels & \# \\ 
\hline
number of clean variables & 2, 4, 6, 8, 10 & 5\\
number of clusters & 2, 3, 4, 5 & 4\\
degree of separation & separated, well-separated & 2\\
percentage noise variables added & 0, 50, 100, 150, 200, 500, 1000, 2000& 8\\
type of noise & Gaussian, uniform & 2\\
percentage of outliers & 0, 5 & 2\\
\hline
\end{tabular} 
\caption{Design factors of the simulation study.}
\label{tab:simsetup}
\end{table}

\noindent All together this gives 1280 different settings and for each of these, we generate 100 datasets. Each generated dataset is scaled using the six scale estimators described above. Afterwards we perform the most popular methods of connectivity-based clustering and centroid-based clustering.
More specifically, we use hierarchical clustering \citep{hartigan1975clustering,anderberg2014cluster} on the Euclidean distances with single, average, complete and Ward's linkage functions \citep{Ward1963} as well as k-means \citep{Lloyd1982, Macqueen1967} and partitioning around medoids using the manhattan distance \citep{Kaufman2009}. For k-means, the algorithm of \cite{Hartigan1979} is used with 100 random starts and 100 maximum iterations for each starting value. \\

\noindent We compare the results using the adjusted Rand index (ARI) \citep{Hubert1985} which lies between -1 and 1 where 1 indicates a perfect clustering. We cluster each dataset for a variety of target clusters $k = 1,\ldots, 3 \times T$, where $T$ denotes the true number of clusters. For k-means this value is a direct input, whereas for hierarchical clustering we cut the dendrogram at these various levels of $k$. We then pick the optimal value of the ARI over these different clusterings. The reason for this procedure is that we want to evaluate the effect of scaling on clustering without any distortion from the question of how to choose the optimal number of clusters. Furthermore, particularly in the case of hierarchical clustering, the clusterings resulting from a higher number of partitions are often more reflective of the real underlying structure than cutting the dendrogram at the true number of clusters, since single outlying observations can distort the dendrogram significantly.\\

\begin{figure}[!ht]
\centering
\begin{subfigure}[b]{0.49\linewidth}
  \centering 
	\includegraphics[width=0.99\textwidth]{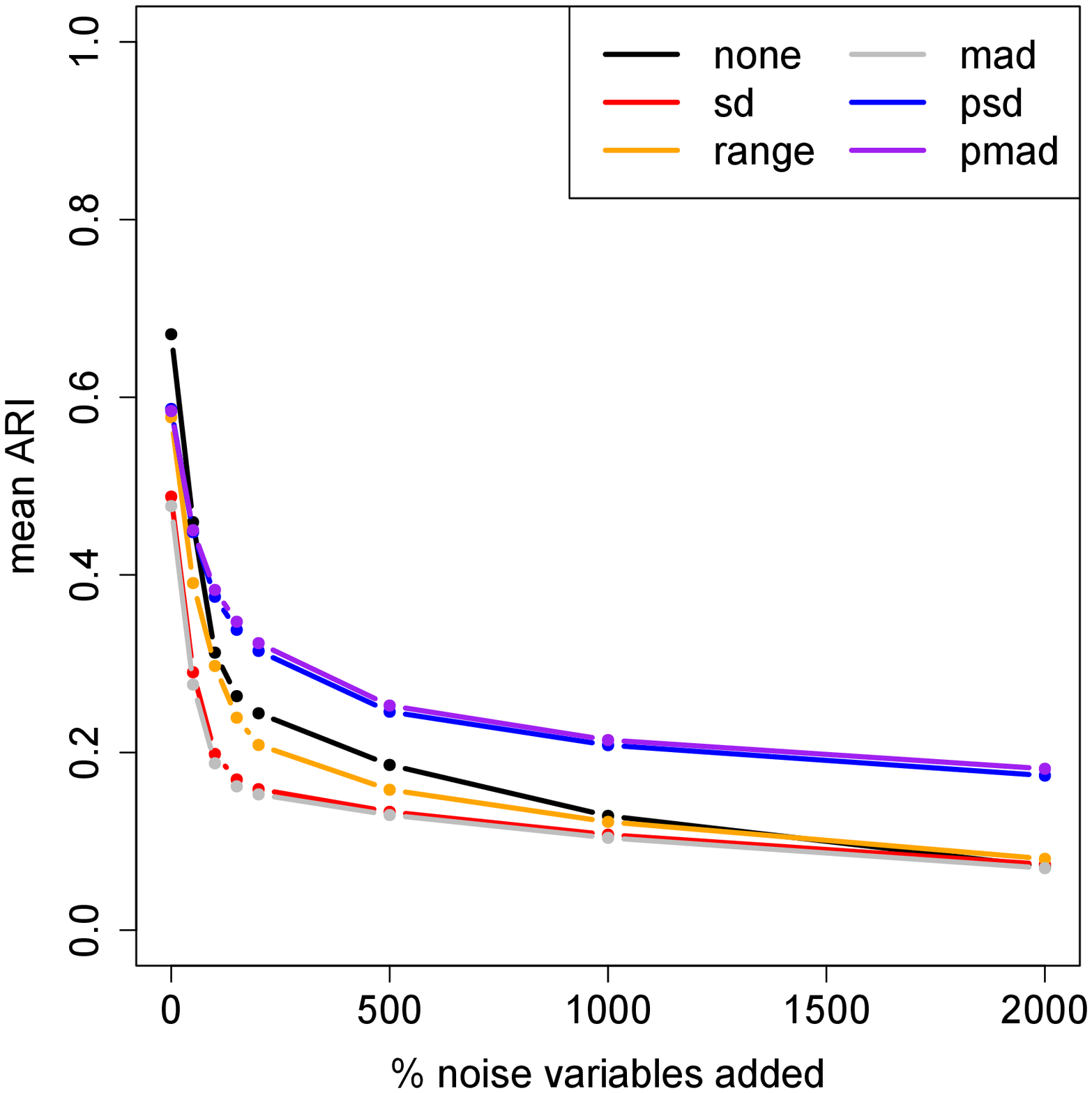}
	\caption{Single linkage}
\vspace{0.25cm}
\end{subfigure}
\begin{subfigure}[b]{0.49\linewidth}
  \centering 
  \includegraphics[width=0.99\textwidth]{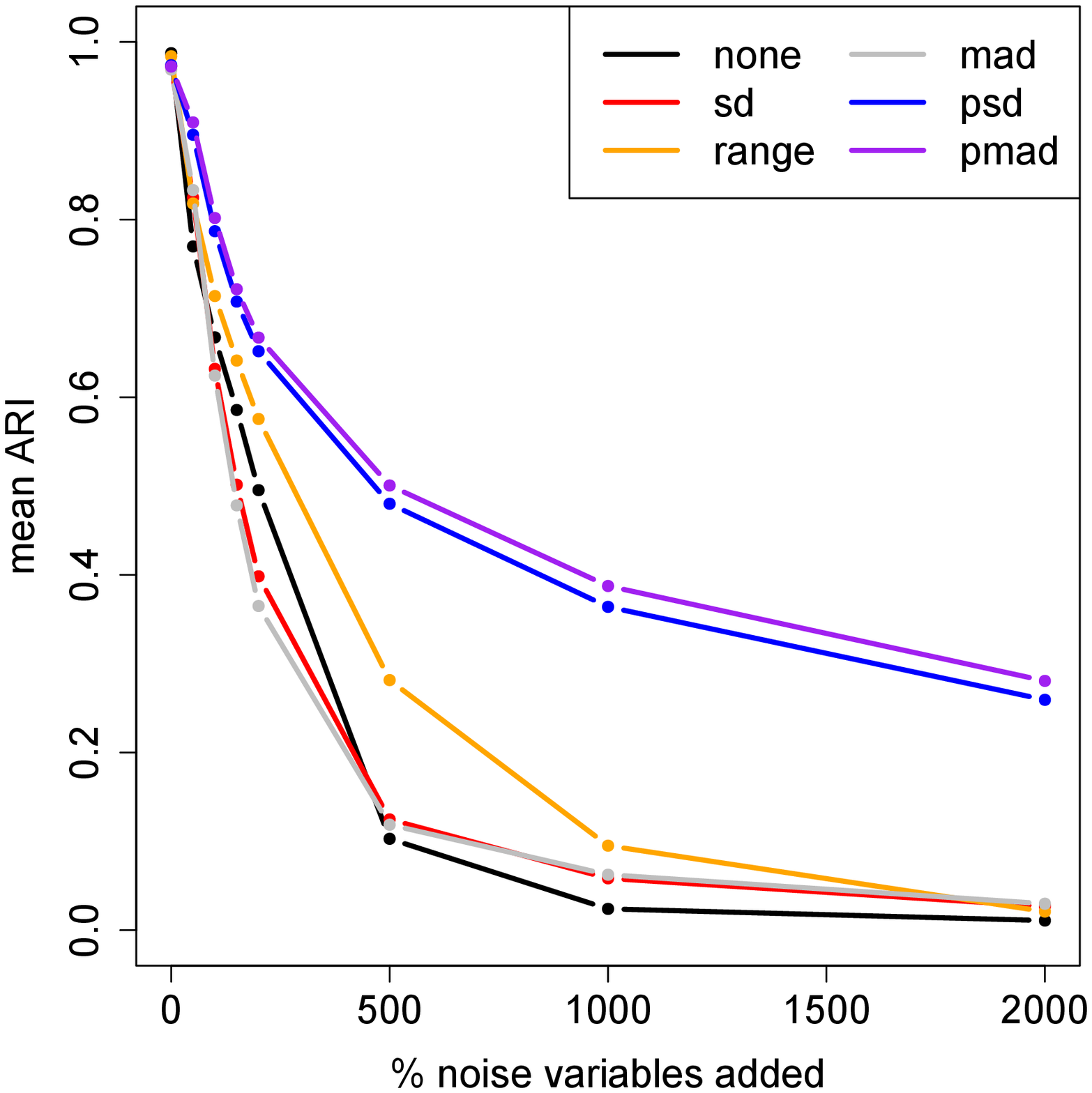} 
	  \caption{Average linkage}
\vspace{0.25cm}
\end{subfigure}
\begin{subfigure}[b]{0.49\linewidth}
  \centering 
  \includegraphics[width=0.99\textwidth]{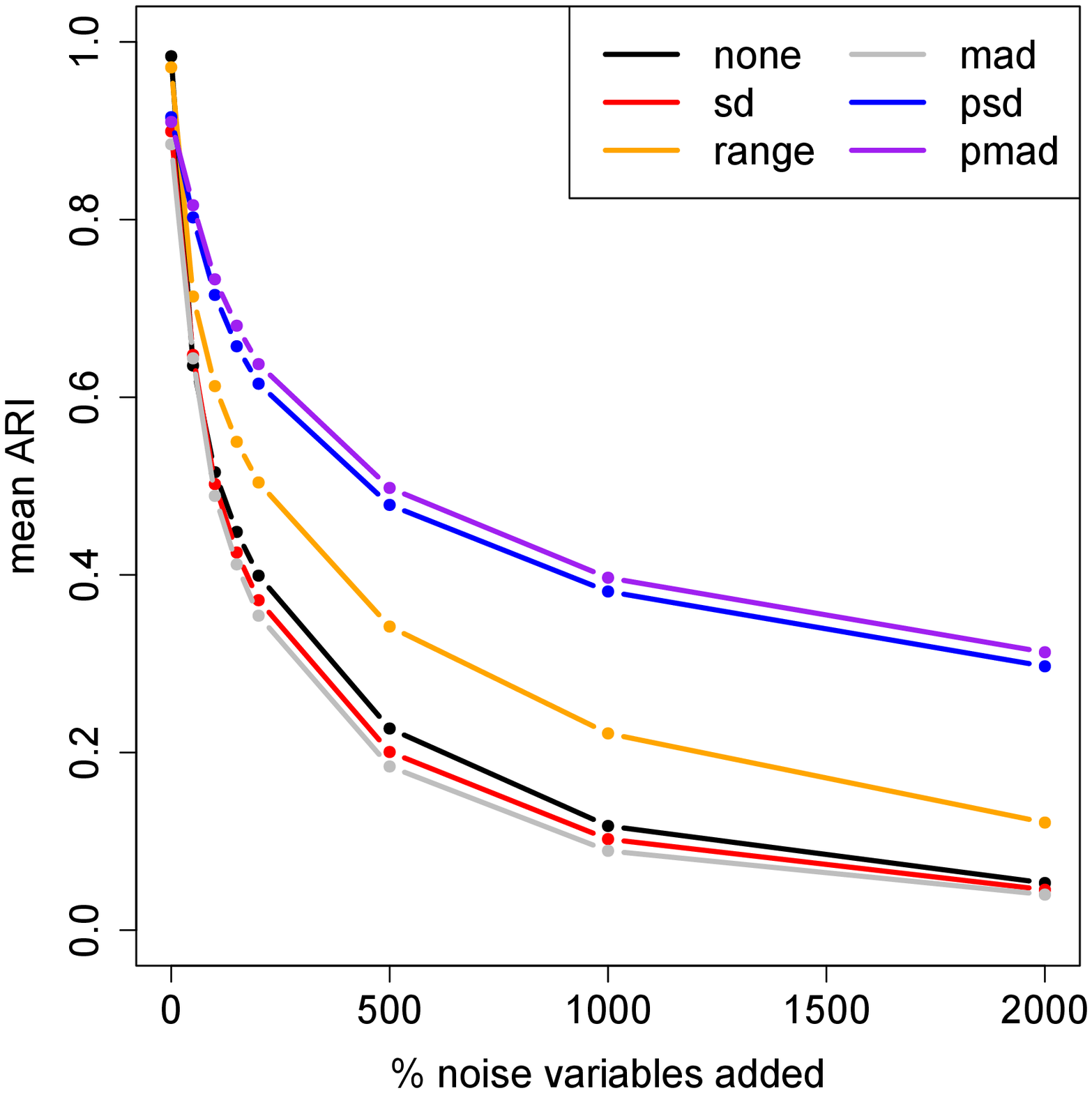} 
	  \caption{Complete linkage}
\end{subfigure}
\begin{subfigure}[b]{0.49\linewidth}
  \centering 
  \includegraphics[width=0.99\textwidth]{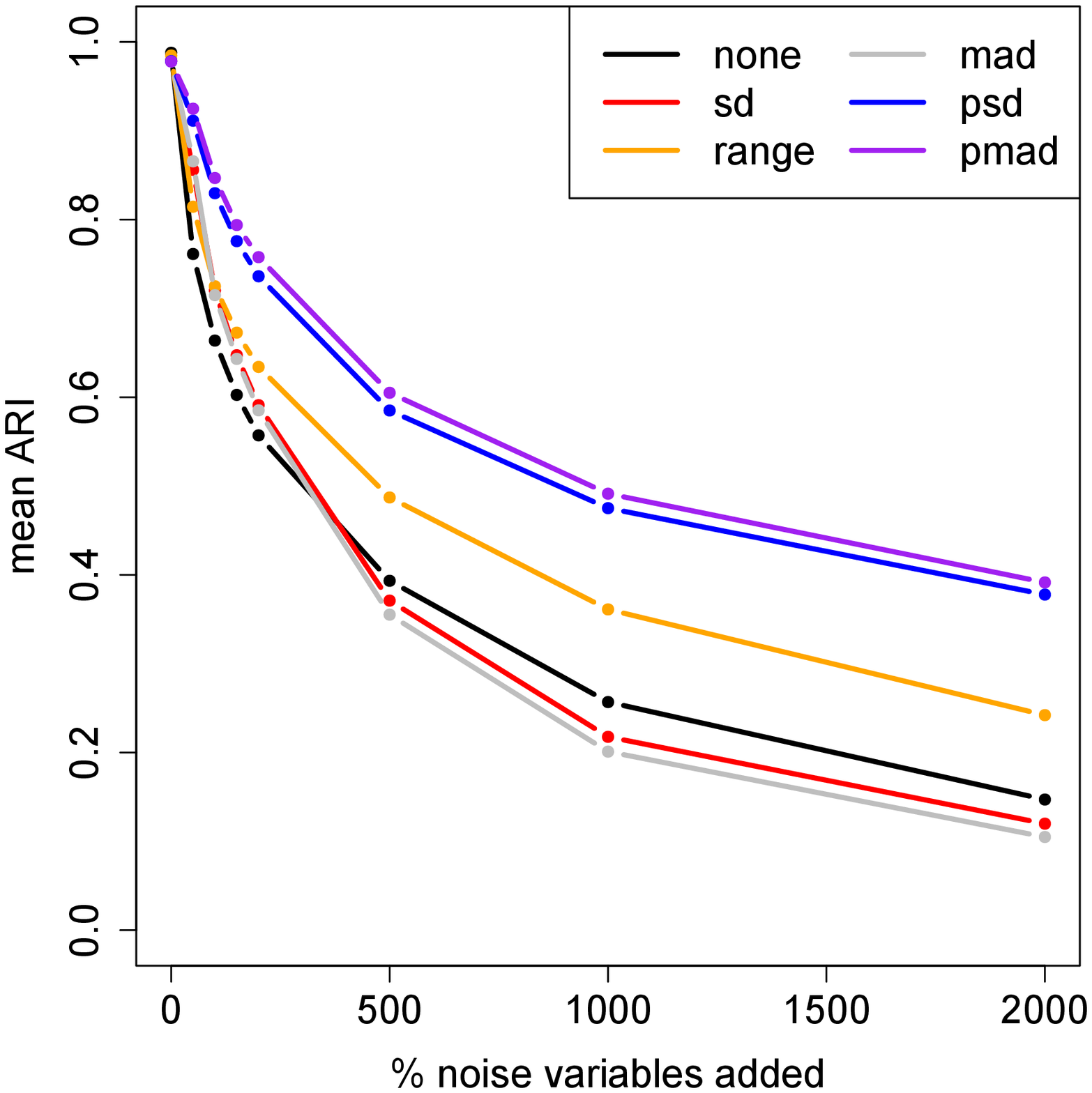} 
	  \caption{Ward linkage}
\end{subfigure}
\caption{Simulation results for hierarchical clustering with single (a), average (b), complete (c) and Ward (d) linkage functions on outlier-free data. The pooled scale estimators are the least sensitive to an increasing number of noise variables.}
\label{fig:sim_hc}
\end{figure}

\noindent Figure \ref{fig:sim_hc} shows the big picture of the simulation results for hierarchical clustering on data without outliers. For each type of linkage, the graphs show the average ARI over all different settings with the increasing number of noise variables on the x-axis. Several interesting observations can be made from these plots. It is clear that as the number of noise variables increases, the clustering gets more difficult resulting in generally lower ARI values. However, scaling with the standard deviation or the mad is clearly more sensitive to noise variables than the other methods. Scaling with the range does fairly well, which is in line with the findings of \cite{Milligan1988}. The pooled scale estimators outperform all the other methods, especially when the number of noise variables is large. The difference between the psd and the pmad seems very small and not significant in these plots. Finally, not scaling appears to perform similarly to scaling with either sd or mad. This is due to the particular simulation setup and one must always take into account that the performance of not scaling the variables can be completely destroyed by taking one noise variable which has a variance which is much larger than the variance of the signal variables.

\begin{figure}[!ht]
\centering
\begin{subfigure}[b]{0.49\linewidth}
  \centering 
	\includegraphics[width=0.99\textwidth]{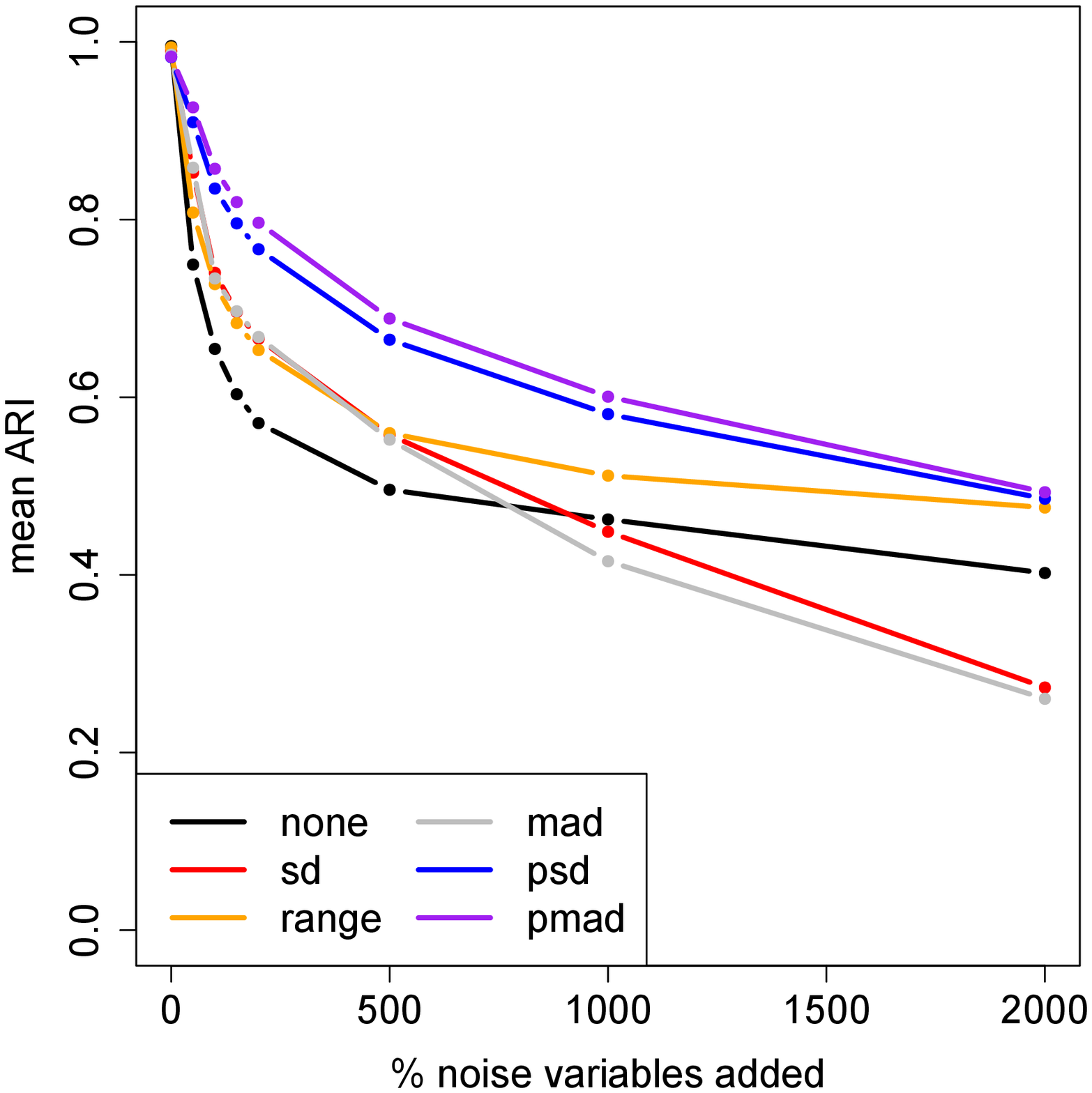}
	\caption{$k$-means}
\end{subfigure}
\begin{subfigure}[b]{0.49\linewidth}
  \centering 
  \includegraphics[width=0.99\textwidth]{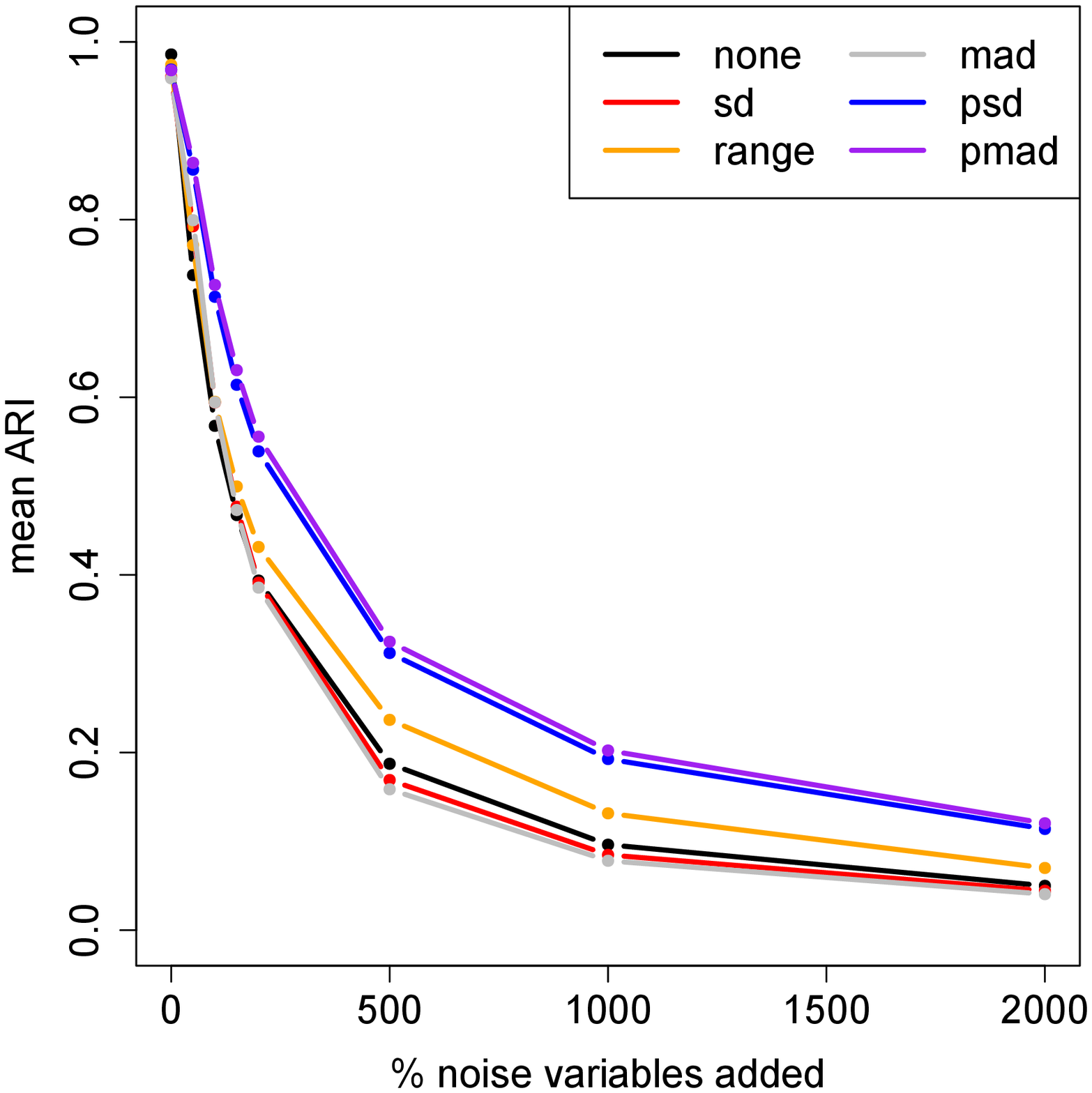} 
	  \caption{Partitioning around medoids}
\end{subfigure}
\caption{Simulation results for $k$-means (a) and partitioning around medoids (b) on data without outliers. The pooled scale estimators are more resistant to the addition of noise variables to the data.}
\label{fig:sim_km}
\end{figure}

\noindent Figure \ref{fig:sim_km} shows the results for k-means clustering and partitioning around medoids. 
The conclusions are largely the same as for the case of hierarchical clustering. When scaling with the standard deviation or mad, the true clustering structure seems more difficult to retrieve. The range preserves more of the cluster structure and performs better than the sd and mad which is in line with \cite{Steinley2004a}. However, the pooled scale estimators again outperform the competitors.

\noindent The simulation results presented above only give a rough overview of the performance of the methods and do not show the performance in the presence of outliers. The most interesting insight from a more detailed analysis of the results is that the performance of the methods is highly dependent on the type of noise variables which are added to the signal variables. Scaling with the range works well when the noise variables are more gaussian but fails when the noise is more uniform. This can be explained by the fact that uniform noise variables have a large variance given their range. As a result, their impact on the clustering is large when scaling with the range. Scaling with the sd and mad work much better when the noise variables are more uniform than the case of gaussian noise. This in turn can be explained by the fact that the uniform noise variables have a high variance for their range compared with gaussian noise variables. Scaling them by their variance pushes the uniform noise more towards the center, which limits their influence.\\

\noindent The effect of outliers is shown in Figures 1 and 2 of the Supplementary Material. These results expose the sensitivity of the range to the presence of outliers. Other than that, the relative performance of the different scaling methods is roughly the same as in the case of outlier free data. An interesting note is that the pmad is clearly more robust to outliers than the psd, which resembles the robustness of the mean absolute deviation versus that of the standard deviation in classical scale estimation.\\

\begin{figure*}
\begin{tabular}{>{\centering\arraybackslash} m{1.7cm} | >{\centering\arraybackslash} m{3.5cm} >{\centering\arraybackslash} m{3.5cm}  >{\centering\arraybackslash} m{3.5cm}}
\hline
& average linkage & complete linkage & ward linkage \\ 
\hline
no scaling &
\includegraphics[width=0.27\textwidth,height=2cm]{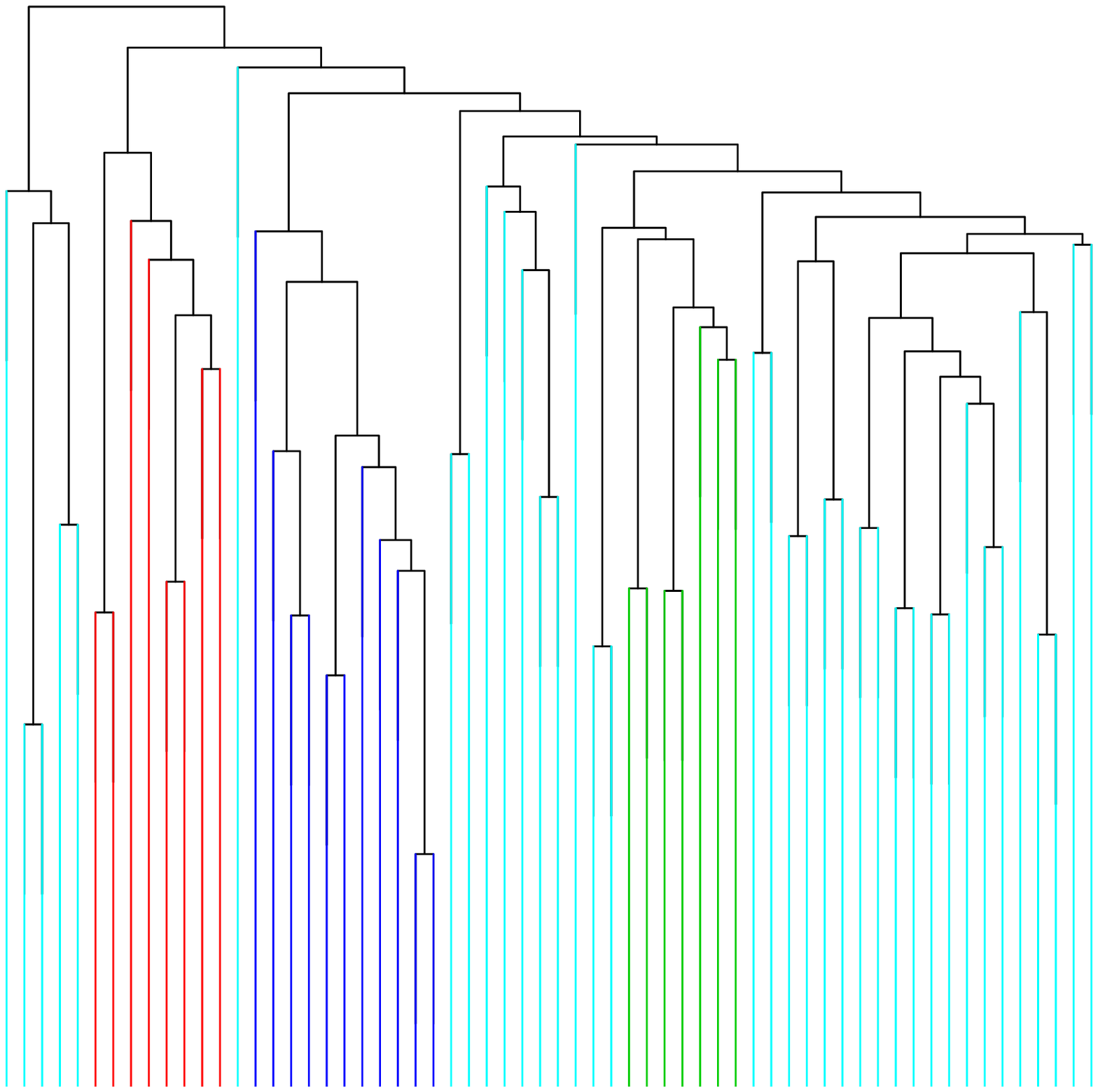} &
\includegraphics[width=0.27\textwidth,height=2cm]{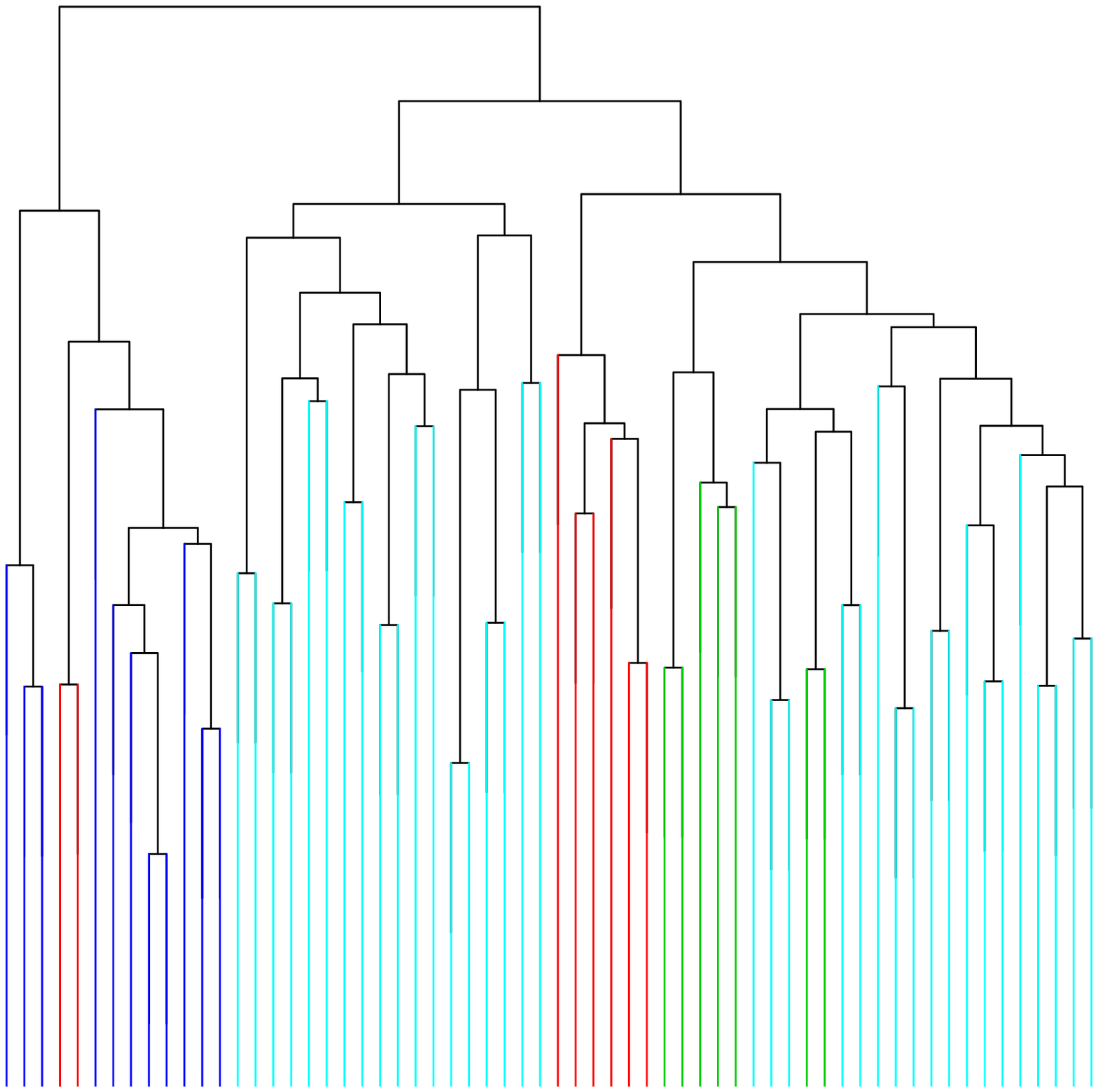} &
\includegraphics[width=0.27\textwidth,height=2cm]{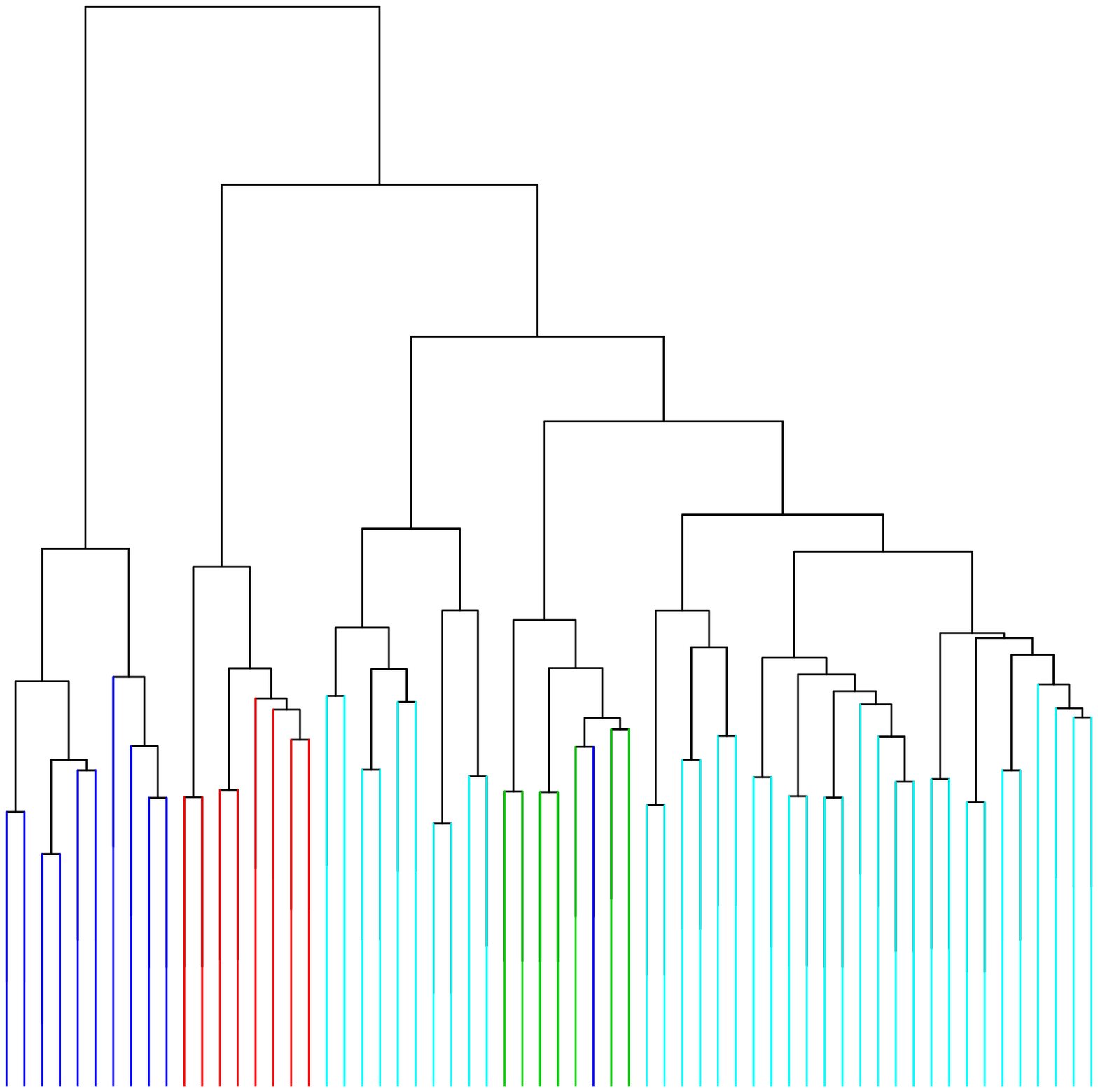} \\
\\[-1em]
standard deviation &
\includegraphics[width=0.27\textwidth,height=2cm]{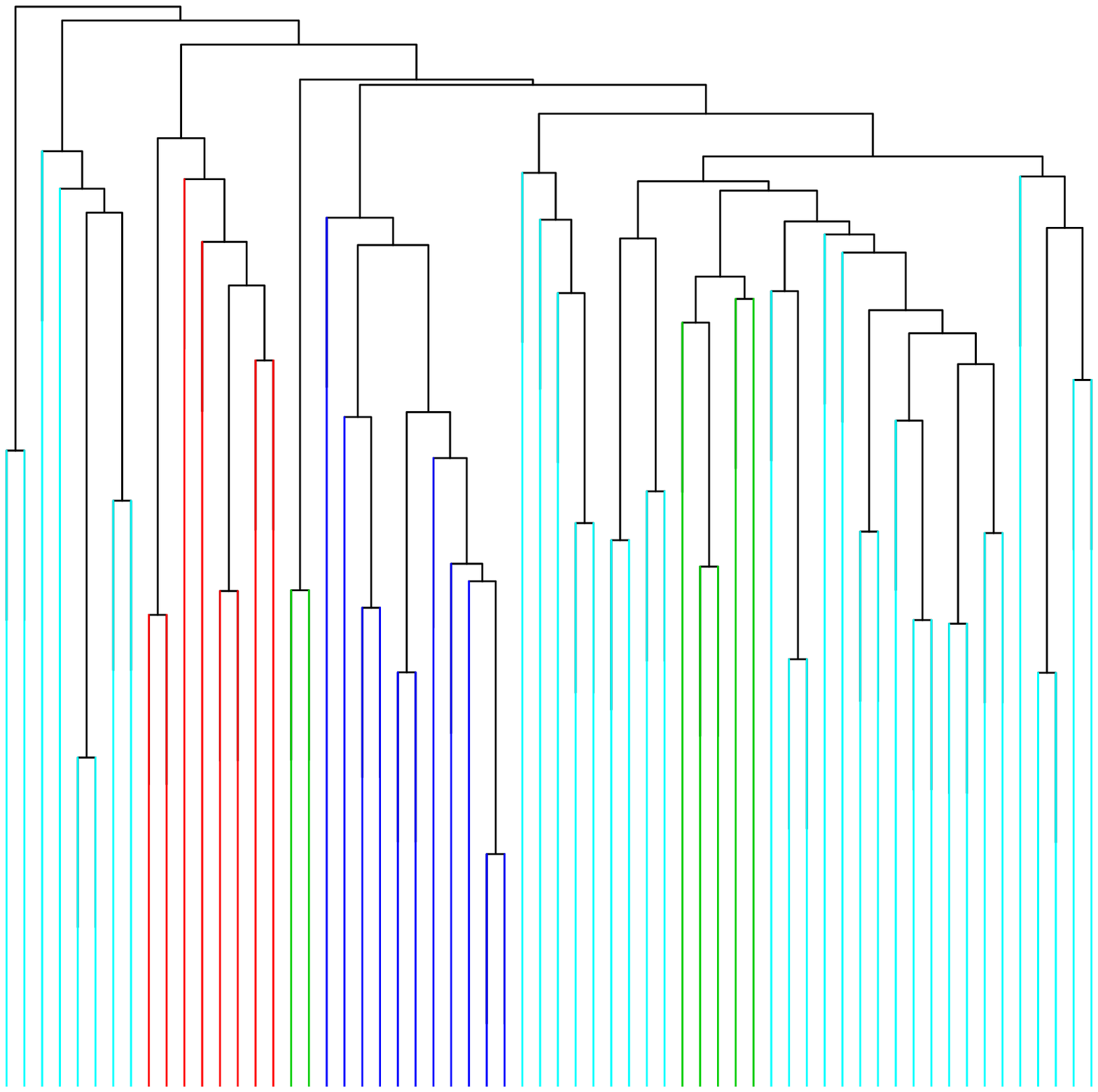} &
\includegraphics[width=0.27\textwidth,height=2cm]{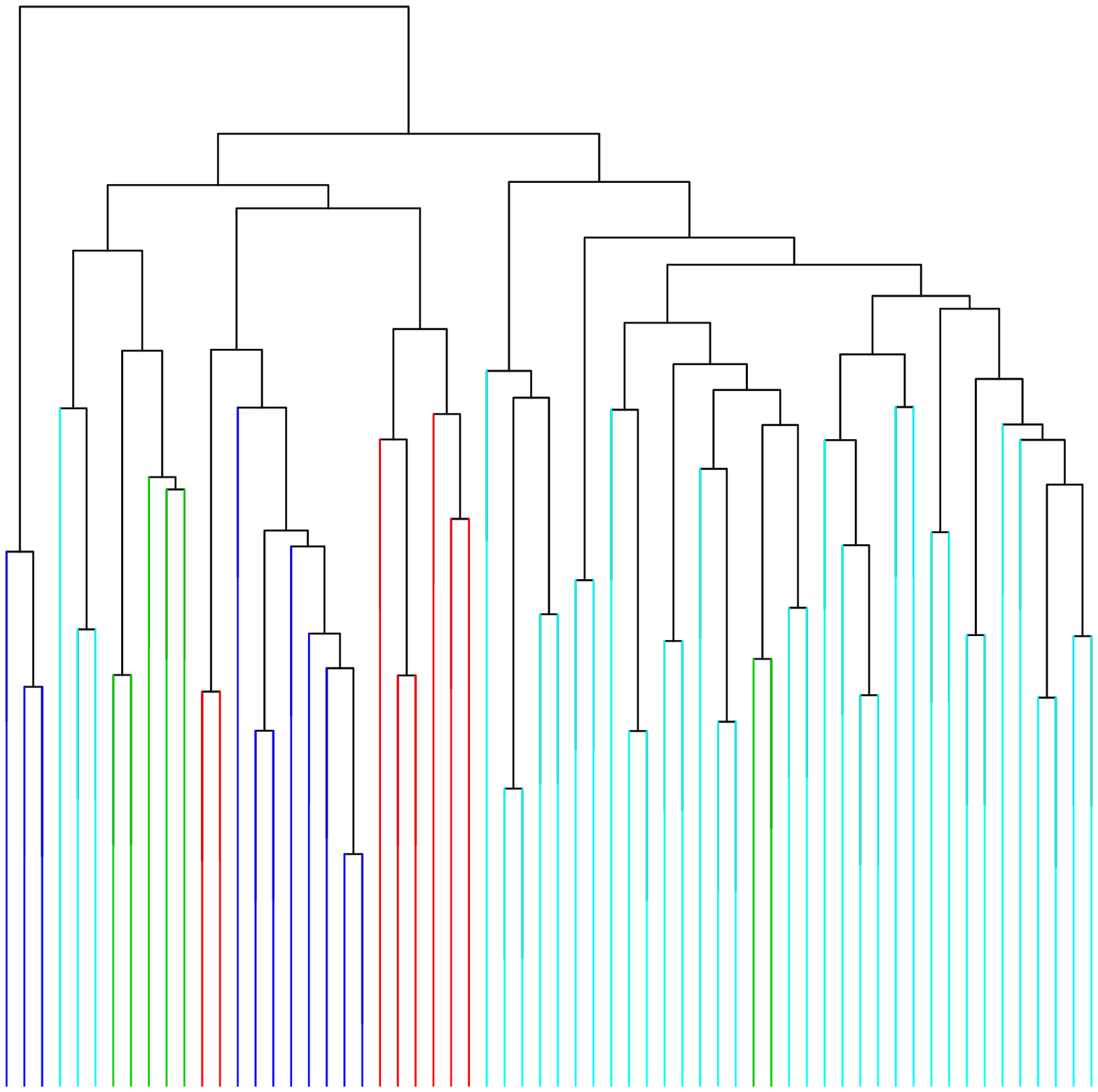} &
\includegraphics[width=0.27\textwidth,height=2cm]{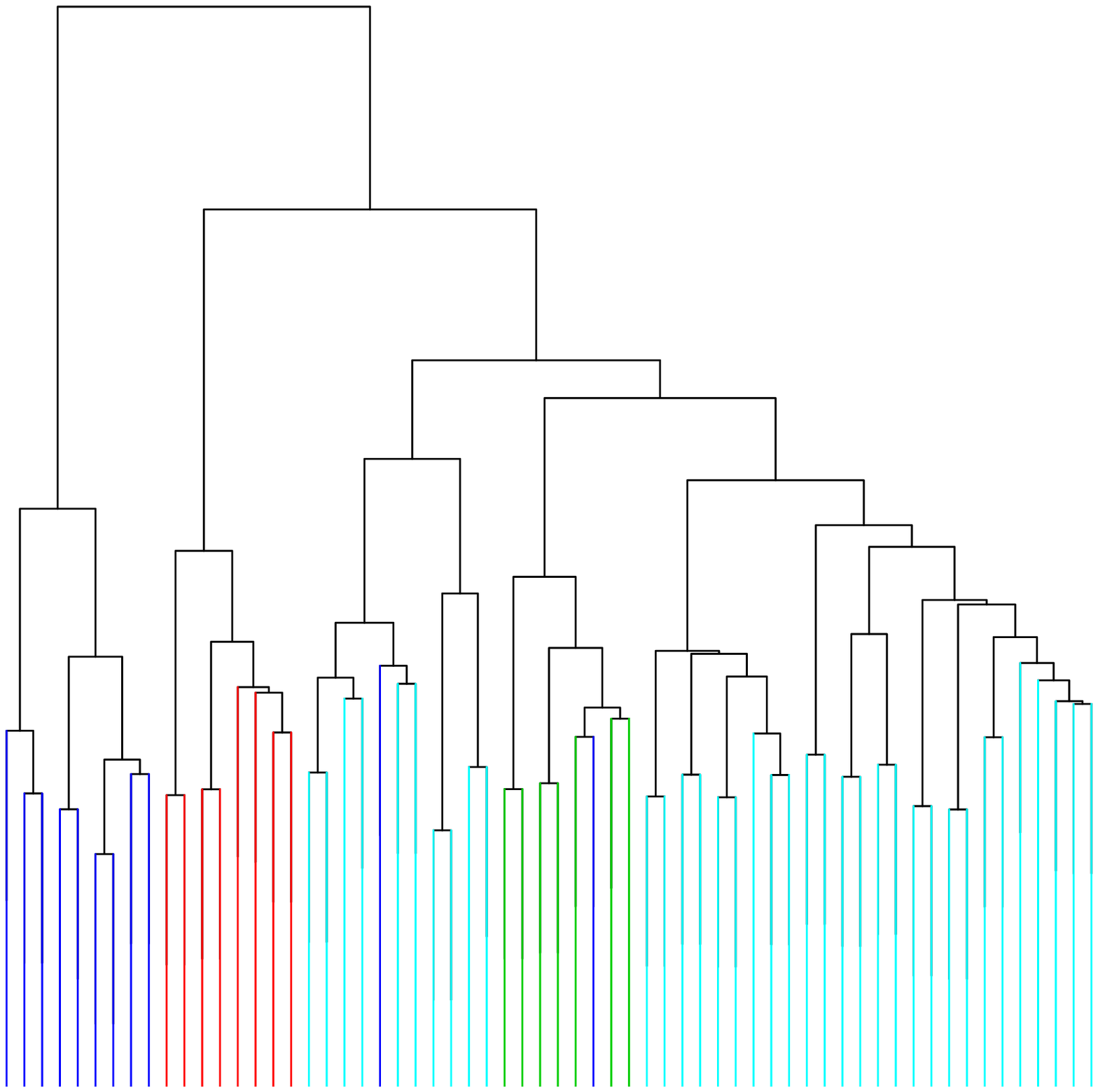} \\
\\[-1em]
range &
\includegraphics[width=0.27\textwidth,height=2cm]{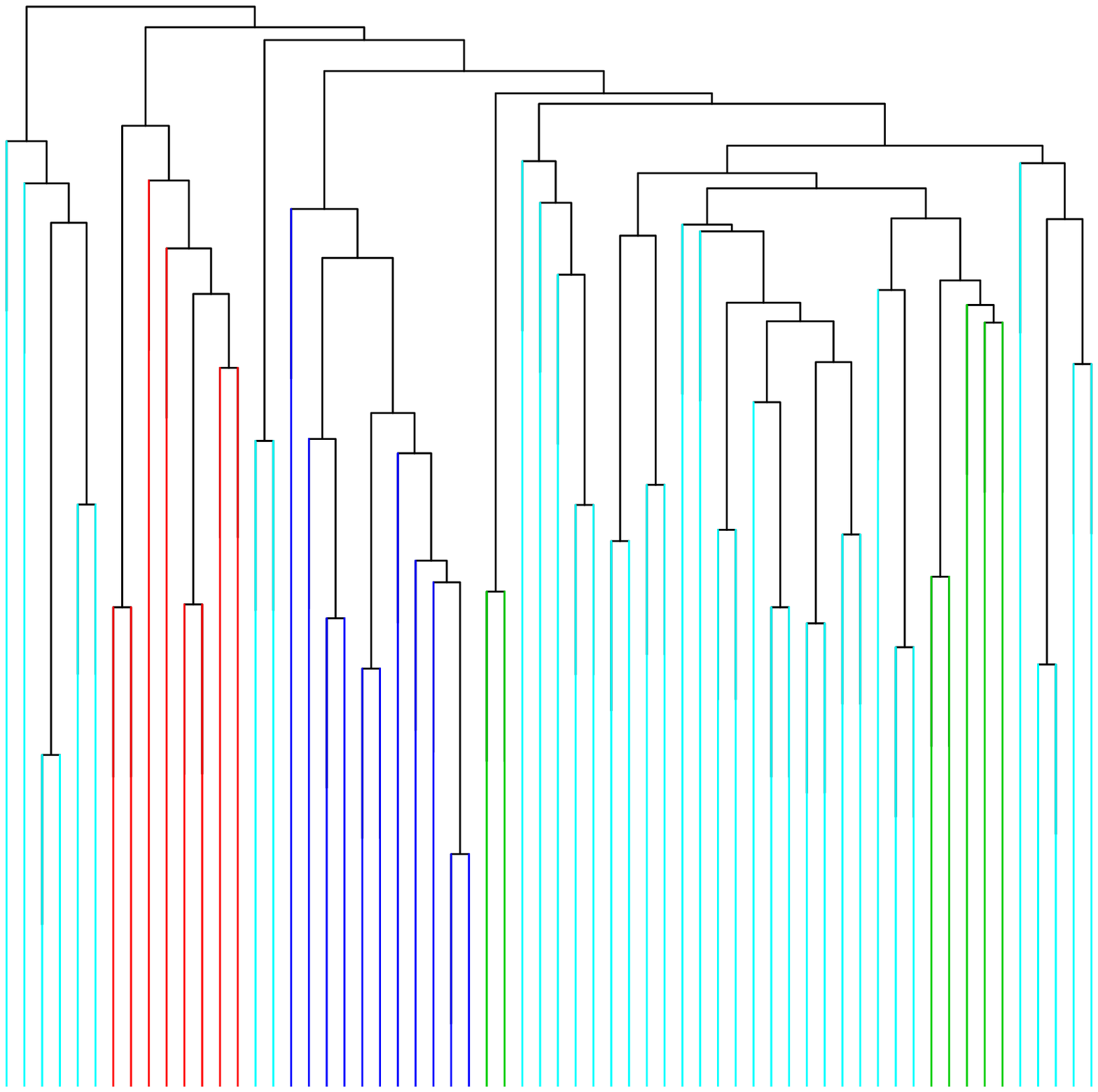} &
\includegraphics[width=0.27\textwidth,height=2cm]{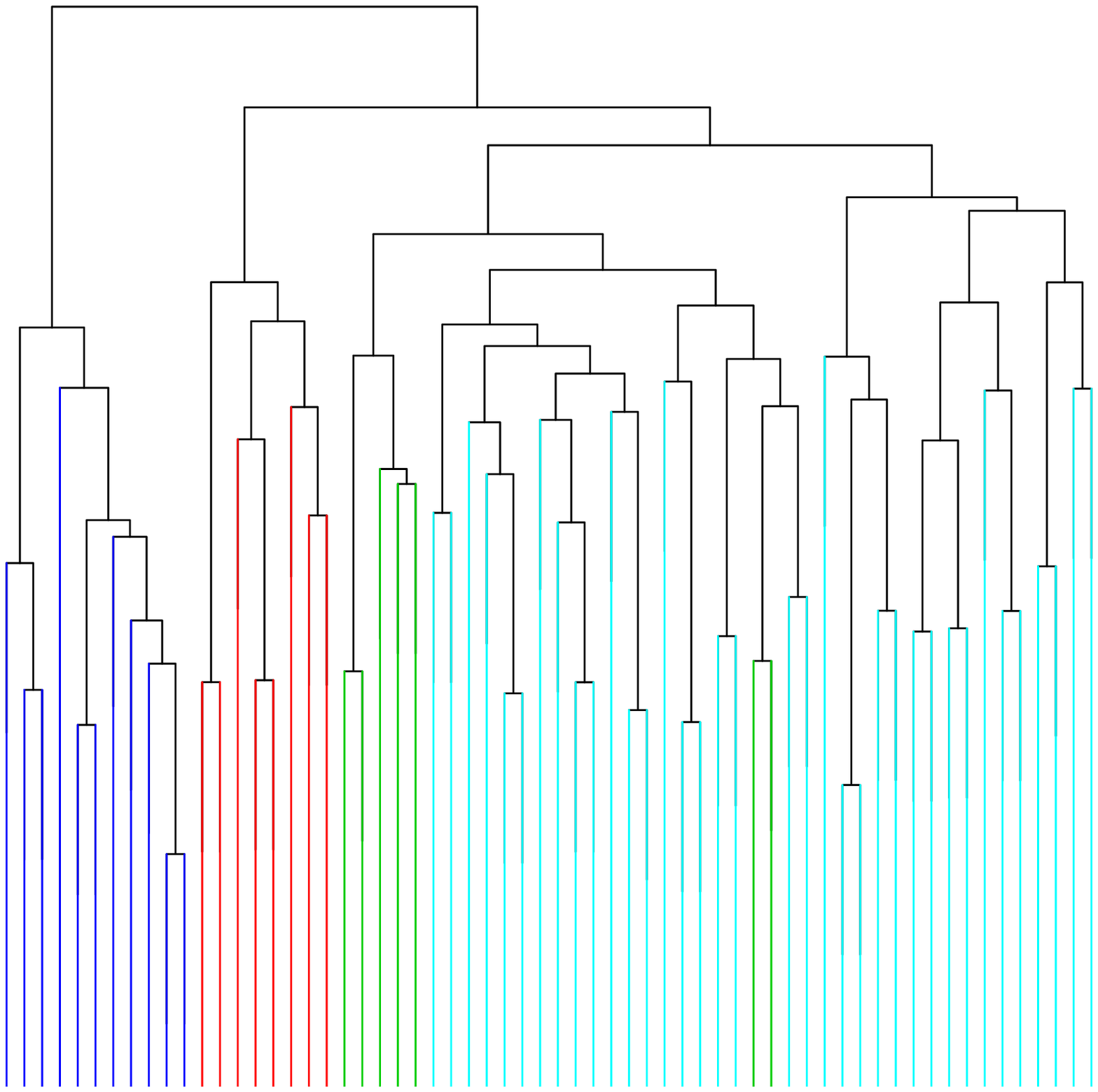} &
\includegraphics[width=0.27\textwidth,height=2cm]{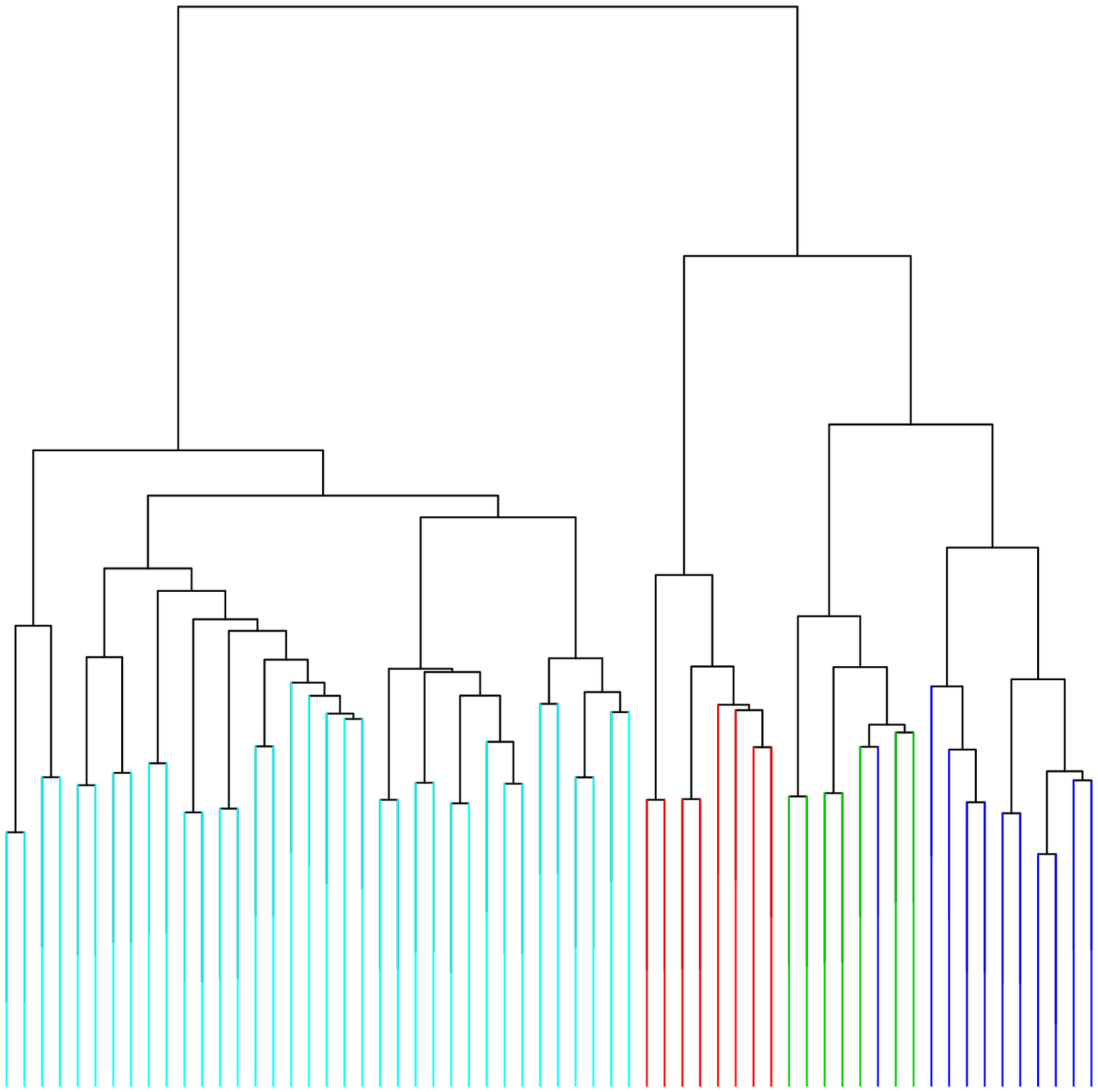} \\
\\[-1em]
pooled standard deviation &
\includegraphics[width=0.27\textwidth,height=2cm]{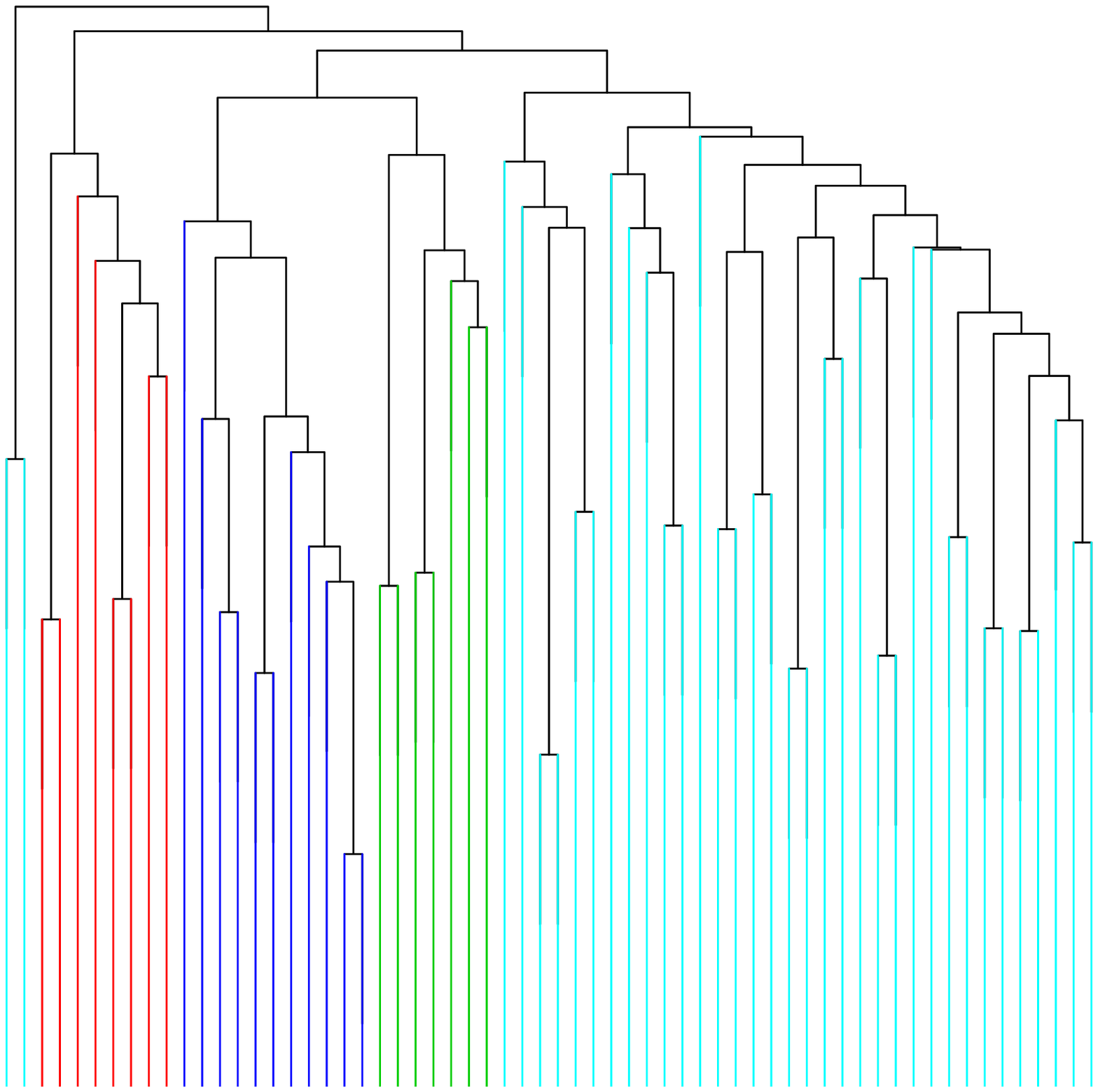} &
\includegraphics[width=0.27\textwidth,height=2cm]{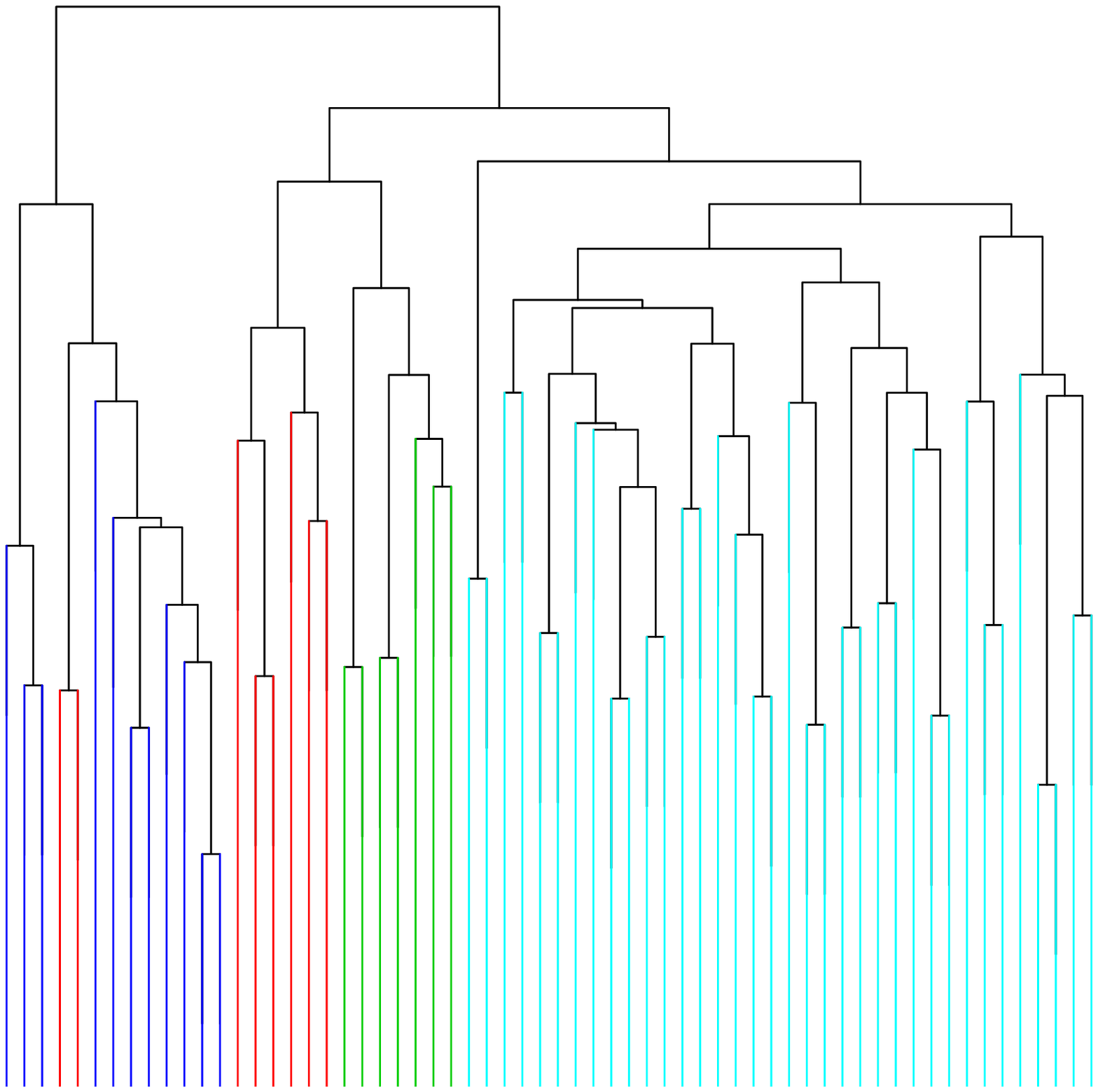} &
\includegraphics[width=0.27\textwidth,height=2cm]{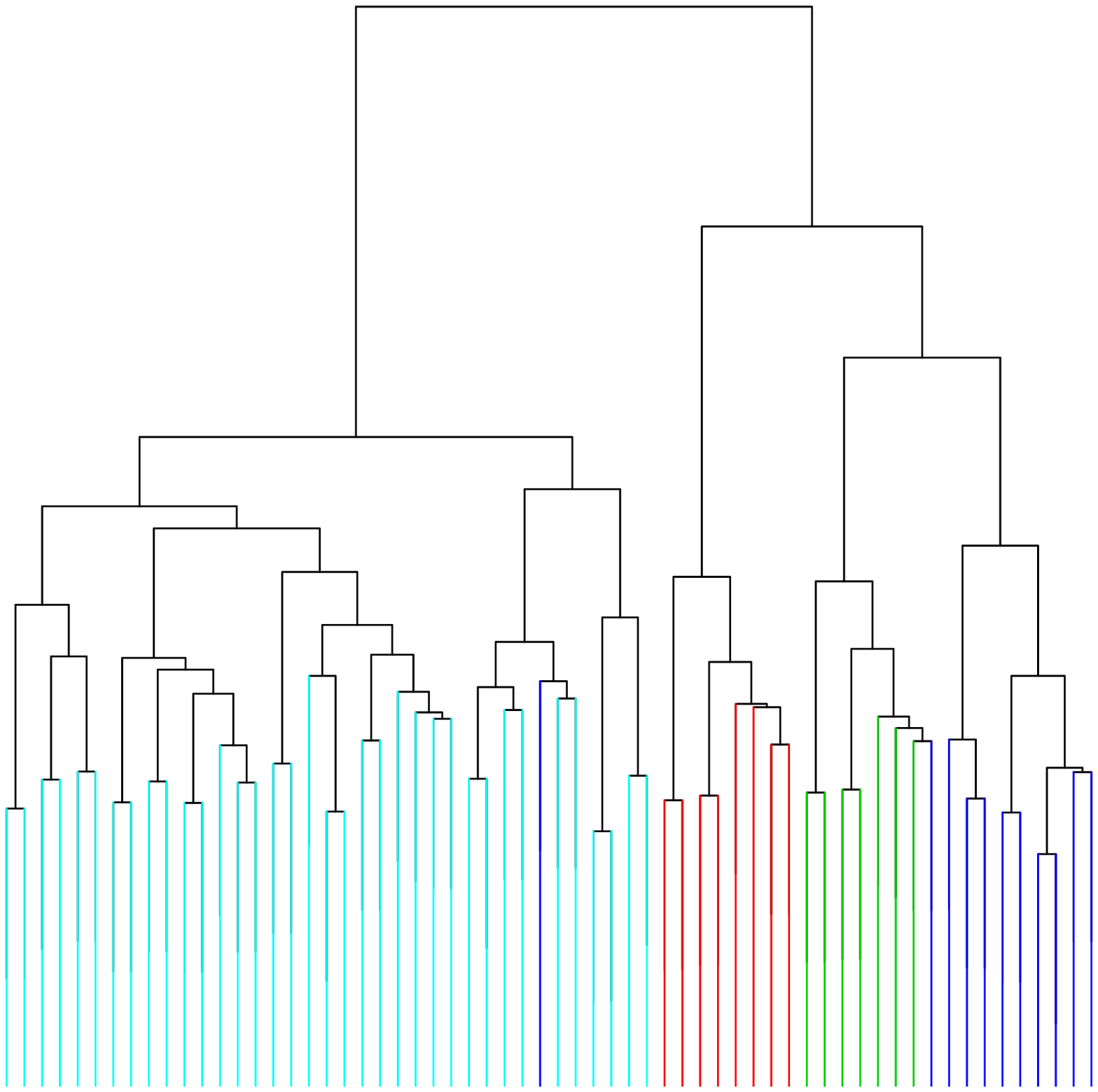} \\
\end{tabular}
\caption{The effect of variable scaling on the gene expression data. The colors correspond to the tumor type: basal-like in red, Erb-B2+ in green, normal-breast-like in dark blue and luminal epithelial/ER+ in cyan. The pooled standard deviation generally yields superior recovery of the true clusters.}
\label{table:genome}
\end{figure*}

\section{Gene expression example}\label{sec:realdata}
In a seminal paper \cite{perou2000} analyzed gene expression patterns of 65 surgical specimens of human breast tumors. The data is publicly available at \url{https://www.omicsdi.org/dataset}. They identified 496 intrinsic genes, which had significantly larger variation between different tumors compared with the variation between paired samples from the same tumor. Using hierarchical clustering with Eisen linkage \citep{Eisen1998}, they clustered the tumors into 4 different types: basal-like, Erb-B2+,  normal-breast-like and luminal epithelial/ER+. We illustrate the effect of the pooled standard deviation in hierarchical clustering with average, complete and ward linkage on these 496 genes.\\

\noindent Table \ref{table:genome} shows the resulting dendrograms when applying each of these clustering algorithms to the dataset after scaling it in various ways. For average linkage, the pooled standard deviation clearly outperforms the other options. It only misclassifies two observations, whereas the other methods split the large group of luminal epithelial/ER+ tumors (in cyan) in two or more clusters and fail to identify the smallest group (in green) which contains the Erb-B2+ tumors. Complete linkage gives better results for all types of scaling. However, without scaling or when using the range, the largest group of tumors get split up into two groups. This does not happen when scaling with the standard deviation, yet quite a few tumors are misclassified in this case. When scaling with the pooled standard deviation, only two tumors are misclassified using complete linkage. Finally, with Ward's linkage, both the pooled standard deviation and scaling with the range work well, with two and one misclassified tumor respectively. Without scaling and with the standard deviation however, the largest cyan cluster gets split up in two smaller clusters. In conclusion, the pooled standard deviation yields better recovery of the true clusters than the other scaling methods.\\

\noindent In addition to improved clustering results, the pooled scaling procedure yields a diagnostic tool in the form of scale ratios. More precisely, we can compare the standard deviations of the variables with their pooled counterparts. Variables for which the ratio of these two scales is large typically show a clear grouping of the data, whereas variables for which this ratio is close to 1 do not distinguish clear groups. 
These ratios can thus be used as a fast and intuitive variable-screening procedure to identify potentially very informative variables, which is often a main research goal in this context.

\begin{figure}[!h]
\centering
\begin{subfigure}[b]{0.49\linewidth}
  \centering 
	\includegraphics[width=0.99\textwidth]{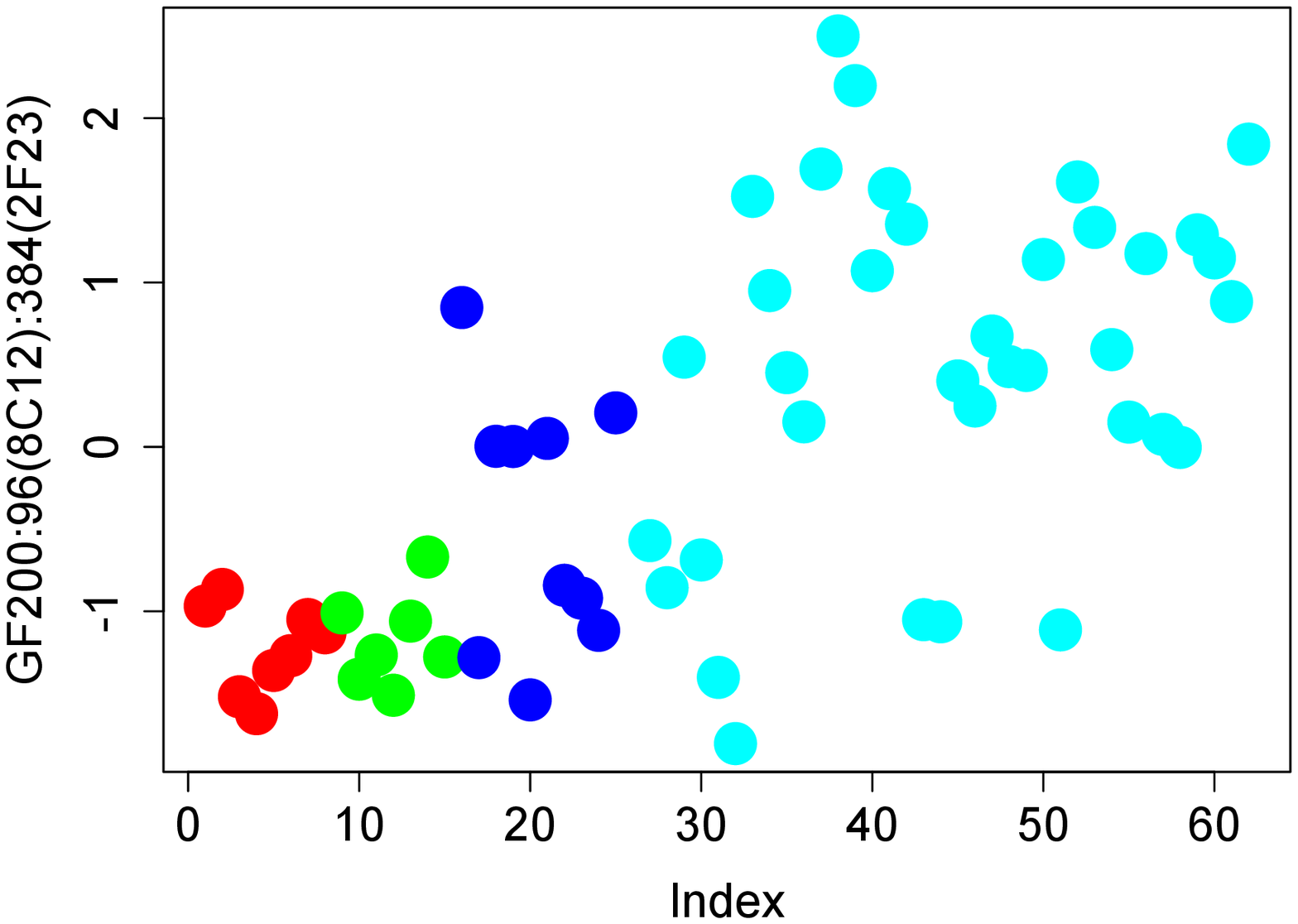}
	\caption{GF200:96(8C12):384(2F23)}
\vspace{0.25cm}
\end{subfigure}
\begin{subfigure}[b]{0.49\linewidth}
  \centering 
  \includegraphics[width=0.99\textwidth]{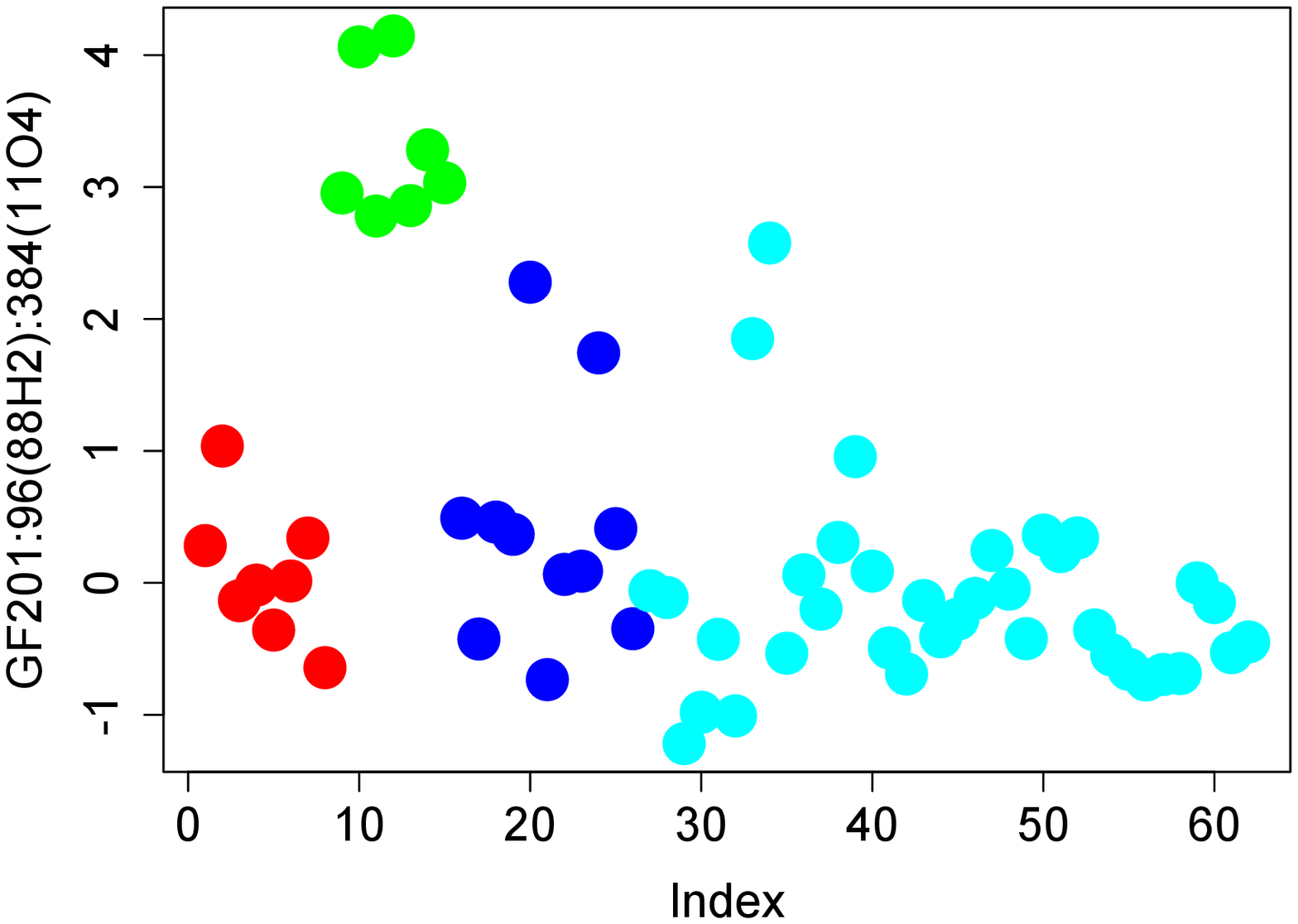} 
	  \caption{GF201:96(88H2):384(11O4)}
\vspace{0.25cm}
\end{subfigure}
\begin{subfigure}[b]{0.49\linewidth}
  \centering 
  \includegraphics[width=0.99\textwidth]{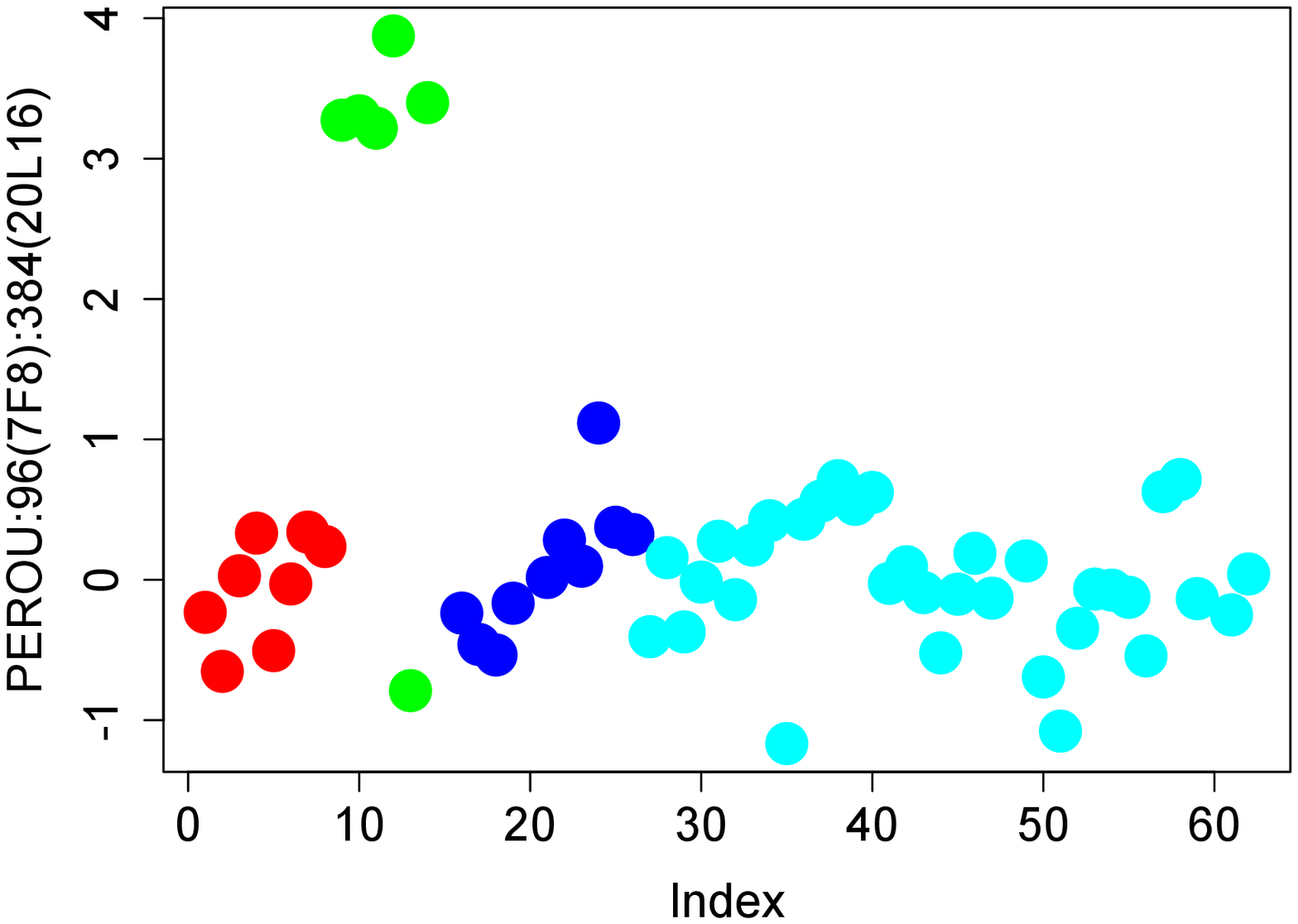} 
	  \caption{PEROU:96(7F8):384(20L16)}
\end{subfigure}
\begin{subfigure}[b]{0.49\linewidth}
  \centering 
  \includegraphics[width=0.99\textwidth]{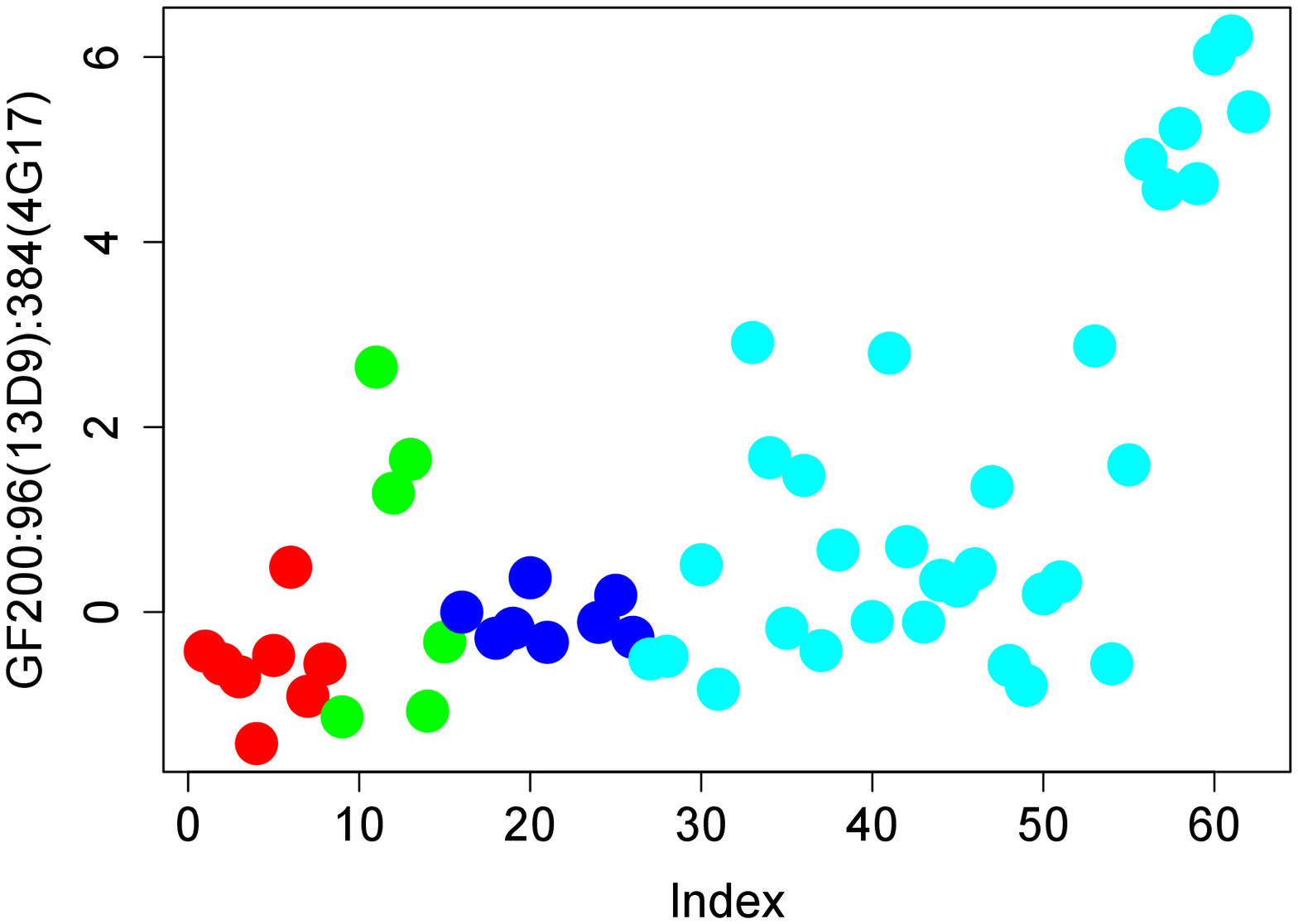} 
	  \caption{GF200:96(13D9):384(4G17)}
\end{subfigure}
\caption{Genes for which the pooled standard deviation is smaller than the standard deviation. These genes generally cluster the majority of the true groups together while sometimes identifying potentially interesting subclusters.}
\label{fig:gene_largeratio}
\end{figure}

\noindent In this example, only 8 of the 496 variables had a scale different from the standard deviation. The information on those 8 variables is presented in table \ref{table:smallscalegenes}. Figure \ref{fig:gene_largeratio} shows the gene expressions for the 4 genes which had the highest scale ratio. The colors again correspond to the 4 types of tumors. The top left panel shows the gene GF200:96(8C12):384(2F23), which clearly groups the red and blue points together and also shows high values for the majority of the cyan group. The top right and bottom left panels show the genes GF201:96(88H2):384(11O4) and PEROU:96(7F8):384(20L16), which distinctly separate the green tumors from the others. In the bottom right panel, the red tumors seem to have slightly lower values than the others, the blue tumors are grouped very tightly together and the cyan tumors appear to contain a sub-cluster with elevated values for this gene.

\begin{table}
\resizebox{\textwidth}{!}{
\begin{tabular}{>{\centering\arraybackslash} m{5cm} >{\centering\arraybackslash} m{8cm} >{\centering\arraybackslash} m{2cm}}
variable ID & Description & sd / psd\\
\hline
GF200:96(8C12):384(2F23) &HUMAN BREAST CANCER, ESTROGEN REGULATED LIV-1 PROTEIN (LIV-1) MRNA, PARTIAL CDS H29407 45 & 3.5\\
GF201:96(88H2):384(11O4) & GROWTH FACTOR RECEPTOR-BOUND PROTEIN 7 H53703 224 & 2.4\\
PEROU:96(7F8):384(20L16)& SWI/SNF RELATED, MATRIX ASSOCIATED, ACTIN DEPENDENT REGULATOR OF CHROMATIN, SUBFAMILY E, MEMBER 1 W63613 228 & 2.4\\
GF200:96(13D9):384(4G17) & CYTOCHROME P450, SUBFAMILY IIA (PHENOBARBITAL-INDUCIBLE), POLYPEPTIDE 7 T73031 61 & 2.2\\
PEROU:96(8A1):384(20B1) &68400 T57034 226 & 2.2	\\
PEROU:96(6A1):384(20A2) &68400 T57034 227 & 2.2\\
GF200:96(14D12):384(4G24) &APOLIPOPROTEIN D H15842 & 2.2\\
PEROU:96(9A9):384(18B18)& IMMUNOGLOBULIN J CHAIN H24896 325 & 2.1\\
\end{tabular}}
\caption{Genes for which the pooled standard deviation is smaller than the standard deviation. The third column shows the ratios of these two scales.}
\label{table:smallscalegenes}
\end{table}

\newpage
\section{Conclusion}\label{sec:conclusion}
We introduced a new approach to variable scaling prior to cluster analysis which we call pooled scale estimators. The performance of pooled scaling is compared with the most common competitors on several popular clustering techniques, with a particular focus on the case of high dimensional data with many uninformative noise variables. Scaling with the pooled scale estimators yields superior cluster recovery on the different clustering methods, in particular when the data contains a lot of noise variables. Since this is a common theme in the clustering of gene expression data, the performance was illustrated on breast cancer gene expressions in which it also outperformed the competition. The pooled scale estimates yield an additional diagnostic tool in the form of the ratio between pooled scales and default scales, which quantify the presence of cluster structure in the individual variables.

\section{Acknowledgements}
This work was supported by internal funds of the KU Leuven [to J.R.] and by the Discovery grant from NSERC [to R.Z.]. The authors would like to thank the Associate Editor, Prof. Dr. Jonathan Wren, and three anonymous reviewers for their useful comments.

\clearpage
\bibliographystyle{Chicago}



\clearpage
\pagenumbering{arabic}
%
\numberwithin{equation}{section} 
\section{Supplementary Material} \label{sec:B}
\renewcommand{\theequation}
   {\thesection.\arabic{equation}}
	
\subsection{Proof of Proposition 1}
\label{A:scaleInvariance}

We prove the result in the case of k-means clustering. For k-medians clustering, the proof is entirely analogous.
\begin{proof}
\textbf{Part 1: effect of scaling on k-means clustering\\}
Let $\bm x = x_1, \ldots, x_n$ be a sample of univariate observations, $k >0$ be a fixed natural number and suppose we have a solution to the k-means clustering problem for this value of $k$. Denote the centers of the clusters in this solution by $\bm \mu = \mu_1, \ldots, \mu_k$, the sets of indices of the clusters by $C_1, \ldots, C_k$ and the value of the objective function by $S_k(\bm \mu) = \sqrt{\frac{1}{n} \sum_{i=1}^{n}{d_i(\bm \mu)}}$ where $d_i(\bm \mu) = \min_{j=1,\ldots, k}{||x_i - \mu_j||_2^2}$. Let $s >0$ be a positive real number, $t\in \mathbb{R}$ a real number and consider the sample $\bm z = z_1, \ldots, z_n$ where $z_i = (x_i-t) / s$.\\

\noindent We show that the clustering given by $C_j' = C_j$ and $\bm \mu' = \mu_1', \ldots, \mu_k'$ where $\mu_j' = (\mu_j - t)/s$ is a solution to the k-means problem on $\bm z$.  Note first of all that if $C_j' = C_j$, we have $\mu_j' = (\mu_j - t) / s$ since the cluster centers are the sample means of the elements in the clusters and the sample mean is affine equivariant. Therefore, we also have that $ s \; S_k'(\bm \mu') = S_k(\bm \mu)$.\\

\noindent Suppose now that the clustering given by $\mu'$ does not solve the k-means problem on $z_1,\ldots, z_n$, i.e. there exists a clustering given by the centers $\bm \theta = \theta_1, \ldots, \theta_k$ such that $S_k'(\bm \theta) < S_k'(\bm \mu')$. Denote $d_i'(\bm \theta) = \min_{j=1,\ldots, k}{||z_i - \theta_j||_2^2}$. Now consider the partition of the original dataset $x_1,\ldots, x_n$ given by the centers $s \bm \theta +t = s \; \theta_1 +t, \ldots, s \; \theta_k+t$. We then have $S_k(s \; \bm \theta+t) = \sqrt{\sum_{i=1}^{n}{d_i(s \;\bm \theta + t)}} = s \sqrt{\sum_{i=1}^{n}{d_i'(\bm \theta)}} = s \; S_k'(\bm \theta) <  s \; S_k'(\bm \mu') = S_k(\bm \mu)$. This is a contradiction since $\bm \mu$ solves the k-means clustering of $x_1,\ldots, x_n$ and thus we cannot have a clustering with a lower objective function.\\

\textbf{Part 2: scale invariance of the gap analysis}\\
Let $k$ be fixed and consider the pooled within-cluster sum of squares $W_k(\bm x) = n S_k^2$ of the k-means clustering of $\bm x$. Let $\bm z = (\bm x - t) / s$ for a scale $s > 0$ and location $t\in \mathbb{R}$ as before. Then   
$W_k(\bm z) = W_k(\bm x) / s^2$ and so 
\begin{align*}
\Gap_{\bm z}(k) &=&  E_{U\left[z_{(1)}, z_{(n)}\right]}\left[\log\left(W_k\right)\right] - \log\left(W_k(\bm z)\right)\\
&=&E_{U\left[x_{(1)}, x_{(n)}\right]}\left[\log\left(W_k\right)- 2\log(s)\right] -\\
&&\left(\log\left(W_k(\bm x)\right)- 2\log(s)\right)\\
&=& E_{U\left[x_{(1)}, x_{(n)}\right]}\left[\log\left(W_k\right)\right] - \log\left(W_k(\bm x)\right) = \Gap_{\bm x}(k).
\end{align*}
Furthermore, we also have
\begin{align*}
\Var_{U\left[z_{(1)}, z_{(n)}\right]}\left[\log\left(W_k\right)\right] &= \Var_{U\left[x_{(1)}, x_{(n)}\right]}\left[\log\left(W_k\right) - 2\log(s) \right]\\ &= \Var_{U\left[x_{(1)}, x_{(n)}\right]}\left[\log\left(W_k\right)\right].
\end{align*}
So for every value of $k$, both the value of the gap statistic and the estimated variance of $\log\left(W_k\right)$ are invariant under the rescaling. Therefore, the optimal number of clusters resulting from the analysis based on the gap statistic is the same for $\bm x$ and $\bm z$.
\end{proof}

\clearpage
\subsection{Additional simulation results}
\label{B:simul}

\begin{figure}[!htb]
\centering
\begin{subfigure}[b]{0.49\linewidth}
  \centering 
	\includegraphics[width=0.99\textwidth]{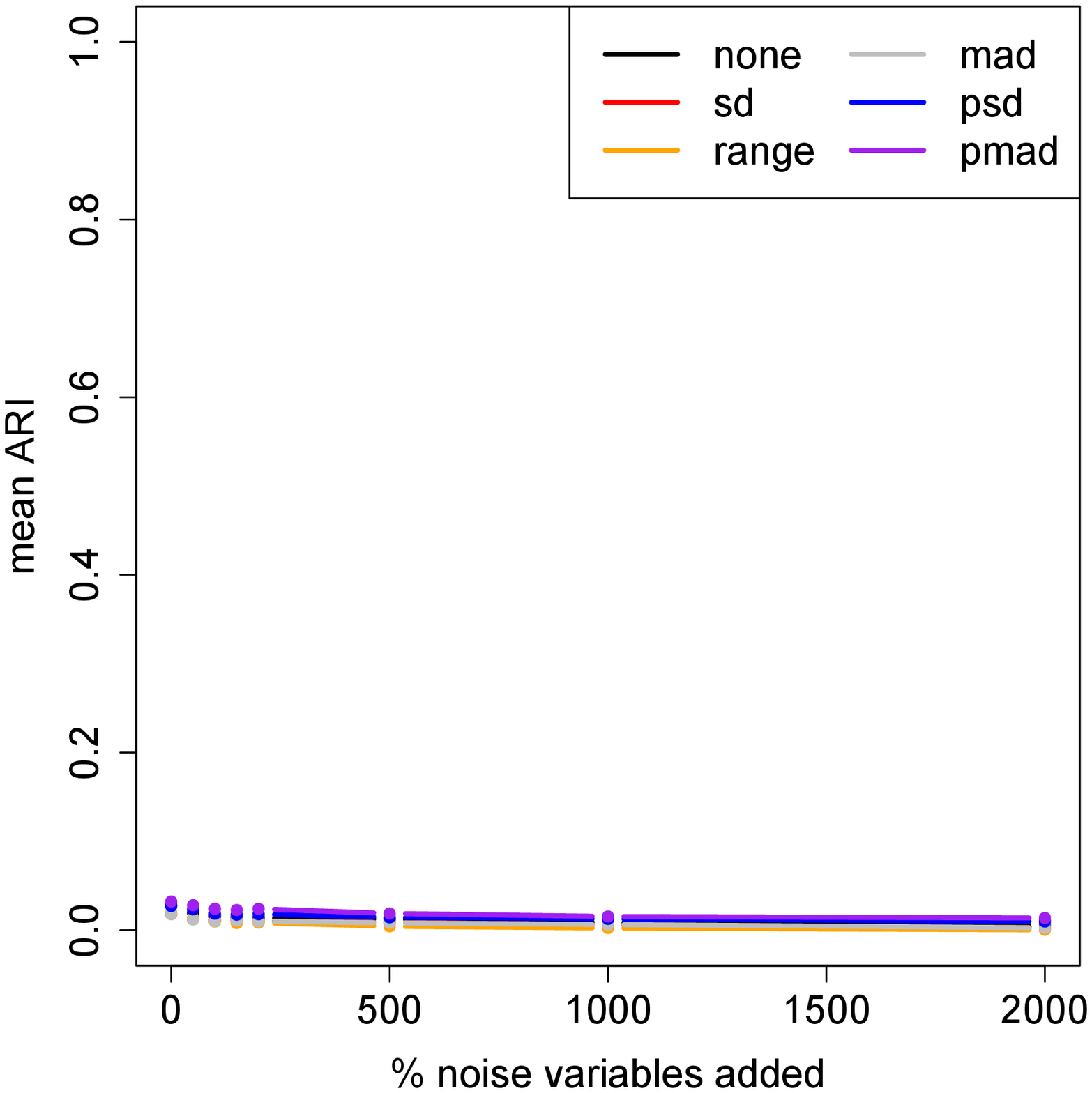}
	\caption{Single linkage}
\vspace{0.25cm}
\end{subfigure}
\begin{subfigure}[b]{0.49\linewidth}
  \centering 
  \includegraphics[width=0.99\textwidth]{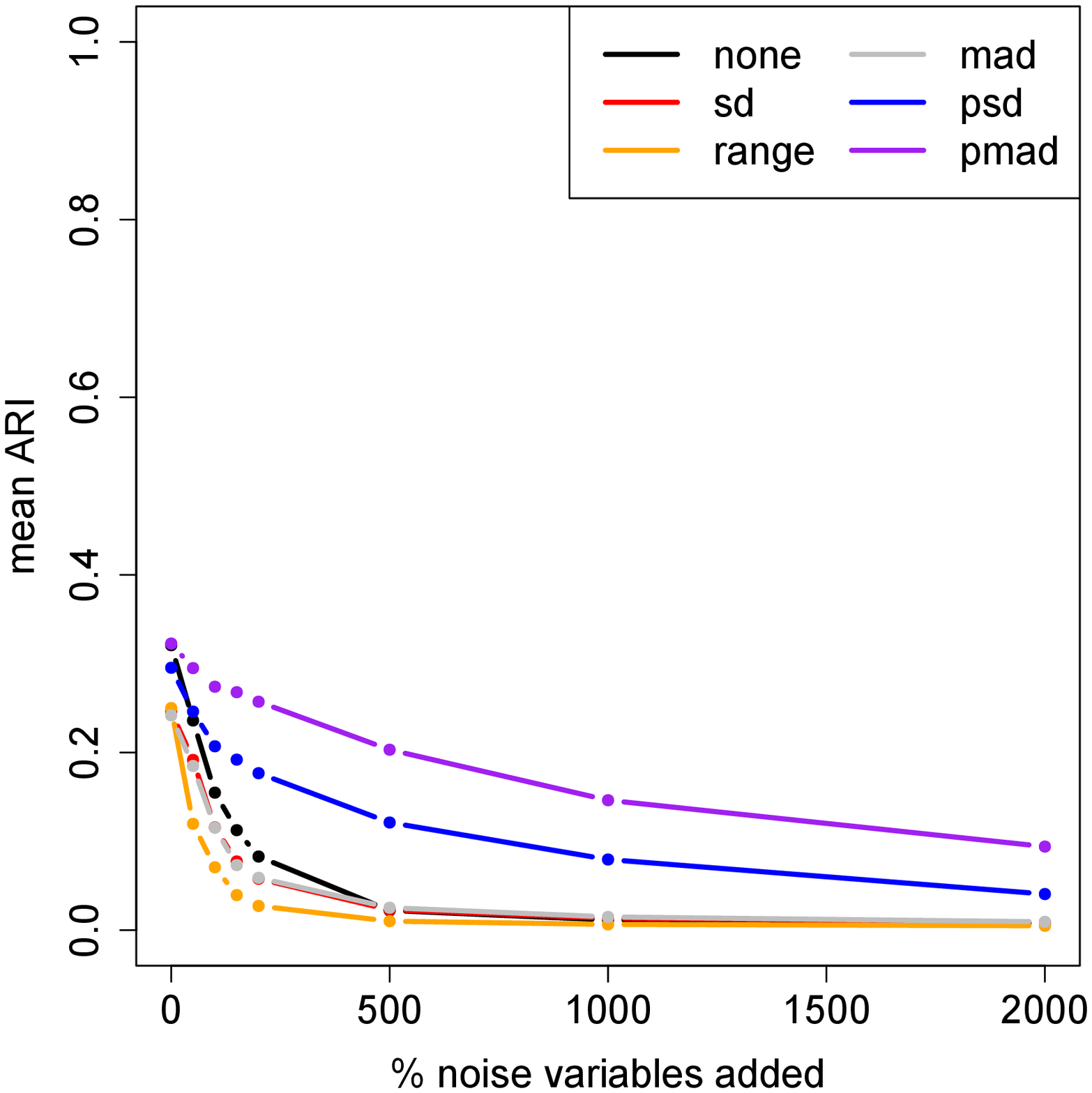} 
	  \caption{Average linkage}
\vspace{0.25cm}
\end{subfigure}
\begin{subfigure}[b]{0.49\linewidth}
  \centering 
  \includegraphics[width=0.99\textwidth]{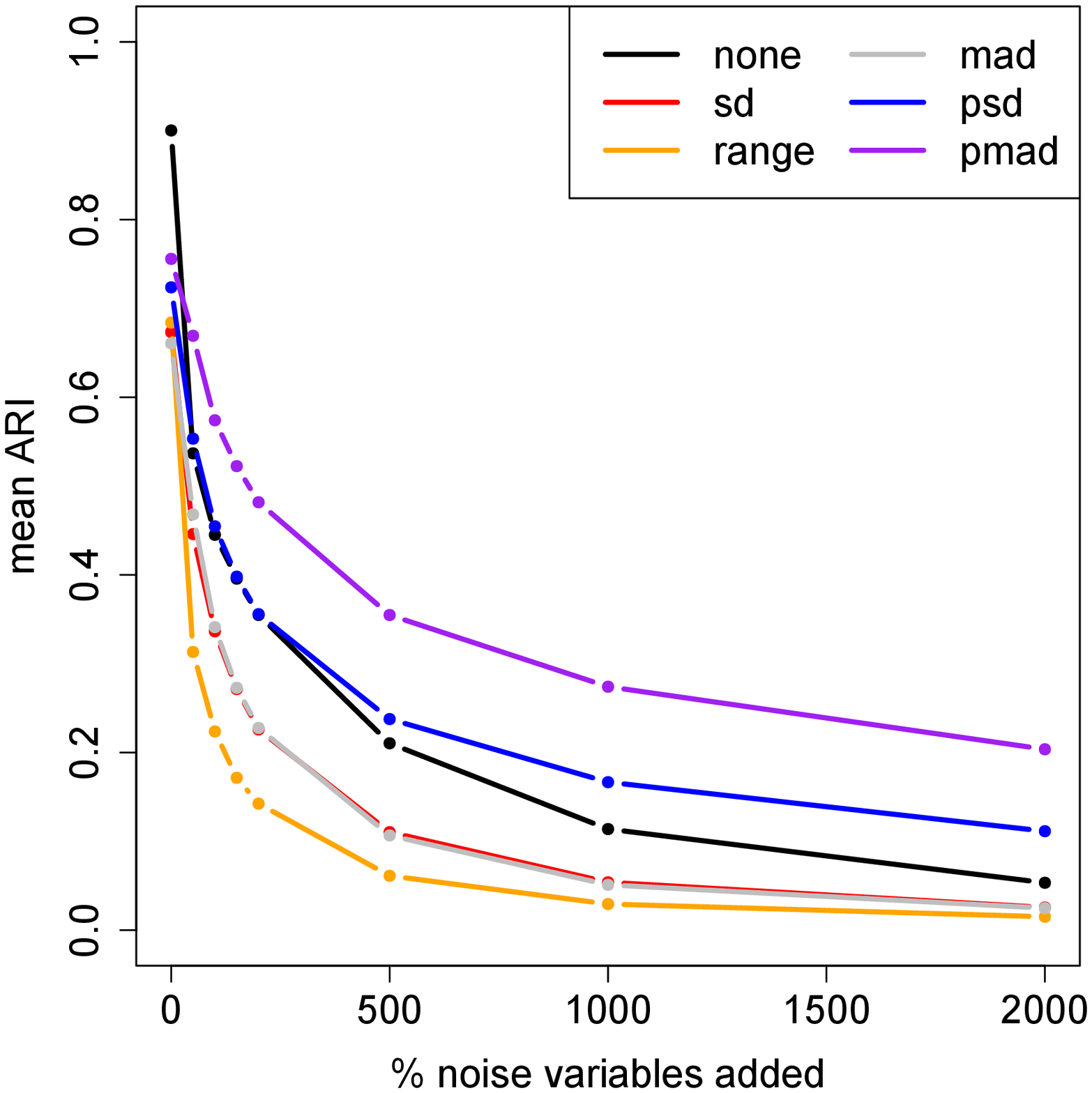} 
	  \caption{Complete linkage}
\end{subfigure}
\begin{subfigure}[b]{0.49\linewidth}
  \centering 
  \includegraphics[width=0.99\textwidth]{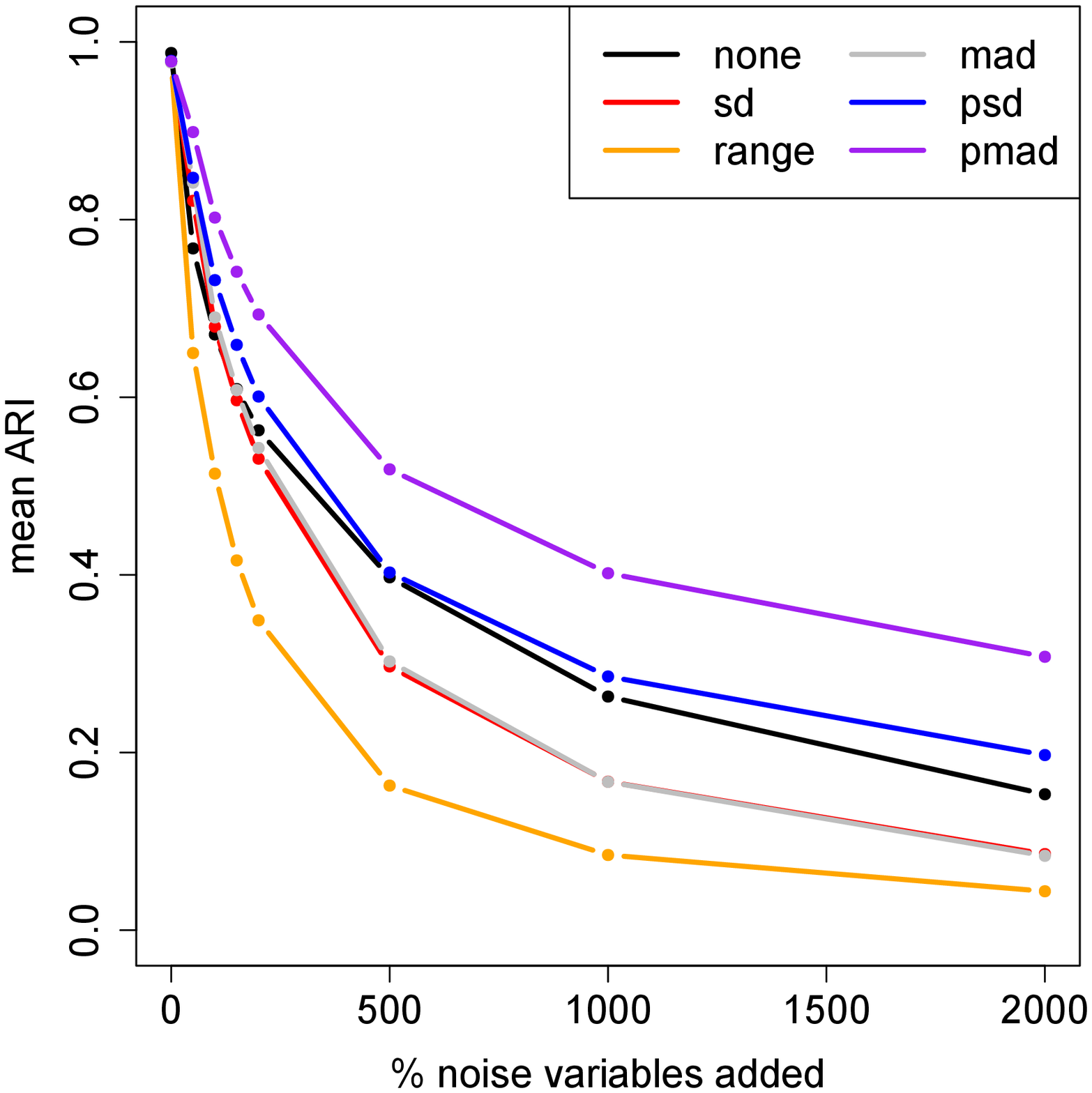} 
	  \caption{Ward linkage}
\end{subfigure}
\caption{Simulation results for hierarchical clustering with single (a), average (b), complete (c) and Ward (d) linkage functions on data with 5 \% outliers. The pooled scale estimators perform the best and the pmad is more robust to outliers than the psd.}
\label{fig:sim_hc_outl}
\end{figure}

\begin{figure}[!htb]
\centering
\begin{subfigure}[b]{0.49\linewidth}
  \centering 
	\includegraphics[width=0.99\textwidth]{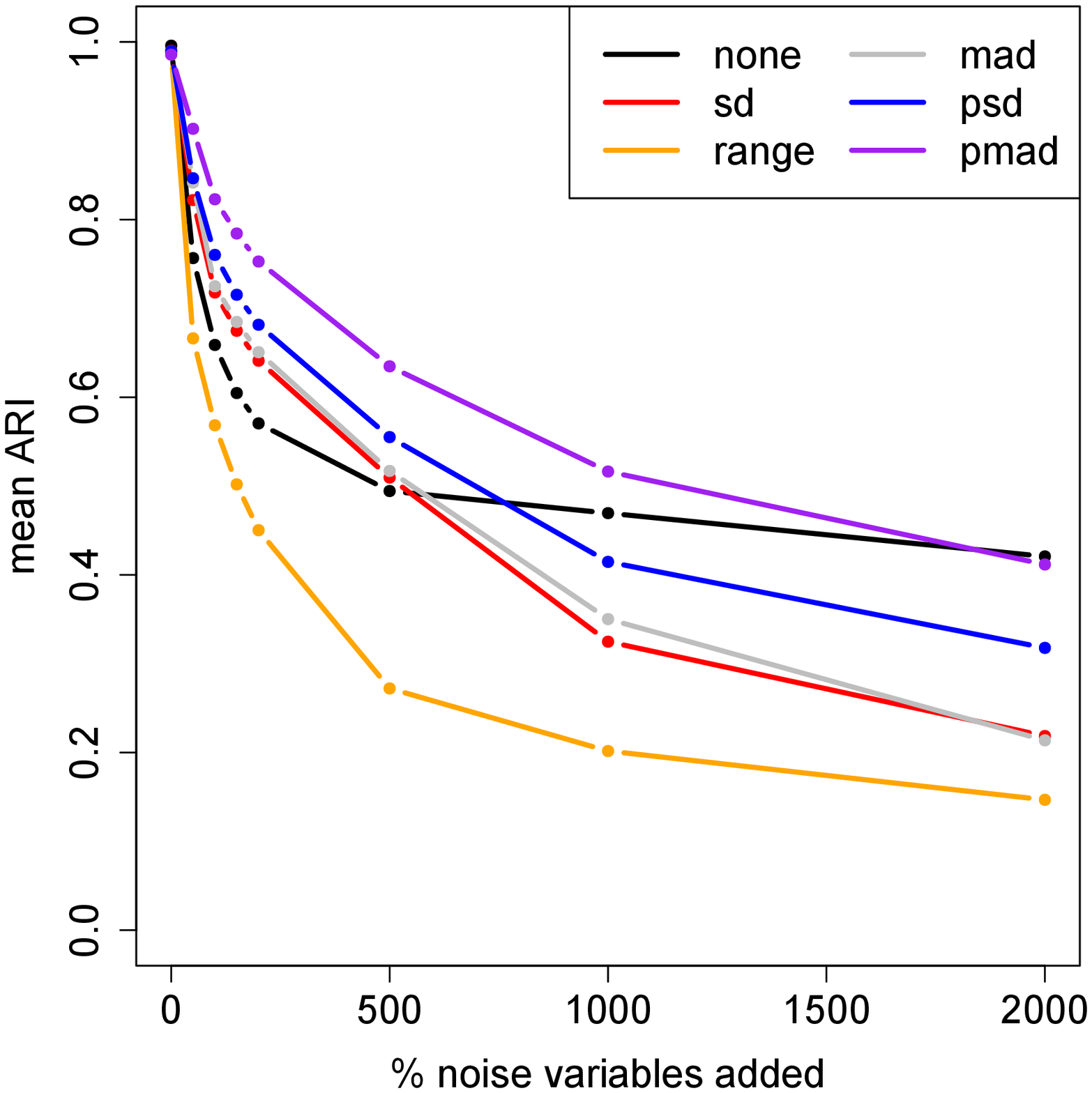}
	\caption{$k$-means}
\end{subfigure}
\begin{subfigure}[b]{0.49\linewidth}
  \centering 
  \includegraphics[width=0.99\textwidth]{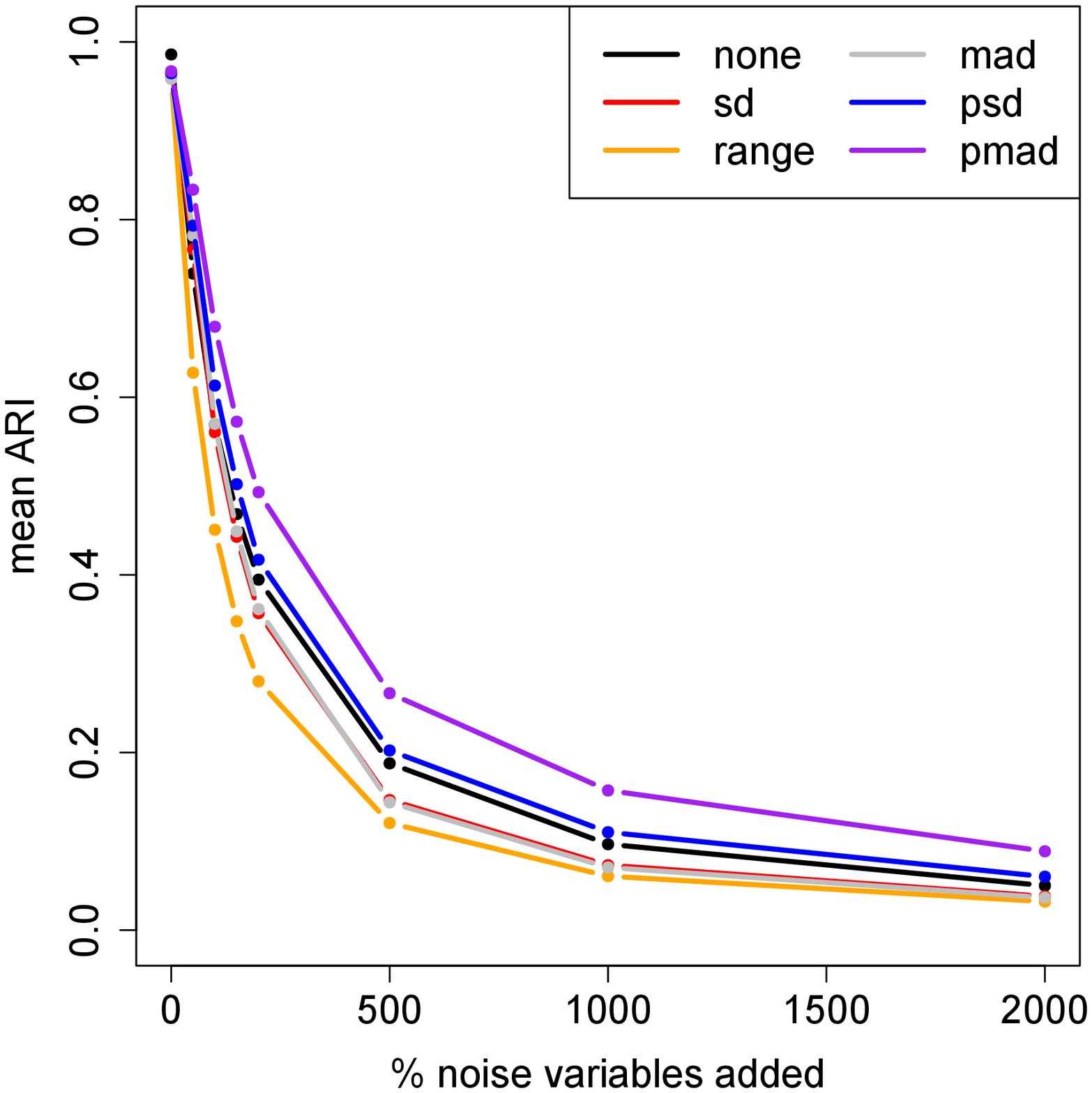} 
	  \caption{Partitioning around medoids}
\end{subfigure}
\caption{Simulation results for $k$-means (a) and partitioning around medoids (b) on data with 5 \% outliers. The pooled scale estimators perform better than the alternatives and the pmad is clearly more robust to outliers than the psd.}
\label{fig:sim_km_outl}
\end{figure}

\clearpage
\subsection{Additional real data examples}

\subsubsection{Lymphoma data}
We analyze the lymphoma dataset first studied in \cite{Alizadeh2000} and publicly available in the \texttt{R}-package \texttt{spls} \citep{SPLS2019}. The dataset contains 4026 gene expression levels for 62 samples of 3 types of lymphoma. More precisely, there are 42 samples of diffuse large B-cell lymphoma, 9 samples of follicular lymphoma (FL), and 11 samples of chronic lymphocytic leukemia (CLL). The data was preprocessed as in \cite{Dettling2002}.\\

We use the sparse hierarchical clustering with complete linkage of \cite{Witten2010} to cluster the data after scaling them with various estimators of scale. The dataset was also studied by \cite{Chung2010} in the context of sparse PLS. The two best performing methods suggested that 50 or 197 features contain most of the information needed to find the groups in the data. We therefore cluster the dataset twice and fix the number of selected features in the sparse hierarchical clustering to be 50 and 197.\\

\begin{figure}[!htb]
\centering
\begin{subfigure}[b]{0.49\linewidth}
  \centering 
	\includegraphics[width=0.99\textwidth]{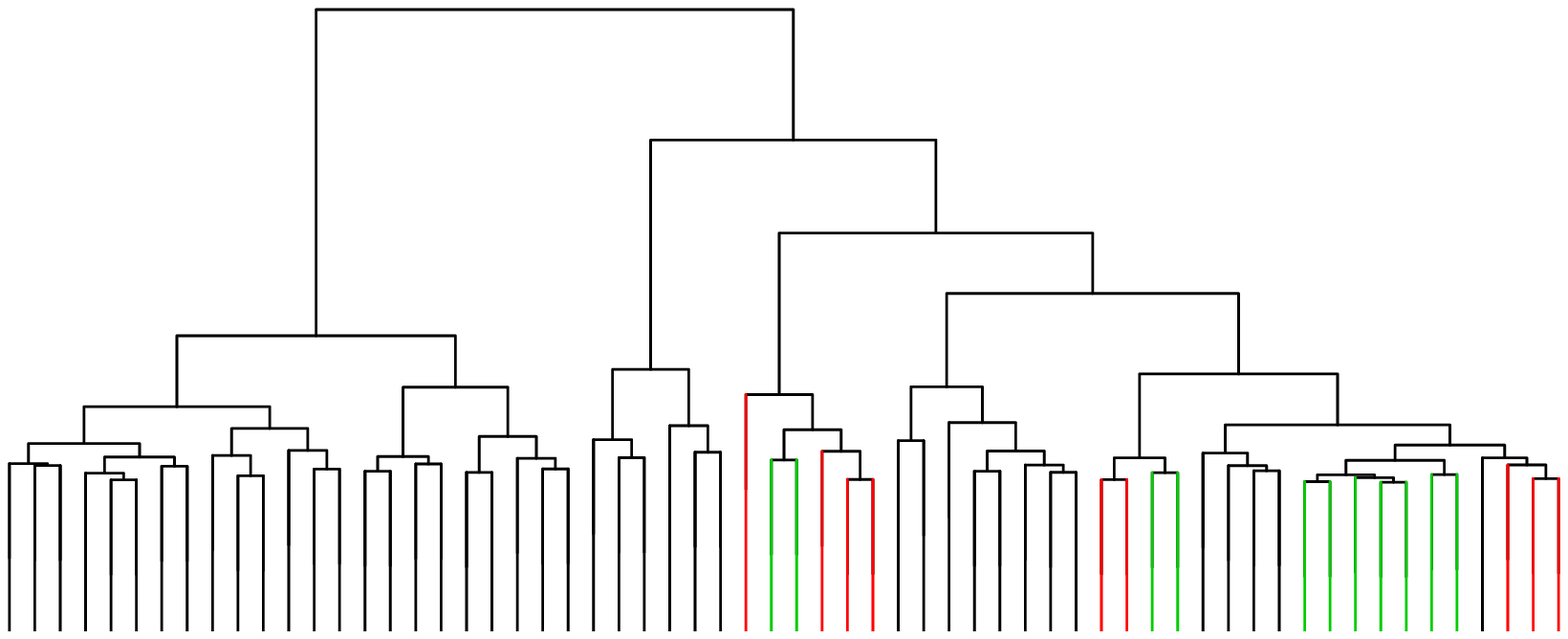}
	\caption{No scaling}
\vspace{0.25cm}
\end{subfigure}
\begin{subfigure}[b]{0.49\linewidth}
  \centering 
  \includegraphics[width=0.99\textwidth]{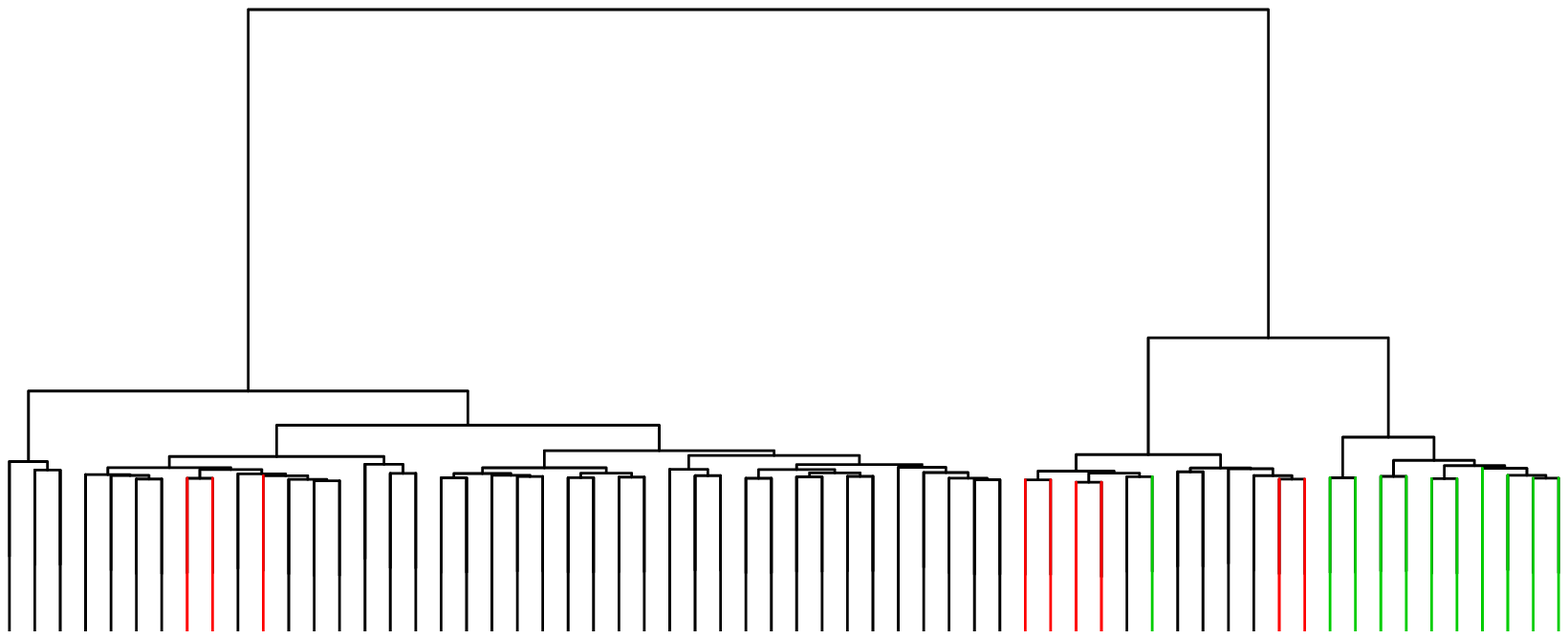} 
	  \caption{Standard deviation}
\vspace{0.25cm}
\end{subfigure}
\begin{subfigure}[b]{0.49\linewidth}
  \centering 
  \includegraphics[width=0.99\textwidth]{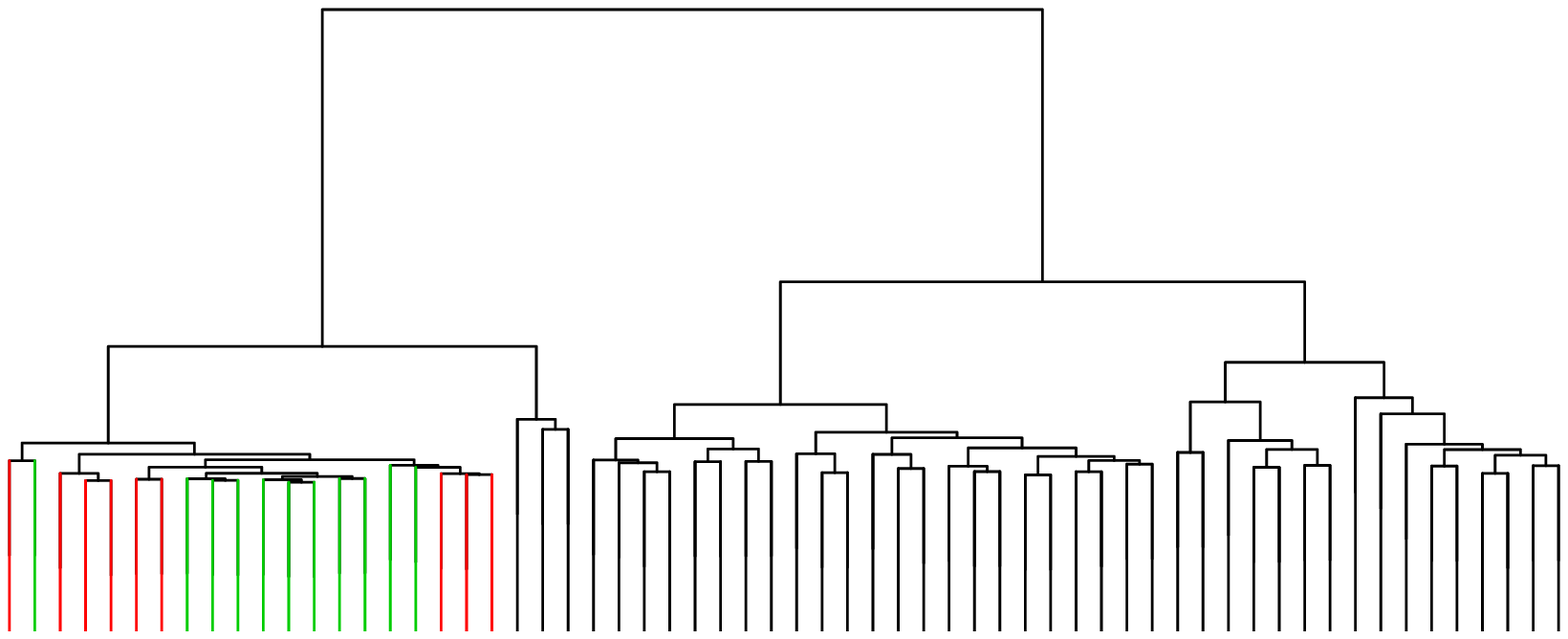} 
	  \caption{Range}
\end{subfigure}
\begin{subfigure}[b]{0.49\linewidth}
  \centering 
  \includegraphics[width=0.99\textwidth]{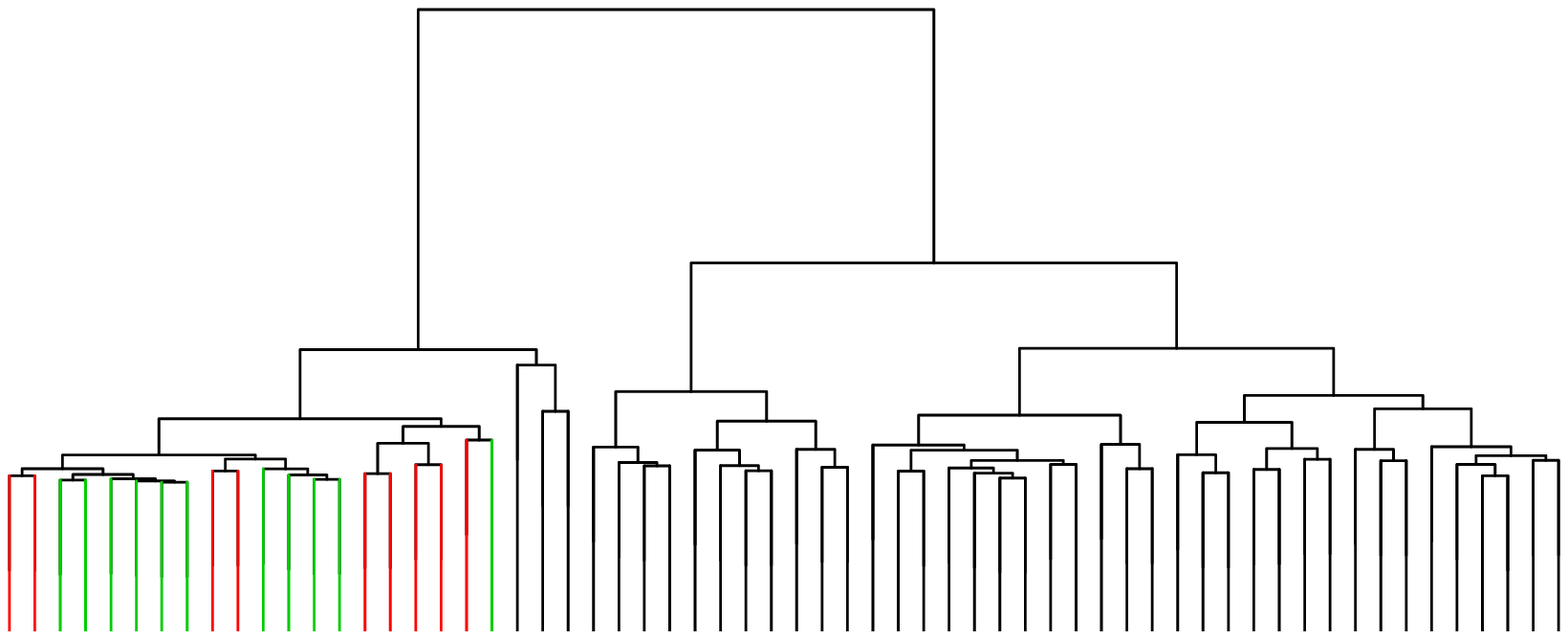} 
	  \caption{Pooled standard deviation}
\end{subfigure}
\caption{Sparse hierarchical clustering with complete linkage on the lymphoma dataset using 50 features. Scaling with the pooled standard deviation yields the lowest number of misclassified observations.}
\label{fig:lymphoma50}
\end{figure}

\begin{figure}[!htb]
\centering
\begin{subfigure}[b]{0.49\linewidth}
  \centering 
	\includegraphics[width=0.99\textwidth]{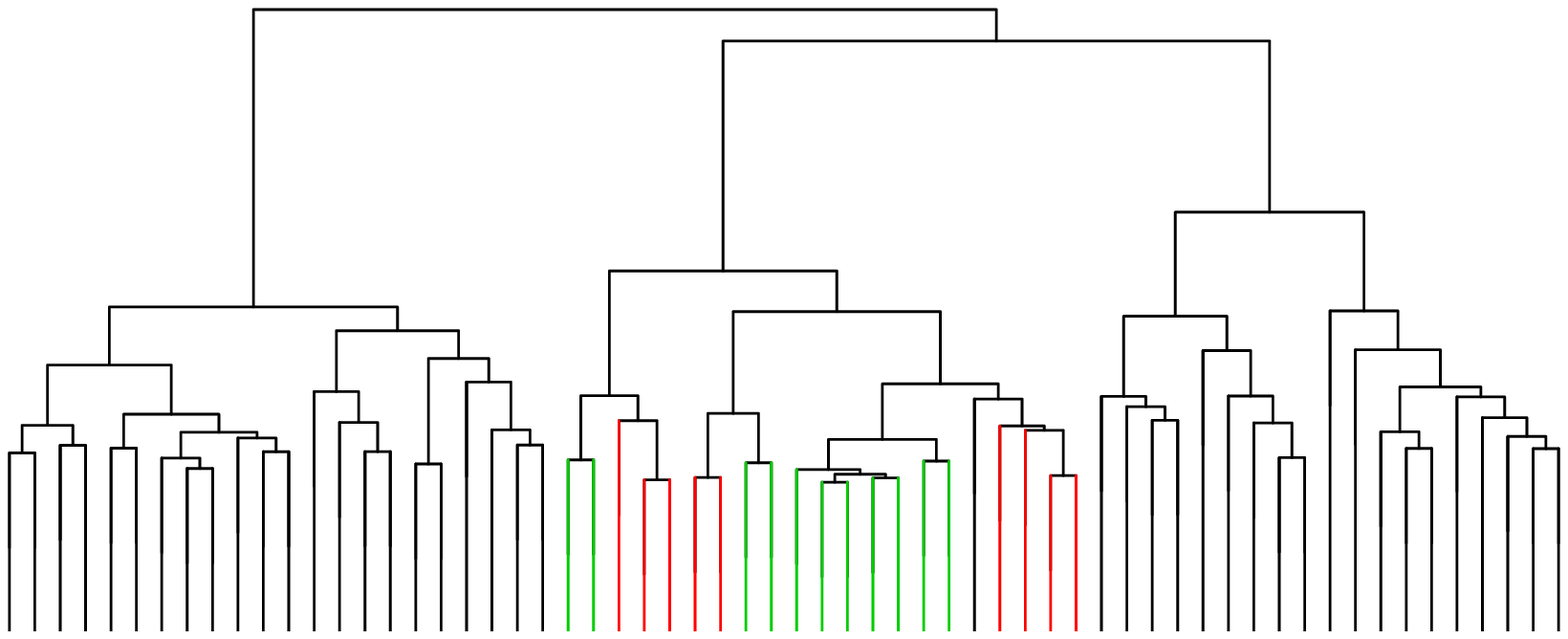}
	\caption{No scaling}
\vspace{0.25cm}
\end{subfigure}
\begin{subfigure}[b]{0.49\linewidth}
  \centering 
  \includegraphics[width=0.99\textwidth]{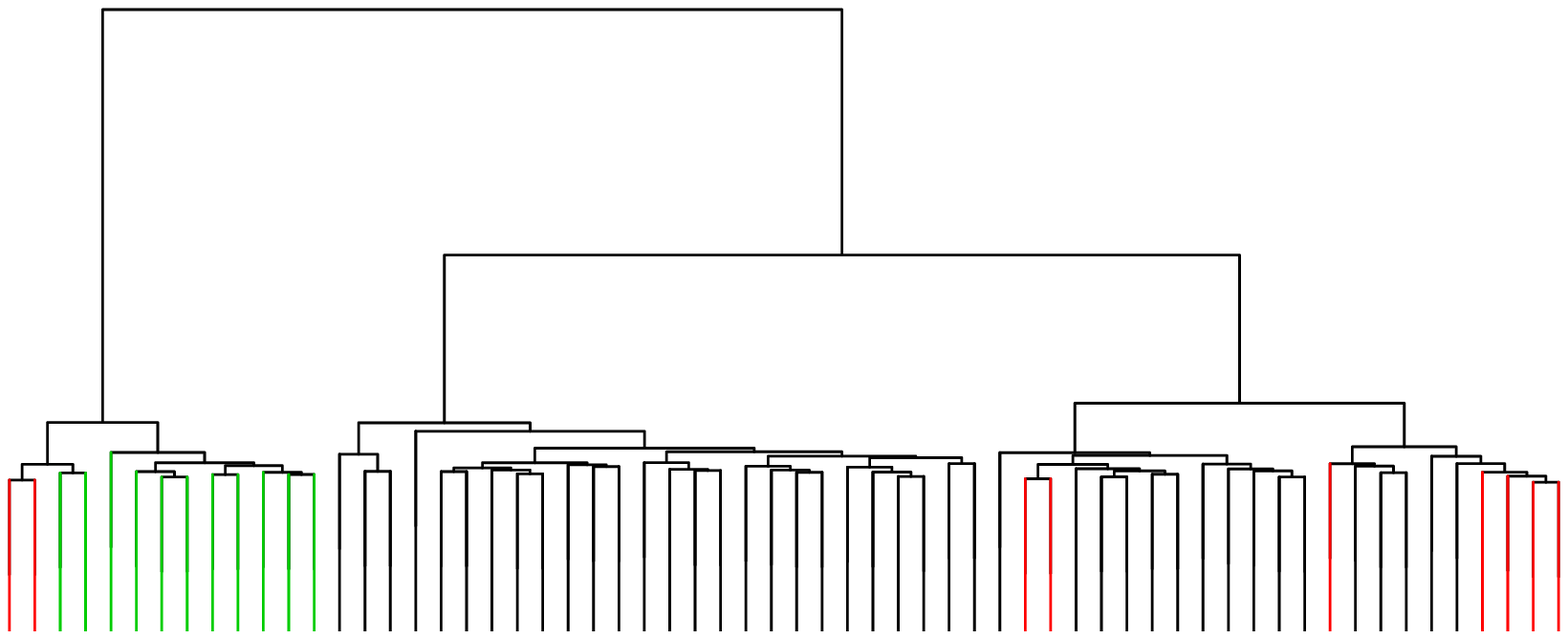} 
	  \caption{Standard deviation}
\vspace{0.25cm}
\end{subfigure}
\begin{subfigure}[b]{0.49\linewidth}
  \centering 
  \includegraphics[width=0.99\textwidth]{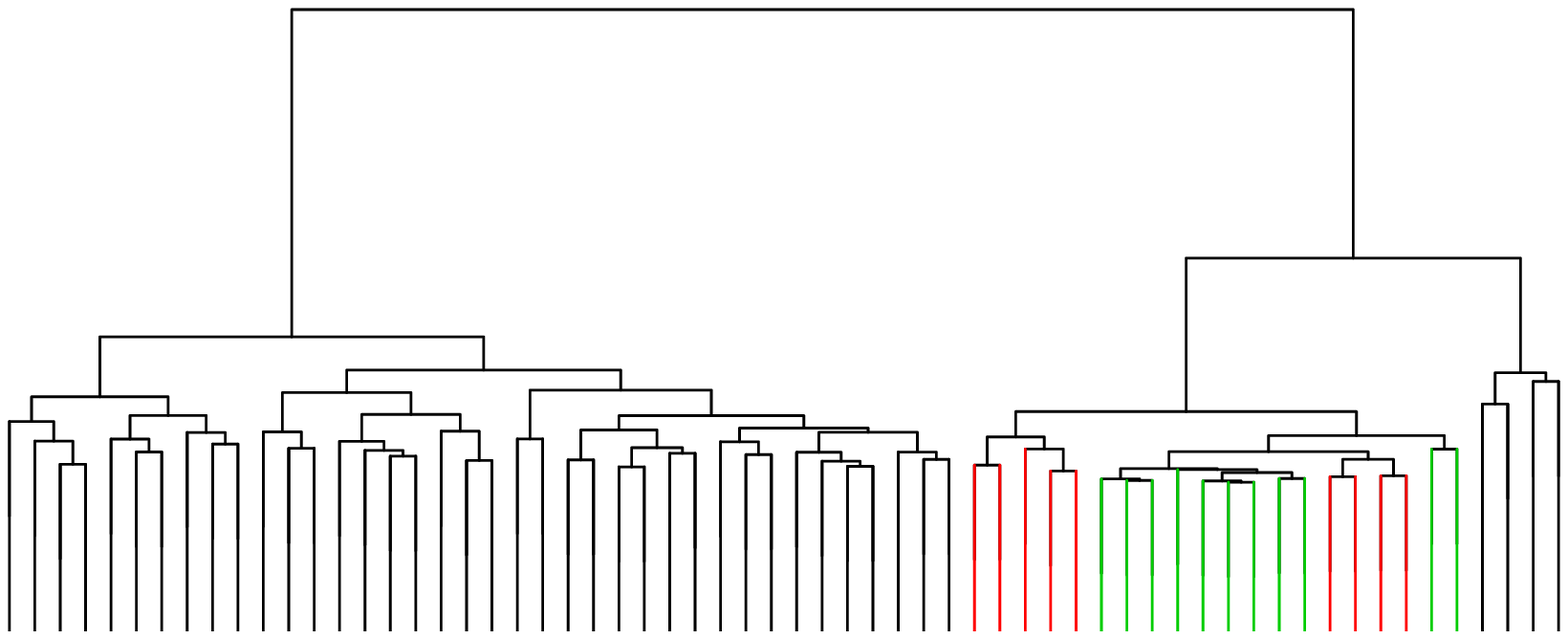} 
	  \caption{Range}
\end{subfigure}
\begin{subfigure}[b]{0.49\linewidth}
  \centering 
  \includegraphics[width=0.99\textwidth]{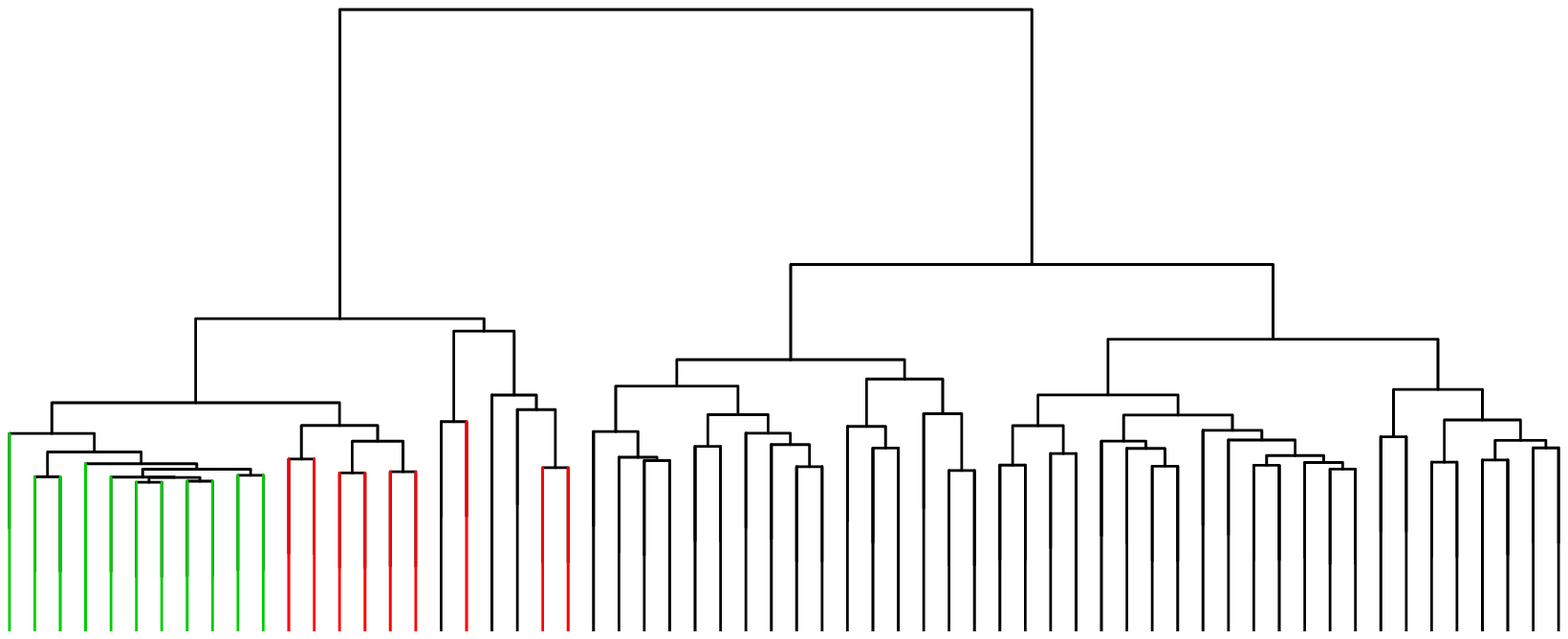} 
	  \caption{Pooled standard deviation}
\end{subfigure}
\caption{Sparse hierarchical clustering with complete linkage on the lymphoma dataset using 197 features. Again, scaling with the pooled standard deviation yields the lowest number of misclassified observations.}
\label{fig:lymphoma197}
\end{figure}

Figure \ref{fig:lymphoma50} presents the result when choosing 50 features. Scaling with the pooled standard deviation yields 8 misclassified samples, the classical standard deviation gives 9 misclassified observations and the range 10. Without scaling, very little of the true clustering structure is recovered. When using 197 genes, the performance improves for the range and the pooled standard deviation, with 8 and 6 misclassified samples respectively (see Figure \ref{fig:lymphoma197}). The classical standard deviation fails to identify the smallest group of follicular lymphoma and not scaling gives mixed clusters. 
In summary, scaling with the pooled standard deviation yields the best cluster recovery.

\subsubsection{Volatile Organic Compound Metabolites data}

\begin{figure}[!htb]
\centering
	\includegraphics[width=0.45\textwidth]{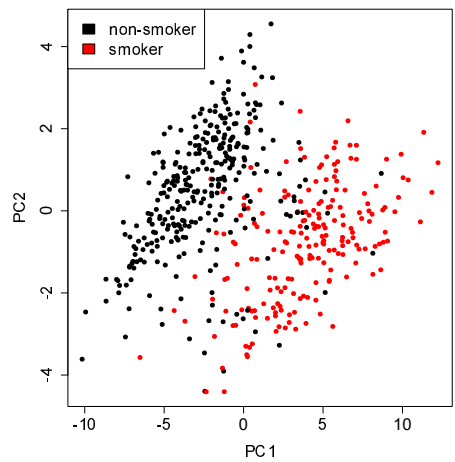}
\caption{Two first score vectors resulting from classical PCA on the VOC data. The two groups of smokers vs. non-smokers are clearly visible.}
\label{fig:PCAVOC}
\end{figure}

This dataset contains levels of volatile organic compounds (VOCs) in human urinary samples. The data was collected for the period 2015-2016 and is publicly available at the website of the National Health and Nutrition Examination Survey (NHANES), see \cite{NHANES}. It is known that long-term exposure to VOCs can lead to cancer and neurocognitive dysfunction. One of the most common causes of suspicious levels of VOCs is exposure to (tobacco) smoke. We therefore matched the patients in the dataset of VOCs with the available background information on their smoking behavior and defined two  ``true clusters'':  heavy smokers and non-smokers. The plot of the first two score vectors from a principal component analysis shown in Figure \ref{fig:PCAVOC} confirms that this grouping is in fact present in the data. Before clustering, we preprocessed the data by removing the observations which contained missing values as well as the variables with extremely low variance. We obtained a dataset of 20 levels of different VOCs for 522 patients. The names of the VOCs are listed in Table \ref{table:VOCs} \\

\begin{table}[h!]
\scalebox{0.9}{
\begin{tabular}{ll}
\hline
Variable Name & VOC name \\
\hline
URX2MH & 2-Methylhippuric acid\\
URX34M & 3- and 4-Methylhippuric acid\\
URXAAM & N-Acetyl-S-(2-carbamoylethyl)-L-cysteine\\
URXAMC & N-Acetyl-S-(N-methylcarbamoyl)-L-cysteine\\
URXATC & 2-Aminothiazoline-4-carboxylic acid\\
URXBMA & N-Acetyl-S-(benzyl)-L-cysteine\\
URXBPM & N-Acetyl-S-(n-propyl)-L-cysteine\\
URXCEM & N-Acetyl-S-(2-carboxyethyl)-L-cysteine\\
URXCYM & N-Acetyl-S-(2-cyanoethyl)-L-cysteine\\
URXDHB & N-Acetyl-S-(3,4-dihydroxybutyl)-L-cysteine\\
URXHEM & N-Acetyl-S-(2-hydroxyethyl)-L-cysteine\\
URXHP2 & N-Acetyl-S-(2-hydroxypropyl)-L-cysteine\\
URXHPM & N-Acetyl-S-(3-hydroxypropyl)-L-cysteine\\
URXIPM1 & N-Acetyl- S- (4- hydroxy- 2- methyl- 2- butenyl)- L-cysteine\\
URXIPM3 & N-Acetyl- S- (4- hydroxy- 2- methyl- 2- butenyl)-L-cysteine\\
URXMAD & Mandelic acid\\
URXMB3 & N-Acetyl-S-(4-hydroxy-2-butenyl)-L-cysteine\\
URXPHE & N-Acetyl-S-(phenyl-2-hydroxyethyl)-L-cysteine\\
URXPHG & Phenylglyoxylic acid \\
URXPMM & N-Acetyl-S-(3-hydroxypropyl-1-methyl)-L-cysteine\\
\hline
\end{tabular}
}
\caption{The volatile organic compounds in the VOC data.}
\label{table:VOCs}
\end{table}

\begin{figure}[!htb]
\centering
	\includegraphics[width=0.45\textwidth]{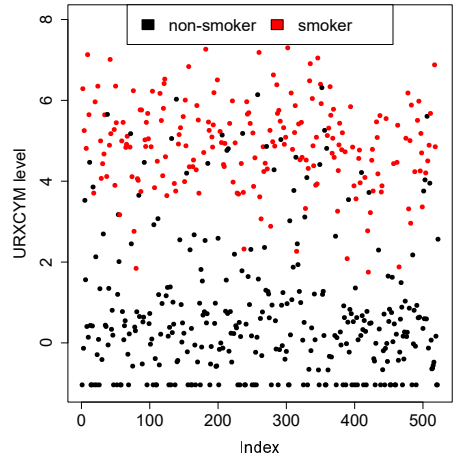}
\caption{The URXCYM variable for which the pooled standard deviation is clearly lower than its classical counterpart plays an important role in the clustering.}
\label{fig:URXCYM}
\end{figure}

The clustering was done using regular k-means clustering with $k = 2$. Without scaling the data, the ARI value for the 2-means is 0.54. When scaling with the standard deviation, this drops down to 0.37. Scaling with the range yields an ARI of 0.47, while scaling with the pooled standard deviation has an ARI of 0.55. Surprisingly, there is only one variable for which the pooled standard deviation was smaller than the classical standard deviation. This variable is ``URXCYM'', which is the variable name for N-Acetyl-S-(2-cyanoethyl)-L-cysteine. It turns out that this is a well-known bio-marker for exposure to smoke, see e.g. \cite{Chen2019}, since it typically results from the metabolization of acrylonitrile, a volatile liquid present in tobacco smoke. As can be seen in Figure \ref{fig:URXCYM}, this variable indeed separates the majority of the smokers from the non-smokers. In conclusion, not only does scaling with the pooled standard deviation yield the best clustering results, it also flags a known important bio-marker for exposure to  smoking.

\subsubsection{Leukemia data}

We analyze the Leukemia data first studied by \cite{Golub1999}. The data consists of the gene expression levels of 3051 genes for 38 patients, 27 with acute myeloid leukemia (AML) and 11 with acute lymphoblastic leukemia (ALL). The dataset is publicly available in the \texttt{R}-package \texttt{plsgenomics} \citep{PLSgen2018}.\\

\begin{figure}[!htb]
\centering
\begin{subfigure}[b]{0.49\linewidth}
  \centering 
	\includegraphics[width=0.99\textwidth]{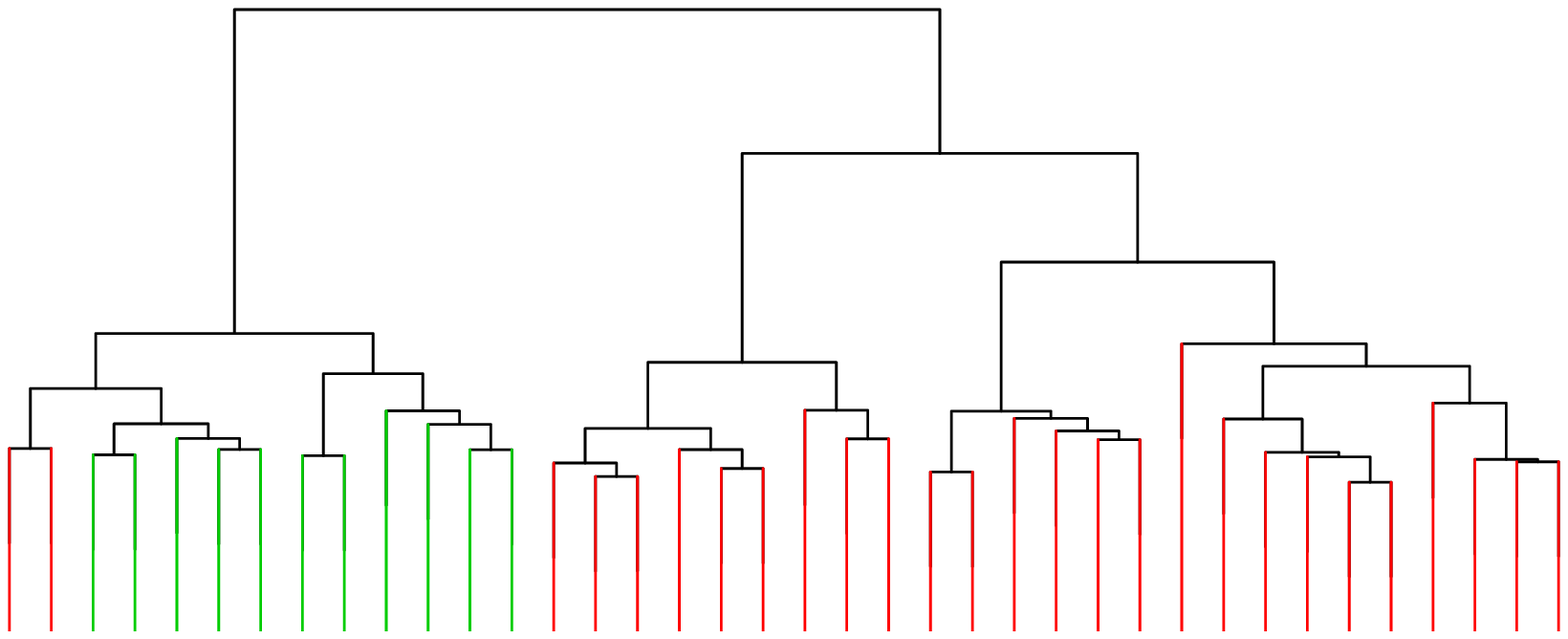}
	\caption{No scaling}
\vspace{0.25cm}
\end{subfigure}
\begin{subfigure}[b]{0.49\linewidth}
  \centering 
  \includegraphics[width=0.99\textwidth]{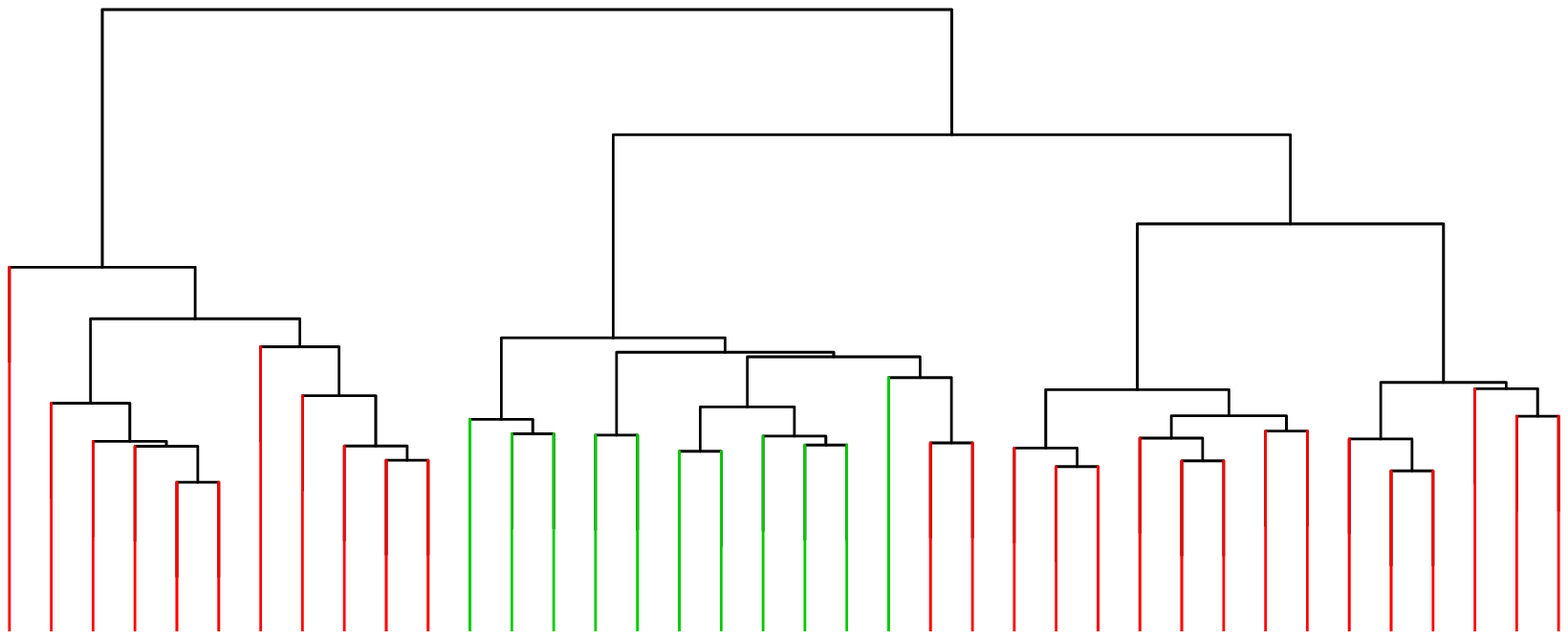} 
	  \caption{Standard deviation}
\vspace{0.25cm}
\end{subfigure}
\begin{subfigure}[b]{0.49\linewidth}
  \centering 
  \includegraphics[width=0.99\textwidth]{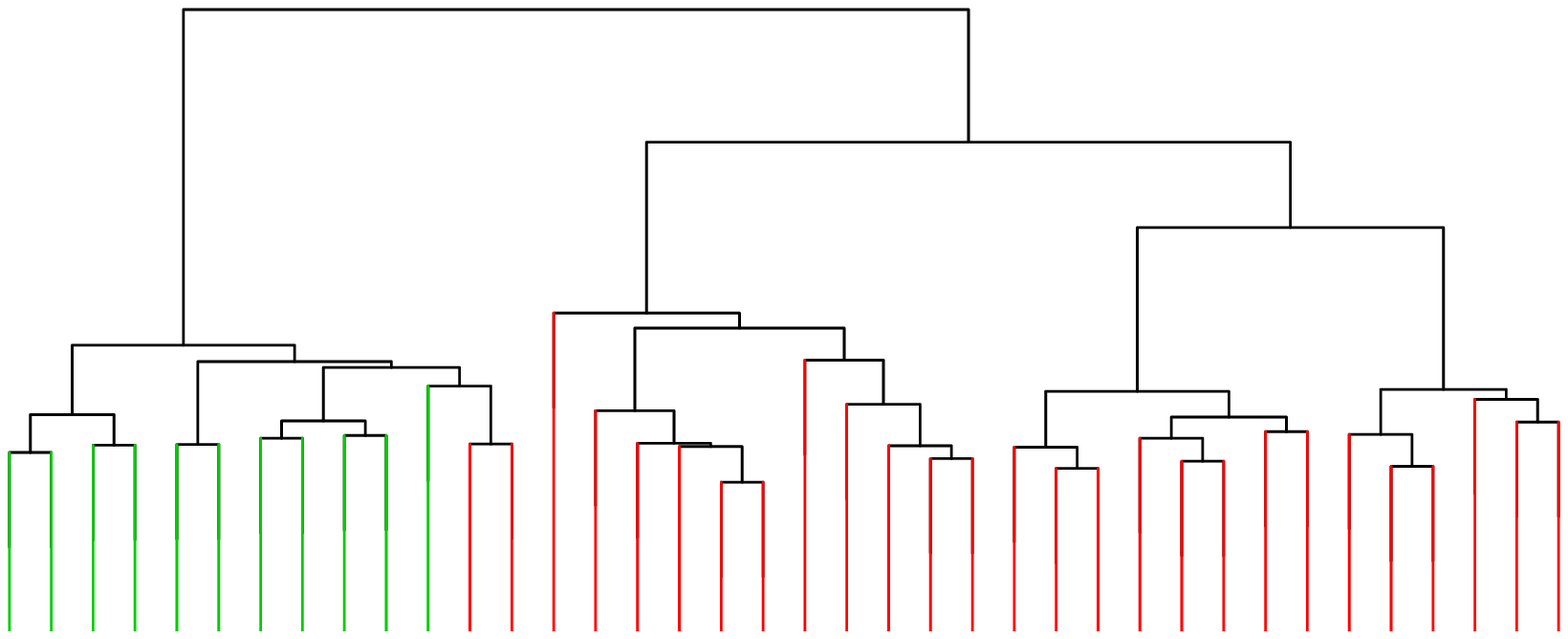} 
	  \caption{Range}
\end{subfigure}
\begin{subfigure}[b]{0.49\linewidth}
  \centering 
  \includegraphics[width=0.99\textwidth]{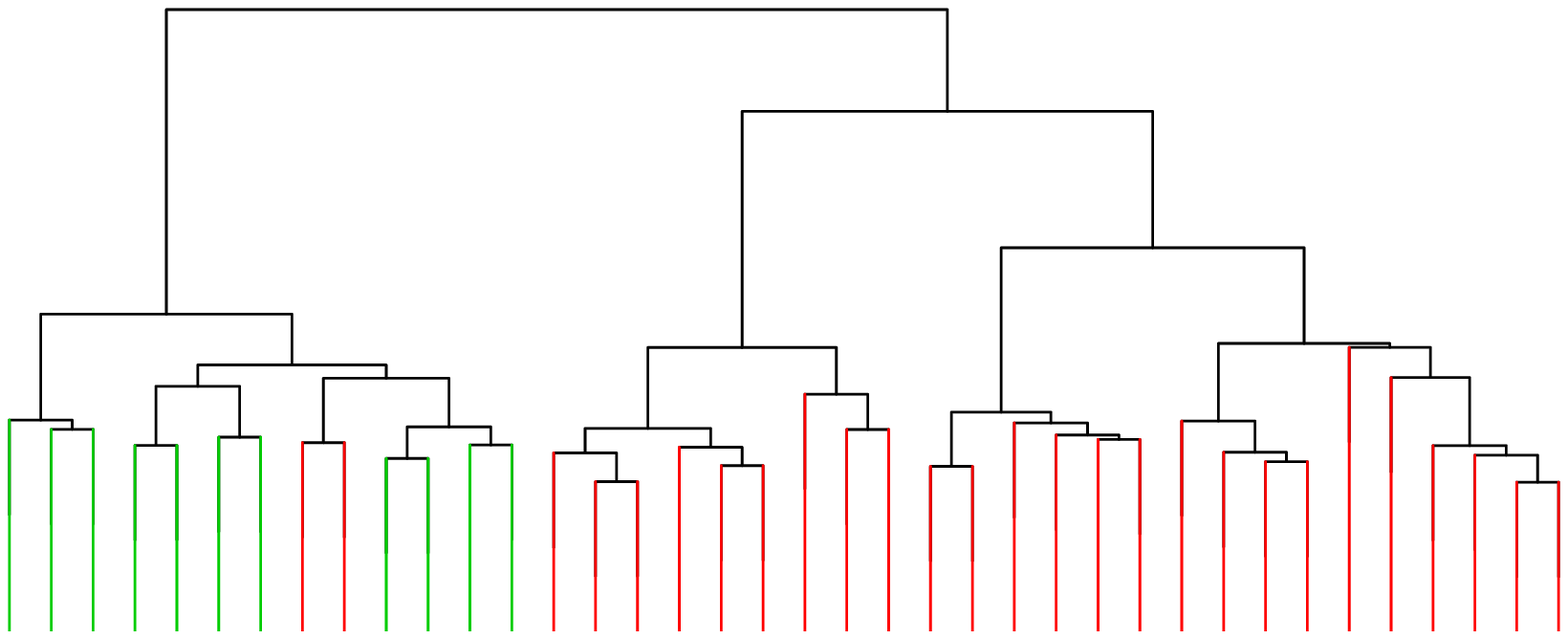} 
	  \caption{Pooled standard deviation}
\end{subfigure}
\caption{Hierarchical clustering with Ward linkage on the leukemia data. All methods work equally well, with the exception of the classical standard deviation.}
\label{fig:leukemiaWard}
\end{figure}

\begin{figure}[!htb]
\centering
\begin{subfigure}[b]{0.49\linewidth}
  \centering 
	\includegraphics[width=0.99\textwidth]{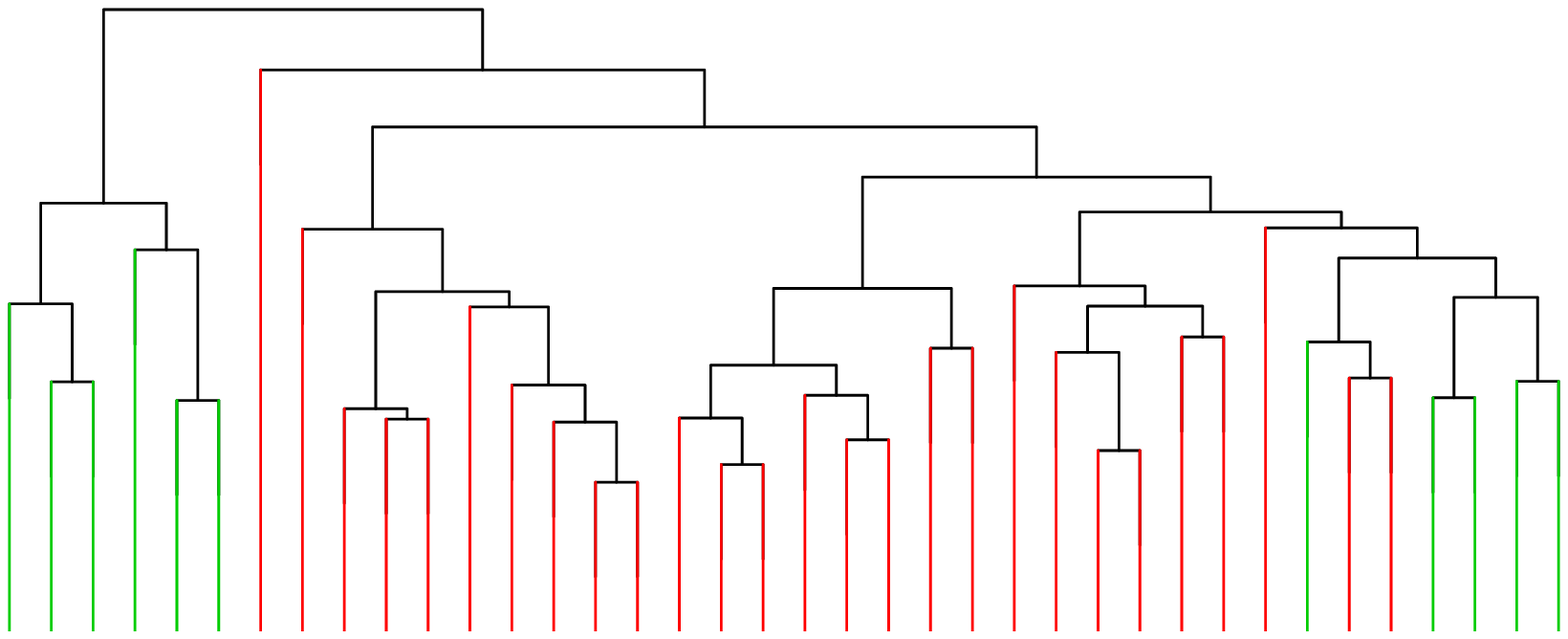}
	\caption{No scaling}
\vspace{0.25cm}
\end{subfigure}
\begin{subfigure}[b]{0.49\linewidth}
  \centering 
  \includegraphics[width=0.99\textwidth]{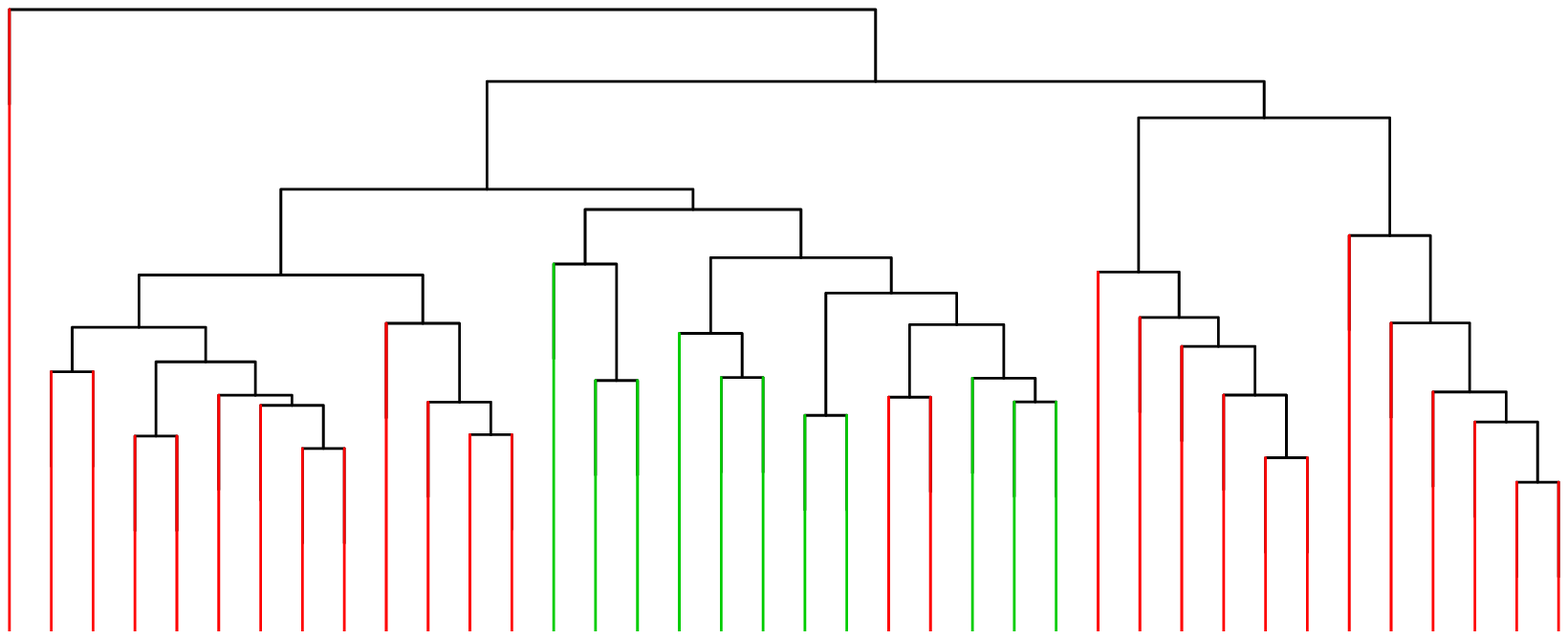} 
	  \caption{Standard deviation}
\vspace{0.25cm}
\end{subfigure}
\begin{subfigure}[b]{0.49\linewidth}
  \centering 
  \includegraphics[width=0.99\textwidth]{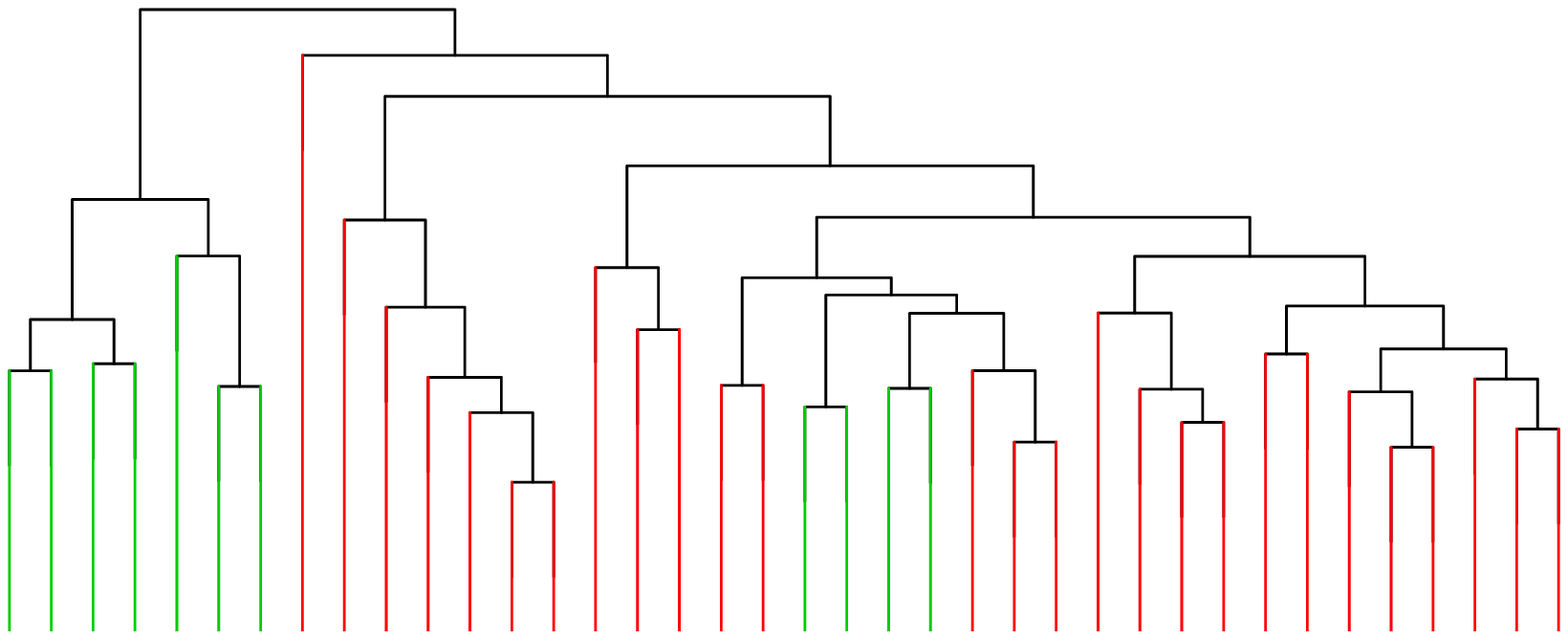} 
	  \caption{Range}
\end{subfigure}
\begin{subfigure}[b]{0.49\linewidth}
  \centering 
  \includegraphics[width=0.99\textwidth]{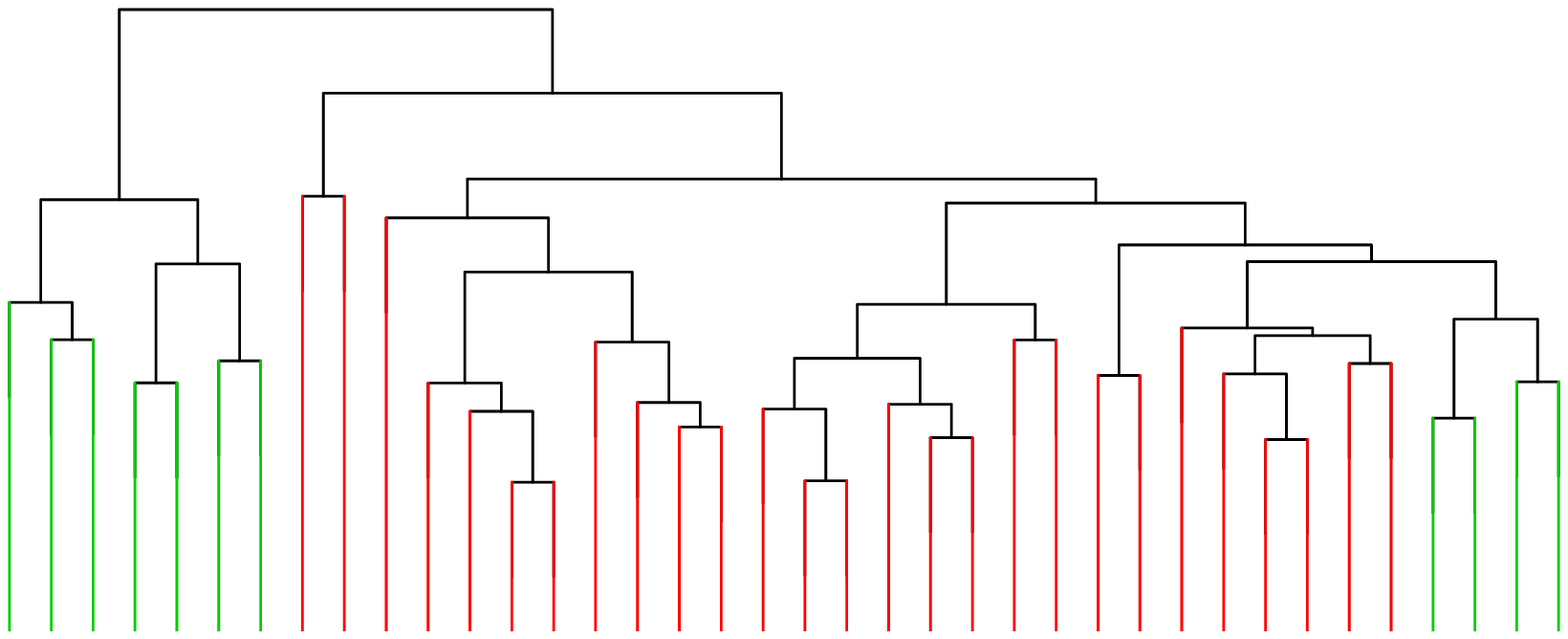} 
	  \caption{Pooled standard deviation}
\end{subfigure}
\caption{Hierarchical clustering with complete linkage on the leukemia data. Scaling with the pooled standard deviation and the range gives the lowest number of misclassified patients.}
\label{fig:leukemiaComplete}
\end{figure}

We apply hierarchical clustering to identify the two groups of patients with a different type of leukemia and again use four different scale estimators to scale the variables before clustering. Figure \ref{fig:leukemiaWard} shows the results of hierarchical clustering with Ward's linkage function. In this case, not scaling, scaling with the range and scaling with the pooled standard deviation perform equally well, misclassifying only 2 observations. Scaling with the standard deviation however splits the group of patients with AML in two and fails to recover the true clustering. Figure \ref{fig:leukemiaComplete} shows the results of clustering the leukemia data using complete linkage. While complete linkage is clearly less suited to cluster these data, there are still noteworthy differences between the different  scaling methods. No scaling gives 5 misclassified samples. Scaling with the range and pooled standard deviation gives 4 misclassified samples, whereas scaling with the standard deviation again messes up the clustering results. In summary, scaling with the pooled standard deviation or the range  yields the best results for this dataset.

\subsubsection{Colon cancer data}

We now analyze a dataset containing gene expressions from normal colon tissue samples as well as colon cancer samples. The data was collected from two datasets in the Gene Expression Omnibus \citep{GEO2002} with IDs GSE8671 and GSE4183, and is publicly available in the \texttt{R}-package \texttt{antiProfilesData} \citep{AP2019}. The probesets annotated to genes within blocks of hypomethylation in colon cancer defined in \cite{Hansen2011}.\\

\begin{figure}[!htb]
\centering
\begin{subfigure}[b]{0.49\linewidth}
  \centering 
	\includegraphics[width=0.99\textwidth]{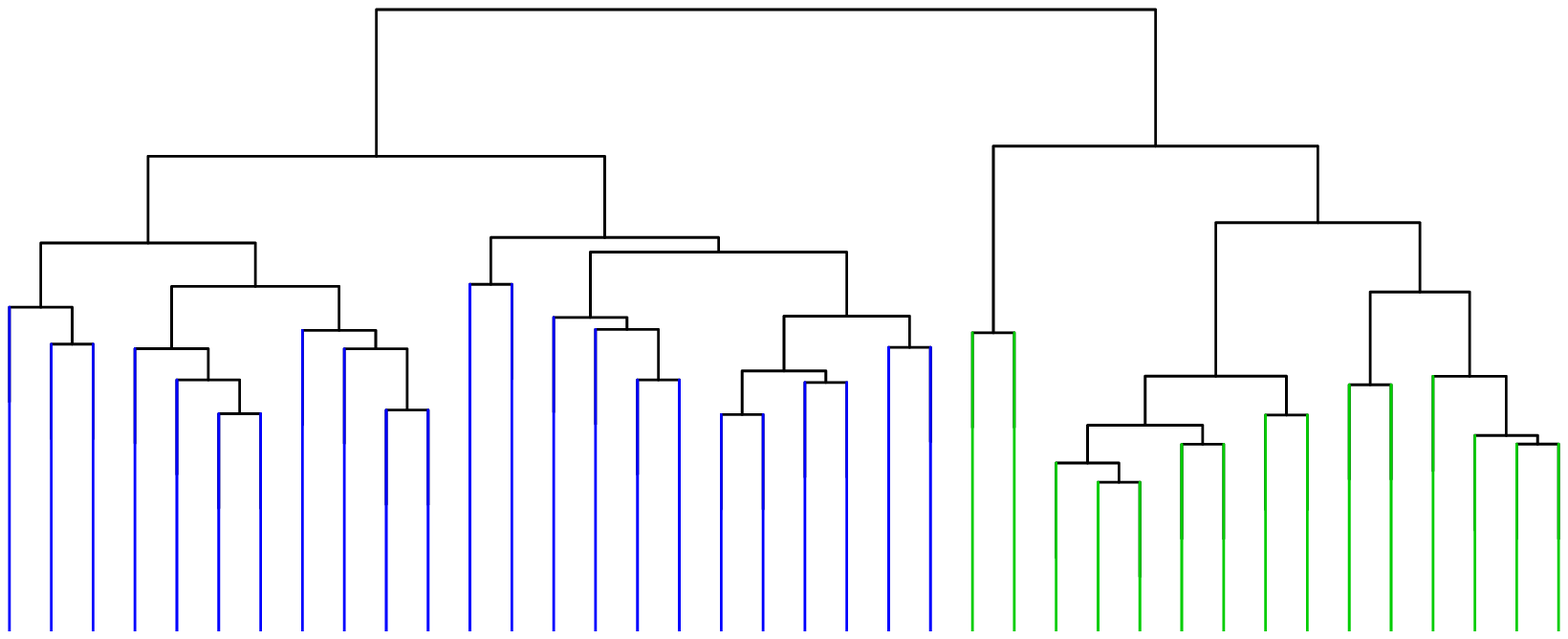}
	\caption{No scaling}
\vspace{0.25cm}
\end{subfigure}
\begin{subfigure}[b]{0.49\linewidth}
  \centering 
  \includegraphics[width=0.99\textwidth]{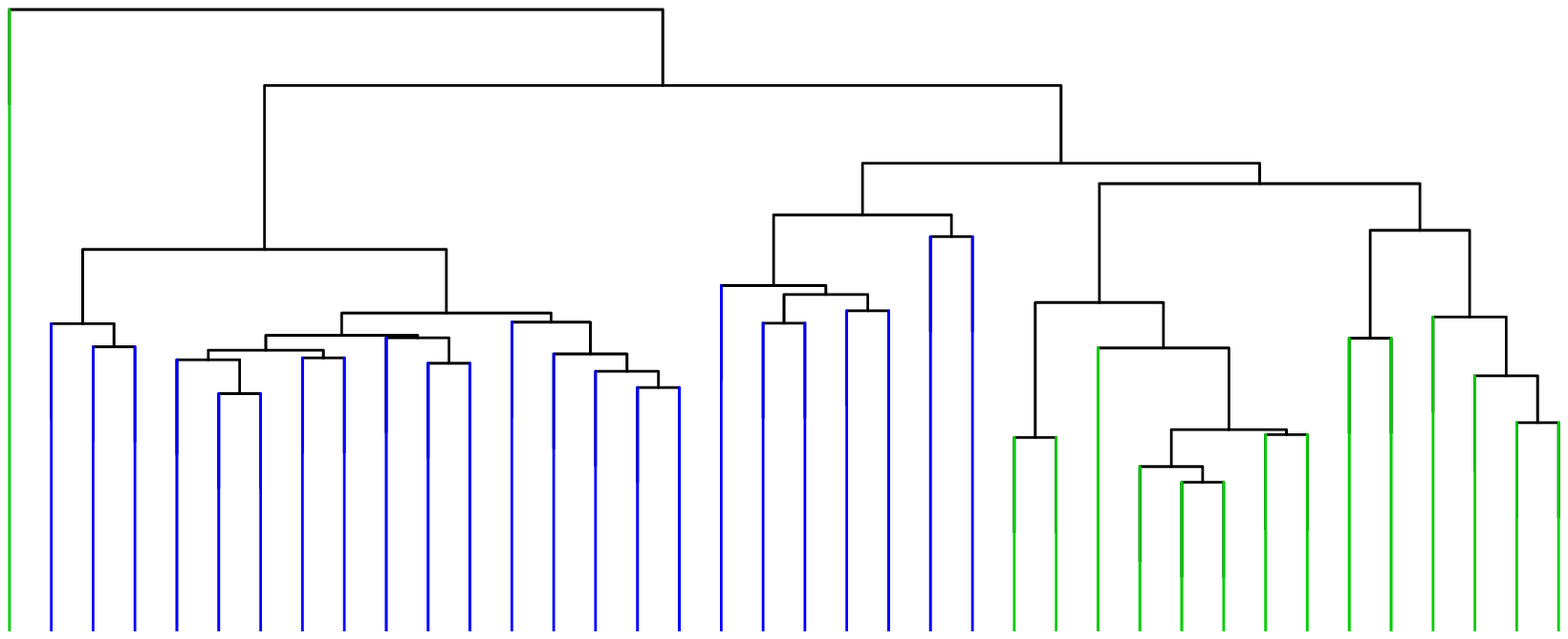} 
	  \caption{Standard deviation}
\vspace{0.25cm}
\end{subfigure}
\begin{subfigure}[b]{0.49\linewidth}
  \centering 
  \includegraphics[width=0.99\textwidth]{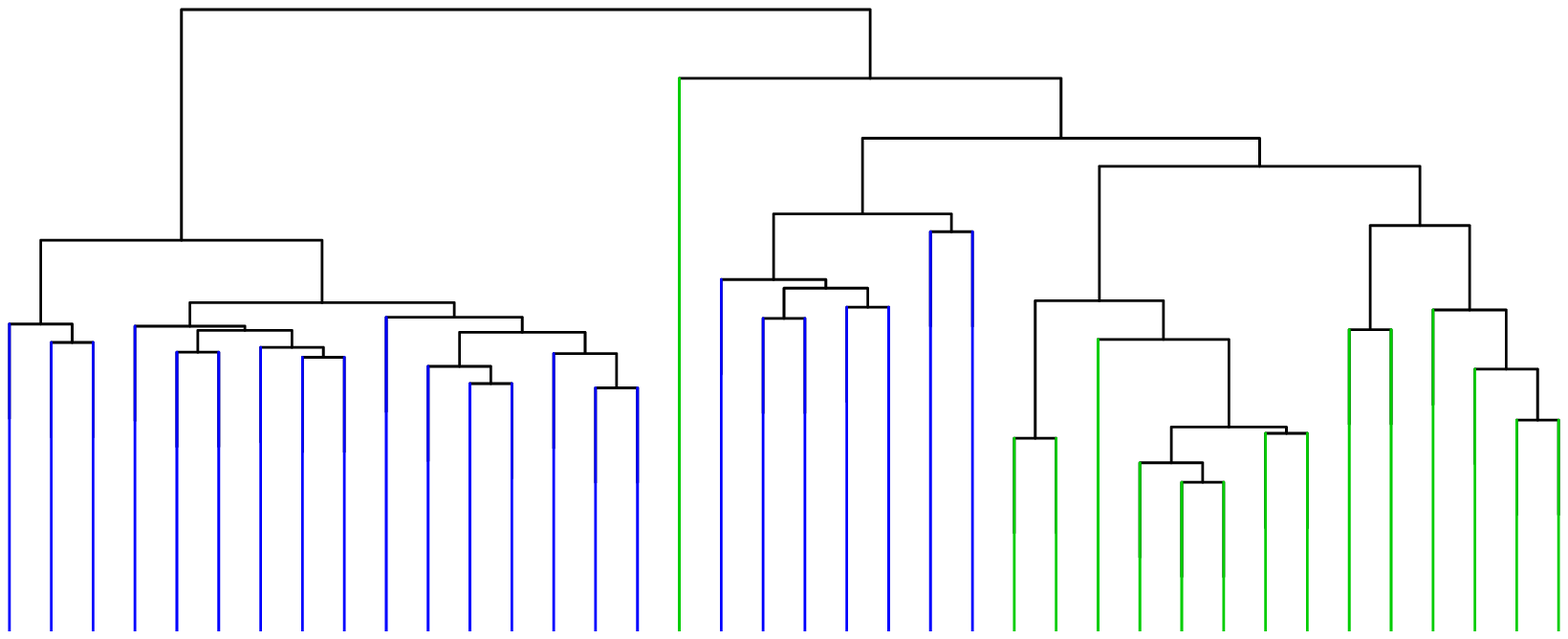} 
	  \caption{Range}
\end{subfigure}
\begin{subfigure}[b]{0.49\linewidth}
  \centering 
  \includegraphics[width=0.99\textwidth]{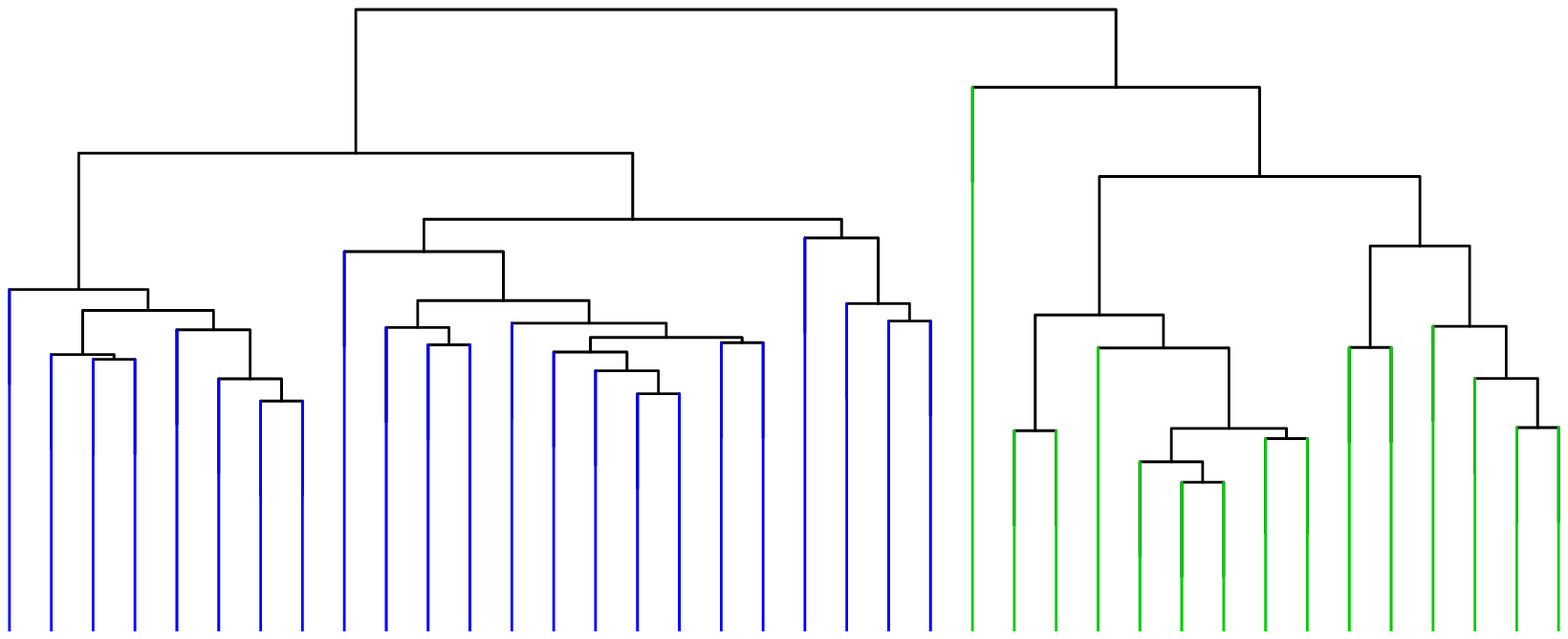} 
	  \caption{Pooled standard deviation}
\end{subfigure}
\caption{Hierarchical clustering with complete linkage on the colon cancer data. Scaling with the pooled standard deviation and not scaling at all yields the best recovery of the true clusters.}
\label{fig:colonComplete}
\end{figure}

The data contains the expression levels of 5339 genes for  15 healthy patients and 23 patients with a tumor. We cluster the patients using hierarchical clustering with complete linkage. 
Figure \ref{fig:colonComplete} shows the resulting dendrograms, indicating that a perfect clustering is achieved without scaling or when scaling using the pooled standard deviation. When scaling with the standard deviation or range however, the true clusters are mixed and the recovery is quite poor.

\begin{figure}[!htb]
\centering
\begin{subfigure}[b]{0.49\linewidth}
  \centering 
	\includegraphics[width=0.99\textwidth]{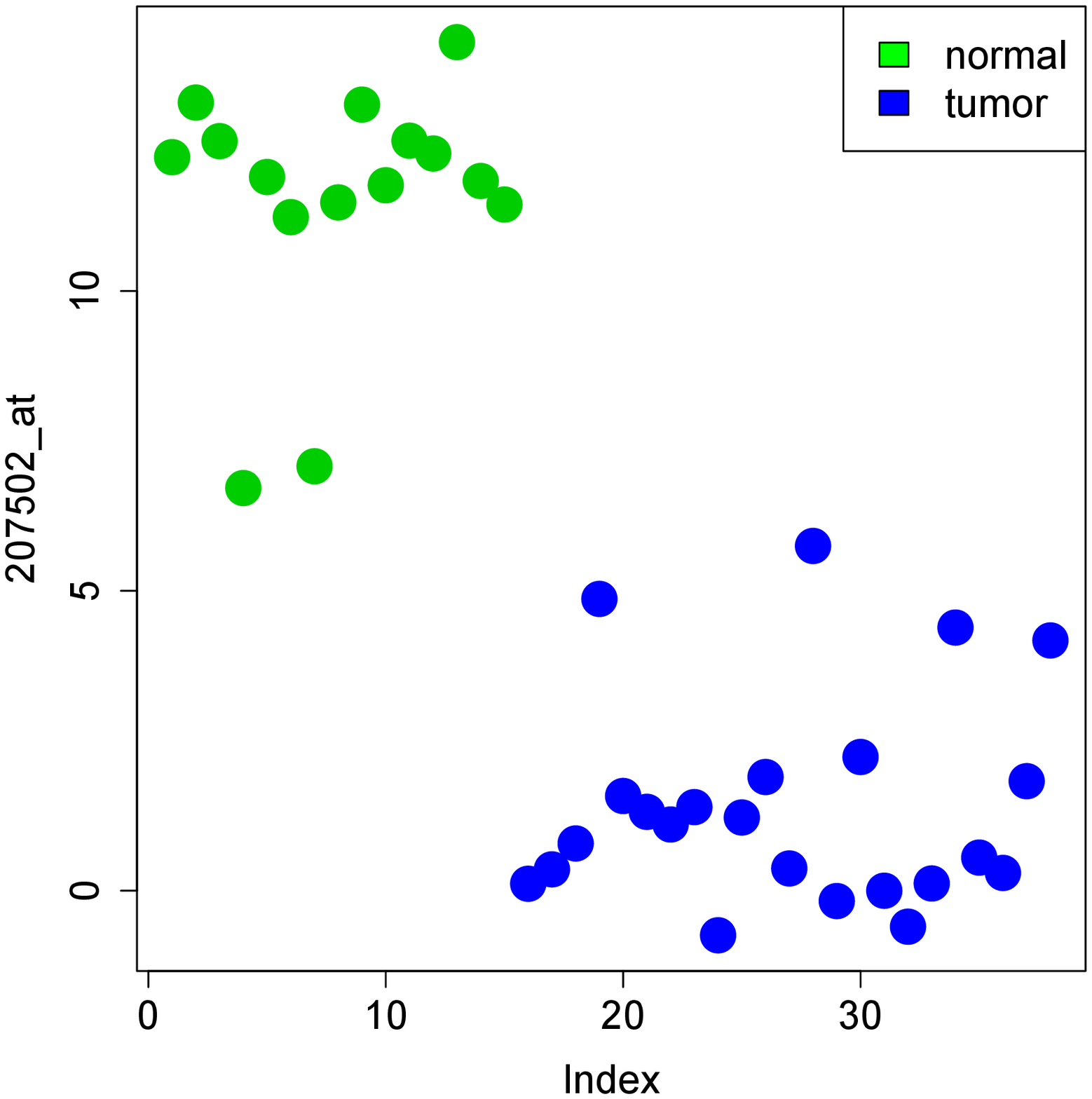}
\vspace{0.25cm}
\end{subfigure}
\begin{subfigure}[b]{0.49\linewidth}
  \centering 
  \includegraphics[width=0.99\textwidth]{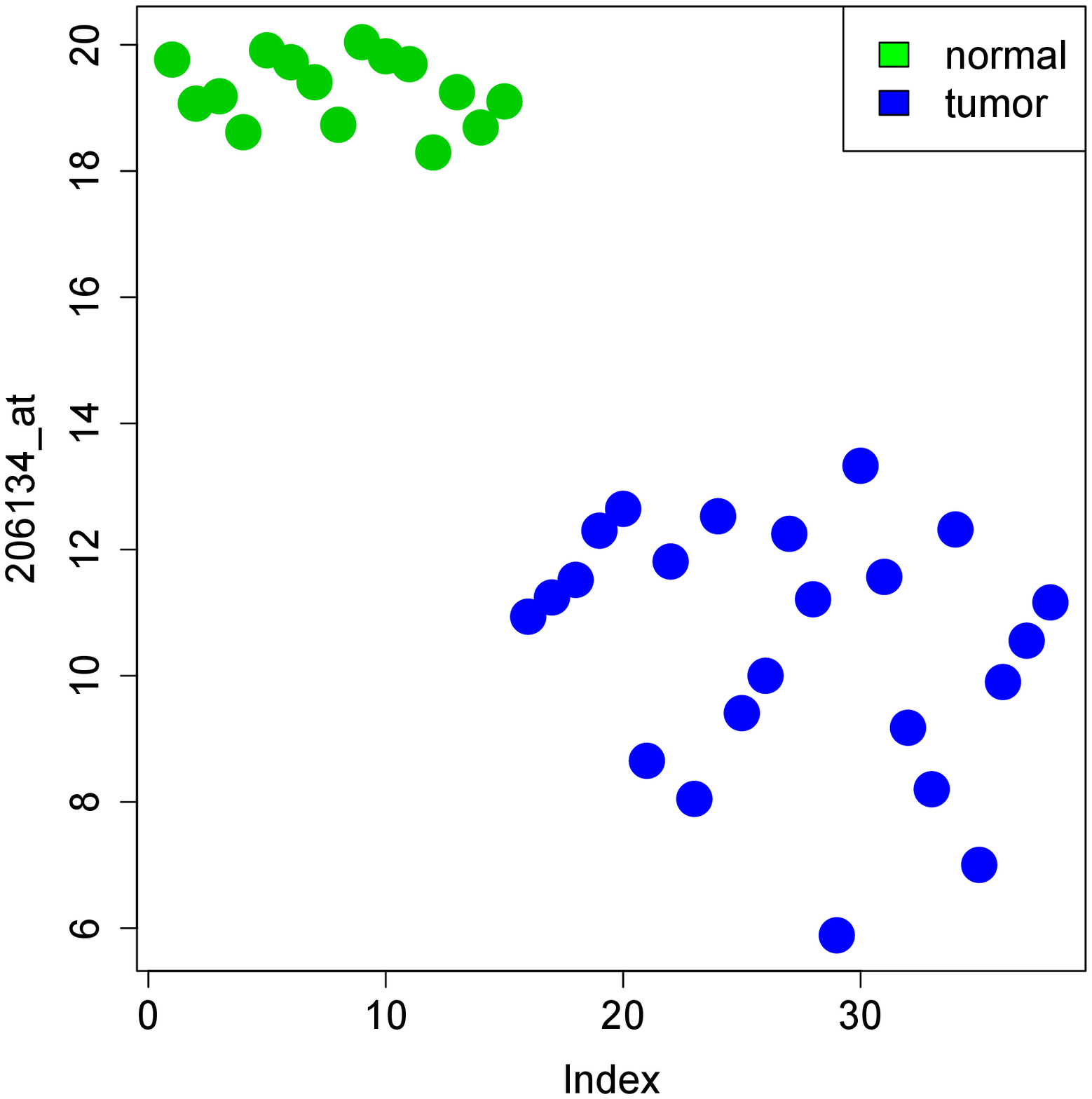} 
\vspace{0.25cm}
\end{subfigure}
\begin{subfigure}[b]{0.49\linewidth}
  \centering 
  \includegraphics[width=0.99\textwidth]{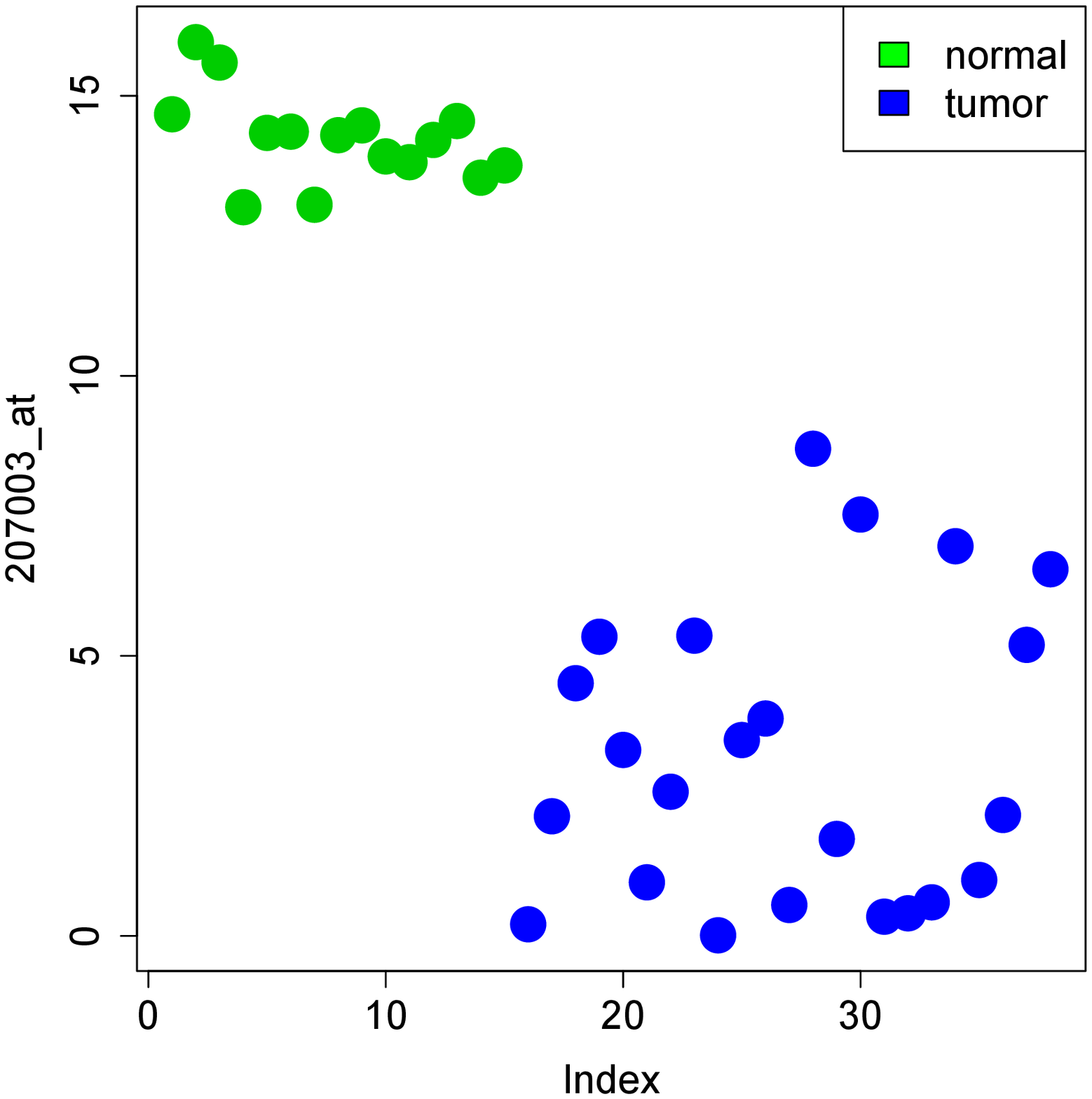} 
\end{subfigure}
\begin{subfigure}[b]{0.49\linewidth}
  \centering 
  \includegraphics[width=0.99\textwidth]{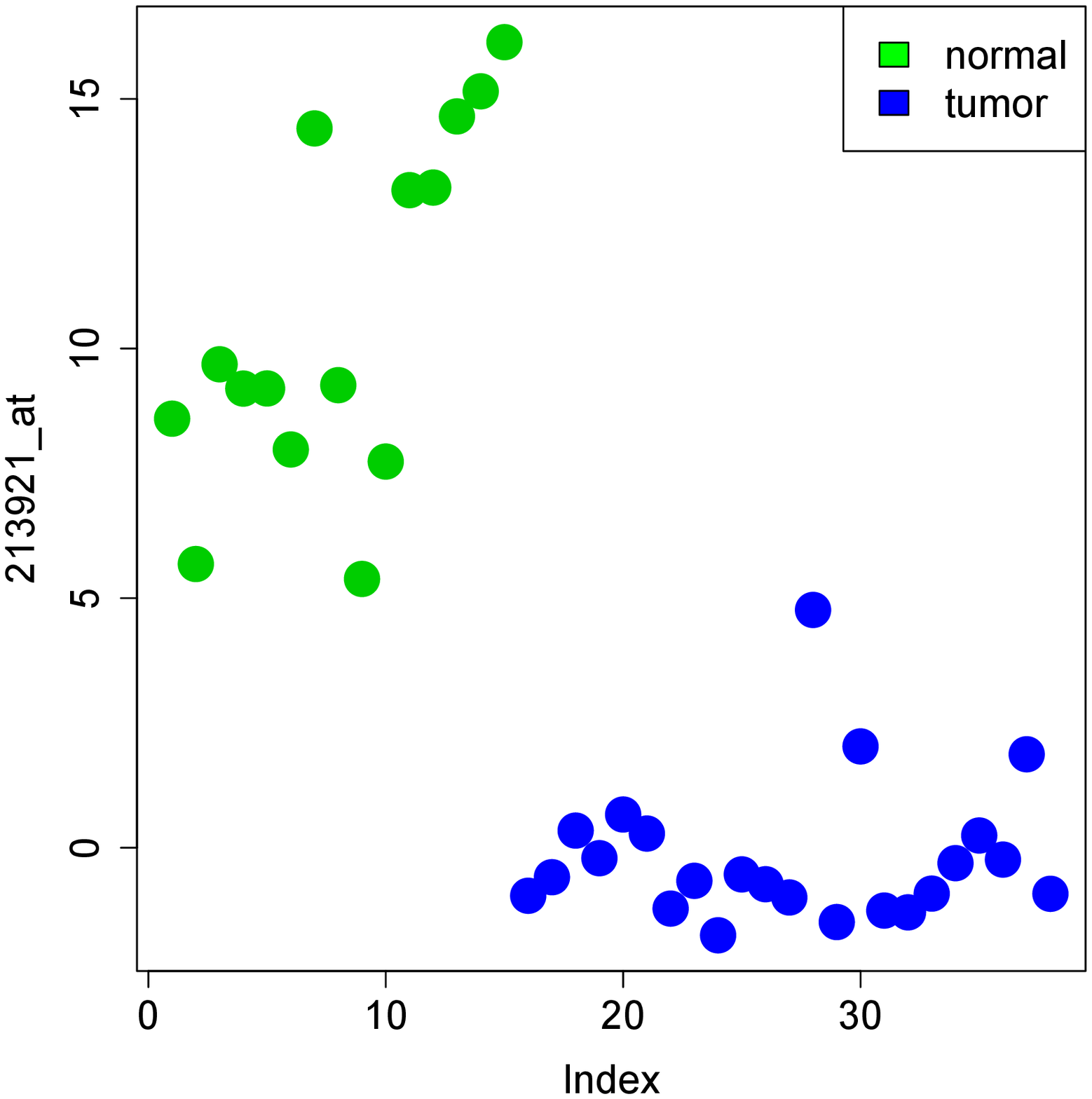} 
\end{subfigure}
\caption{The 4 genes in the colon data with the highest ratio of classical over pooled standard deviation. There is a clear separation between the healthy patients and those with a tumor.}
\label{fig:colonGenes}
\end{figure}

As an added feature of computing the pooled standard deviations, we consider the 4 genes with the highest ratio of classical over pooled standard deviation. The Affy ID's of these genes are \texttt{207502\_at}, \texttt{206134\_at}, \texttt{207003\_at} and \texttt{213921\_at} and they are displayed in Figure \ref{fig:colonGenes}. It is clear that all 4 of these genes clearly separate the patients with tumors from the healthy ones, indicating their biological importance.

\end{document}